\documentclass{emulateapj}
\usepackage{lscape}
\usepackage{amsmath}

\slugcomment{Accepted by ApJ}

\shorttitle{A backward evolution model for infrared surveys}
\shortauthors{Valiante et al.}

\begin{document}

\title{A backward evolution model for infrared surveys: the role of AGN- and Color-$L_{\rm TIR}$ distributions}


\author{E. Valiante\altaffilmark{1,2,\dag}, D. Lutz\altaffilmark{1},
E. Sturm\altaffilmark{1}, R. Genzel\altaffilmark{1}, E. Chapin\altaffilmark{2}}
\altaffiltext{1}{Max-Planck-Institut f\"ur extraterrestrische Physik, Postfach 1312, 85741 Garching, Germany}
\altaffiltext{2}{Dept. of Physics and Astronomy, Univ. of British Columbia, 6224 Agricultural Road, Vancouver, B.C., V6T 1Z1, Canada}
\altaffiltext{\dag}{\url{valiante@phas.ubc.ca}, \url{www.physics.ubc.ca/$\sim$valiante/model}}

\begin{abstract}

Empirical ``backward'' galaxy evolution models for infrared bright galaxies are constrained using multi-band infrared surveys. We developed a new Monte-Carlo algorithm for this task, implementing luminosity dependent distribution functions for the galaxies' infrared spectral energy distributions (SEDs) and for the AGN contribution, allowing for evolution of these quantities. The adopted SEDs take into account the contributions of both starbursts and AGN to the infrared emission, for the first time in a coherent treatment rather than invoking separate AGN and star-forming populations. 

In the first part of the paper we consider the quantification of the AGN contribution for local universe galaxies, as a function of total infrared luminosity. It is made using a large sample of LIRGs and ULIRGs for which mid-infrared spectra are available in the {\itshape Spitzer} archive. We find the ratio of AGN $6\,\mu\rm{m}$ luminosity and total infrared luminosity to rise with $L_{\rm TIR}^{1.4}$ over the infrared luminosity range $10^{11}$ to $10^{13}\,L_\odot$ and estimate its spread. Judging from the modest number of distant sources with {\itshape Spitzer} spectroscopy, the relation changes at high $z$.

In the second part we present the model. Our best-fit model adopts very strong luminosity evolution, $L=L_0(1+z)^{3.4}$, up to $z=2.3$, and density evolution, $\rho=\rho_0(1+z)^2$, up to $z=1$, for the population of infrared galaxies. At higher $z$, the evolution rates drop as $(1+z)^{-1}$ and $(1+z)^{-1.5}$ respectively. To reproduce mid-infrared to submillimeter number counts and redshift distributions, it is necessary to introduce both an evolution in the AGN contribution and an evolution in the luminosity-temperature relation. At a given total infrared luminosity, high redshift infrared galaxies have typically smaller AGN contributions to the rest frame mid-infrared, and colder far-infrared dust temperatures than locally. We also suggest an extension of the local infrared galaxy population towards lower dust temperatures. 

Our models are in plausible agreement with current photometry-based estimates of the typical AGN contribution as a function of mid-infrared flux, and well placed to be compared to upcoming {\itshape Spitzer} spectroscopic results. As an example of future applications, we use our best-fitting model to make predictions for surveys with {\itshape Herschel}.

\end{abstract}


\section{Introduction} \label{section:intro}

The discovery of the cosmic infrared background (CIB) (see \citealt{hauser01} for a review), together with recent deep cosmological surveys in the infrared (IR) and submillimeter bands, have offered new perspectives on our understanding of galaxy formation and evolution. The surprisingly large amount of energy contained in the CIB showed that it is crucial to probe its contributing galaxies to understand when and how the bulk of stars formed in the Universe. Thanks to the deep cosmological surveys carried out by ISO (\citealt{kessler96}, e.g.~\citealt{aussel99,oliver00}), SCUBA (\citealt{holland99}, e.g.~\citealt{hughes98}), MAMBO (\citealt{kreysa98}, e.g.~\citealt{bertoldi00,dannerbauer04}) and {\itshape Spitzer} (\citealt{werner04}, e.g.~\citealt{papovich04,frayer06,dole06}) it is now possible, to various degrees, to resolve the CIB into discrete sources. The source counts are high when compared to no-evolution or moderate-evolution models for infrared galaxies \citep{guiderdoni98, franceschini98}. The striking results of these surveys concerning the evolution of the infrared and submillimeter galaxy population, and the constraints from the measurements of the CIB, require the development of new models to explain the high rate of evolution of infrared galaxies.

The problem of describing the number distribution of galaxies in the Universe is usually tackled via one of two methods. In the first method, known as {\itshape forward evolution}, the calculation assumes some initial conditions and physical processes for chemical and photometric evolution of the stellar populations that heat the dust. The so-called {\itshape semi-analytical} approach combines these assumptions about the chemical/photometric evolution of galaxies with models for the dissipative and non-dissipative processes affecting galaxy formation within dark matter halos \citep{baugh06} and has provided a reasonable fit to the source counts in the infrared \citep{guiderdoni98,lacey08}. Arguably, the {\itshape forward evolution} approach has the advantage of being based on a more fundamental set of assumptions. However the obvious disadvantage is the large number of free/unknown parameters assumed in these models. Compared to other wavelengths, there is one further complication for forward modelling of infrared surveys. Even if global properties of the evolving galaxies like star formation rates or AGN activity are correctly modelled, additional assumptions about the dust content and structure of the galaxy have to be invoked to convert them into a prediction of the luminosity and spectral energy distribution of the re-emitted infrared radiation. 

The alternative method, often called the {\itshape backward evolution} approach, takes the observed, present day ($z=0$) luminosity function (LF) and evolves it in luminosity and/or density out to higher redshifts assuming some parametrization of the evolution (e.g.~\citealt{franceschini88,pearson96,xu01,franceschini01,rowanrobinson01,lagache03}). This method has the advantage of being both direct and relatively simple to implement. The disadvantage in the past was that the information on which the measurements of evolution were often made with IRAS or ISO data which extended out only to relatively low redshifts. Because of their simplicity, backward evolution models have traditionally played a strong role in the planning of new infrared surveys. A backward evolution model fitting the main constraints provided by previous missions can be easily modified to predict the results of new missions and help in the first steps of interpreting their results. Thanks to the latest deep and comprehensive observations cited above, the tools to constrain the  {\itshape backward evolution} methodology to significantly higher redshifts are now available.

Very recently, several empirical approaches have been proposed to model the high rate of evolution of infrared galaxies, in particular to reproduce source counts of the mid-IR surveys made with {\itshape Spitzer} (e.g.~\citealt{lagache03,lagache04,gruppioni05,rowanrobinson09}).

The ``classical'' backward evolution model starts from different populations of galaxies, typically cirrus, starburst, active galactic nuclei (AGN) and ultraluminous infrared galaxies (ULIRGs) (e.g.~\citealt{rowanrobinson01}), or a subset of those. Each population is  assigned a spectral energy distribution (SED) and local luminosity function, and evolved independently, assuming it has a proper cosmic evolution rate. These hypotheses are not always satisfied. First of all, the SEDs adopted for starburst galaxies are usually either represented by a single SED or come from a set of templates where a unique relation between temperature and luminosity is assumed. This aspect was already discussed by \citet{chapman03} and \citet{chapin09}: analyzing a sample of local infrared galaxies, they found that the luminosity-temperature ($L-T$) relation presents a significant spread. The latter, in particular, showed evidence for some evolution of the relation with redshift, comparing the local trend with high-redshift data from SHARC-II $350\,\mu\rm{m}$ observations of SMGs (e.g.~\citealt{kovacs06,coppin08}). Second, and perhaps more important, different populations of galaxies are not as distinct as assumed in the models: very often AGN and starbursts co-exist in the same object, can be predominant at different times depending of the evolutionary stage of the object itself and can influence each other (e.g.~AGN feedback on star formation).

Backward evolution models based on single SEDs or the simple SED family approach described above have been quite successful in fitting the number counts from the IRAS and ISO missions, the first submillimeter counts, the global CIB level, and for making predictions for {\itshape Spitzer} and {\itshape Herschel}. The number and quality of observed constraints is increasing, however, and already the first {\itshape Spitzer} results have led to on-the-spot modifications (c.f. the modified SEDs adopted by \citealt{lagache04}) that may either genuinely represent improved knowledge, or reflect limitations of the simple assumptions made previously. Furthermore, questions gaining increased importance, like the co-existence of AGN and star-formation in infrared galaxies, cannot be addressed by this generation of models, not even in the simple sense of fitting existing data and extrapolating to new observations. The goal of this work is hence to take the next step and develop a backward evolution model that considers realistic spreads in far-infrared SEDs, in AGN contributions at different luminosities, and their possible evolution with redshift. To that end, we first have to describe the local situation that is the starting point of the backward evolution scenario.

Luminous (LIRGs: $10^{11}<L_{\rm TIR}\equiv L_{8-1000\mu{\rm m}}<10^{12}\,L_\odot$) and ultraluminous (ULIRGs: $L_{\rm TIR}>10^{12}\,L_\odot$) infrared galaxies \citep{sanders88} have been studied extensively in the local Universe, for instance  with the {\itshape Infrared Astronomical Satellite} (IRAS; e.g.~\citealt{soifer87,saunders90}), the {\itshape Infrared Space Observatory} (ISO; e.g.~\citealt{lutz98,genzel00,tran01}), and more recently, with the {\itshape Infrared Spectrograph} (IRS; \citealt{houck04}) on board of \textit{Spitzer} (e.g.~\citealt{weedman05,brandl06,armus07,desai07,veilleux09}). These galaxies exhibit a large range of properties in the mid-IR, some showing strong PAH emission features characteristic of powerful (up to $\approx 1000\,M_{\odot}{\rm yr}^{-1}$) star formation rates (e.g.~\citealt{brandl06,smith07}), and exhibiting a large range in $9.7\,\mu{\rm m}$ silicate absorption or emission strengths (e.g.~\citealt{weedman05,desai07,imanishi07}). \textit{Spitzer} IRS is now enabling the study of mid-IR spectra of infrared galaxies to much higher redshifts ($z\gtrsim 2$; e.g.~\citealt{houck05,yan07,menendez07,valiante07,pope08}). Although locally rare, infrared galaxies are an important population at high redshifts and account for an increasing fraction of the star-formation activity in the universe \citep{lefloch05}. By studying their infrared properties, we are just starting to estimate the extent to which AGN and star-formation contribute to their infrared luminosities, and therefore determine a correct census of starbursts and AGN at epochs in the Universe when their luminosity density was at its maximum.

The local luminosity function of IR-bright galaxies can be described considering two different populations: normal galaxies, dominating the ``low-luminosity'' part of LF, and starburst galaxies, dominant in the ``high-luminosity'' part. The AGN contribution appears dominant only at very high luminosity ($L_{\rm{TIR}}\gtrsim 2\times 10^{12}L_{\odot}$). Nevertheless, a small fraction of the total infrared luminosity can be due to the presence of an active galactic nucleus even in star formation dominated cases, and vice versa \citep{genzel98}. There are several studies of the AGN content for luminous galaxies at ULIRG and HYLIRG (hyperluminous infrared galaxies, $L_{\rm TIR}>10^{13}\,L_\odot$) levels, both using X-ray \citep{franceschini03,teng05} and mid-IR (e.g.~\citealt{genzel98,lutz98,tran01}) emission, and it is generally believed that AGN are typically less important at lower infrared luminosities. Still missing is a comprehensive study including lower luminosities ($<10^{12}\,L_{\odot}$), with the aim of quantifying the distribution of AGN luminosity respect to the infrared luminosity of the host and the fraction of infrared luminosity due to accretion. Such a study is made in the following section (see \S~\ref{section:AGN}).

This work is organized as follows: in the first part (\S~\ref{section:AGN}), starting from an IRAS-based sample of LIRGs and ULIRGs observed with IRS, we derive local distribution functions for the AGN contribution at different infrared luminosities. Using a small sample of distant galaxies, we also try to explore this relation at higher $z$. In the second part (\S~\ref{section_model}), we present a new backward evolution model whose new element is to treat infrared galaxies as a single population with a single local luminosity function, but with realistic spreads in both SED properties and AGN content. An SED including the contribution due to starbursts and AGN is associated with each source, following the relations derived in the first part for local and distant objects. This model, besides reproducing existing source counts, redshift distributions and CIB intensity, is also able for the first time to quantify the contribution due to starbursts and/or AGN to the total infrared luminosity and how this contribution evolves for different luminosity classes.

\section{The AGN contribution in local ULIRGs and LIRGs} \label{section:AGN}

In order to be useful for the backward modelling project outlined above, a local calibration of the distribution of AGN contributions at given infrared luminosity has to span the highest luminosities down to at least the low end of the LIRG regime ($10^{11}L_{\odot}$). Given the evidence for the major contribution of objects above this luminosity threshold to the CIB and to the cosmic star formation rate at $z\gtrsim 1$ (e.g.~\citealt{elbaz02,perez05}), covering this luminosity range is essential. Historically, many of the references cited above have focussed on the ULIRG ($>10^{12}L_{\odot}$) regime and even its upper end where the AGN contribution is highest, while studies at lower luminosities mostly focussed on individual interesting objects. What is needed is a quantification of AGN content for an unbiased far-IR selected sample reaching down to $10^{11}L_{\odot}$. Of the two principle routes towards this observational goal, X-ray observations and mid-IR spectroscopy, we make use of the second for two reasons. First, there is not yet a full unbiased and deep X-ray dataset for such a sample available. There are XMM and Chandra data of the depth required for good X-ray spectral analysis for many local LIRGs/ULIRGs, but still with a tendency to target known AGN. Second, since the implicit goal of our modelling effort is to characterize the AGN effect on the infrared SED, mid-IR data can provide useful constraints quite directly even when adopting simple analysis methods, while evidence from other wavelengths would have to go through a conversion to the mid-IR range, or even in two steps via the AGN bolometric power, a process that will be subject to the considerable scatter of AGN SEDs.

Our basic approach is to determine the distribution of AGN contributions for local LIRGs and ULIRGs from the IRAS Revised Bright Galaxy Sample \citep{sanders03} for which archival Spitzer spectroscopic coverage is essentially complete. We adopt a simple quantification of the AGN contribution from the rest frame $6\,\mu{\rm m}$ continuum which, while not making use of the full details of the best S/N spectra, is applicable to the entire sample and allows for straightforward use in the later modelling.

In mid-IR spectra of AGN, the $6\,\mu{\rm m}$ flux after subtraction of the starbursts contribution ($f_{6\rm{AGN}}$), is related to the nuclear activity, as indicated by the intrinsic hard X-ray flux \citep{lutz04}. Hard X-rays, unless extremely obscured in fully Compton-thick objects, can provide a direct view to the central activity, while the infrared continuum is due to AGN emission reprocessed by dust, either in the torus or on somewhat larger scales. The mid-IR continuum has the advantage of showing no significant differences between type 1 and type 2 AGN and of being a good tracer of nuclear activity even in those cases where hard X-rays are strongly absorbed. Low resolution mid-IR spectra of galaxies can be decomposed into three components \citep{laurent00}: a component dominated by the aromatic ``PAH'' features arising from photodissociation regions or from the diffuse interstellar medium of the host, a continuum rising steeply toward wavelengths beyond $10\,\mu{\rm m}$ due to HII regions and a typically flatter thermal AGN dust continuum present in active galaxies. 

In the range covered by the low resolution modules of IRS (SL1 and SL2, from $\sim$5 to $\sim$14$\,\mu{\rm m}$ as described in \citealt{houck04}), the AGN emission is most easily isolated shortwards of the complex of aromatic emission features \citep{laurent00}. We determine the continuum at $6\,\mu{\rm m}$ rest wavelength and eliminate non-AGN emission by subtracting a star formation template scaled with the strength of the aromatic ``PAH'' features arising from the host or from circumnuclear star formation. This method does not require spatially resolving the AGN from the host and, thanks to the high sensitivity of IRS, allows the detection of a weak AGN in the presence of strong star formation in many nearby galaxies.

For the purposes of this work, the use of mid-IR spectra, in particular of $f_{6\rm{AGN}}$, is the best way to quantify the AGN fraction for several reasons: (1) in order to derive a local luminosity function, we need a large sample: the number of objects for which mid-IR spectra are available is much greater than the number of sources with X-ray data of sufficient quality; (2) we are interested in the contribution of AGN to the infrared emission: $L_{6\rm{AGN}}$ is more strongly correlated with the AGN's infrared emission than, for example, $L_{2-10\,\rm{KeV}}$; (3) $f_{6\rm{AGN}}$ is the only consistent way to constrain the AGN contribution in our comparison sample at high redshifts, due to the limitations (in wavelength coverage and S/N) of the high-z spectra; (4) as explained above, $f_{6\rm{AGN}}$ is also sensitive to Compton-thick AGN, not detected in the X-rays: these kinds of galaxies become increasingly important at high-$z$, as shown both in mid-IR \citep{daddi07} and X-ray \citep{gilli07} surveys. 

\subsection{Sample selection and data reduction} \label{section:AGNdatared}





The starting  point for this study is the IRAS Revised Bright Galaxy Sample (RBGS), a complete flux-limited survey of all extragalactic objects with $60\,\mu{\rm m}$ flux $f_{60}>5.24\,{\rm Jy}$ \citep{sanders03}. The galaxies with $L_{\rm TIR}>10^{11}\,L_{\odot}$ were selected from this sample and the mid-IR IRS data were obtained from the \textit{Spitzer} archive for all of the them, with the exception of 6 sources that were not yet observed or for which there were no observation programs at all. Most of our observations have been obtained within the program PID 30323 (PI L.Armus).  A comprehensive study of LIRGs and ULIRGs is being carried on by the Great Observatory All-sky LIRG Survey (GOALS\footnote{http://goals.ipac.caltech.edu/}) and a more complete approach to derive AGN contributions from mid-IR spectra will be presented in forthcoming studies by the GOALS team. Additional data have been retrieved from other {\itshape Spitzer} programs\footnote{PIDs 14, 105, 666 (PI J.R.Houck), PID 1096 (PI H.Spoon), PID 3237 (PI E.Sturm), PID 3247 (PI C.Struck), PID 20549 (PI R.Joseph), PID 20589 (PI C.Leitherer)}.

Because of the limit in flux, this sample misses the most luminous infrared galaxies that are rare in the local universe. In fact, it does not include any galaxy with $L>10^{12.5}L_{\odot}$, excluding all HYLIRGs, for example, from the analysis. In order to enlarge our range in luminosity, in particular including representatives of the most luminous tail of the local galaxy population, we added {\itshape Spitzer} spectra for 14 galaxies from the sample of 16 by \citet{tran01}. These latter galaxies were selected from surveys at different limits in $f_{60}$ and with a preference for high-luminosity targets. They do not introduce any bias in mid-IR spectral properties or AGN content. In particular, no AGN-related IRAS color criteria, like the $f_{25}/f_{60}$ ratio, were applied in the selection.

\begin{figure}[p!]
\centering

\includegraphics[width=13.cm,height=9.3cm]{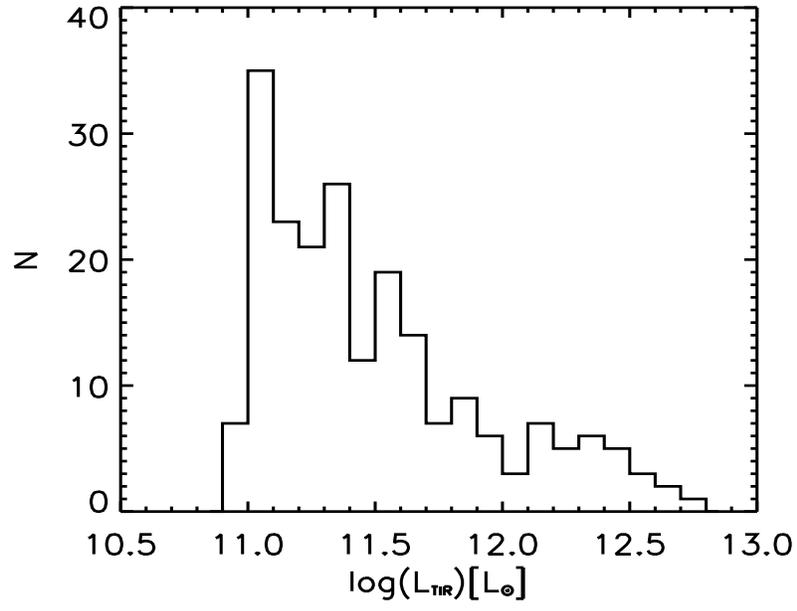}
\caption{Total infrared luminosity distribution of the galaxies of our sample, including galaxies from RBGS catalog \citep{sanders03} and sources from \citet{tran01}.
}
\label{RBGShist}
\end{figure}

The total sample includes 211 sources spanning a luminosity range $10^{10.9}\leq L_{\rm TIR}\leq 10^{12.8}\,L_{\odot}$ (see Fig.~\ref{RBGShist} and Tab.~\ref{AGNdata}).

We reduced the data as follows. We subtracted, for each cycle, the two nod positions of the basic calibrated data (BCD) frames. In the difference just calculated, we replaced deviant pixels by values representative of their spectral neighbourhoods. We subtracted residual wavelength dependent backgrounds, measured in source-free regions of the two dimensional
difference spectra. In averaging all the cycles of the 2-dimensional subtracted frames, we excluded values more than three times the local noise away from the mean. The calibrated 1-dimensional spectra for the positive and the negative beams were extracted and the two 1-dimensional spectra averaged in order to obtain the final spectrum, using the SMART analysis package \citep{higdon04}.

\begin{figure}[p!]
\centering

{\includegraphics[width=6.5cm,keepaspectratio]{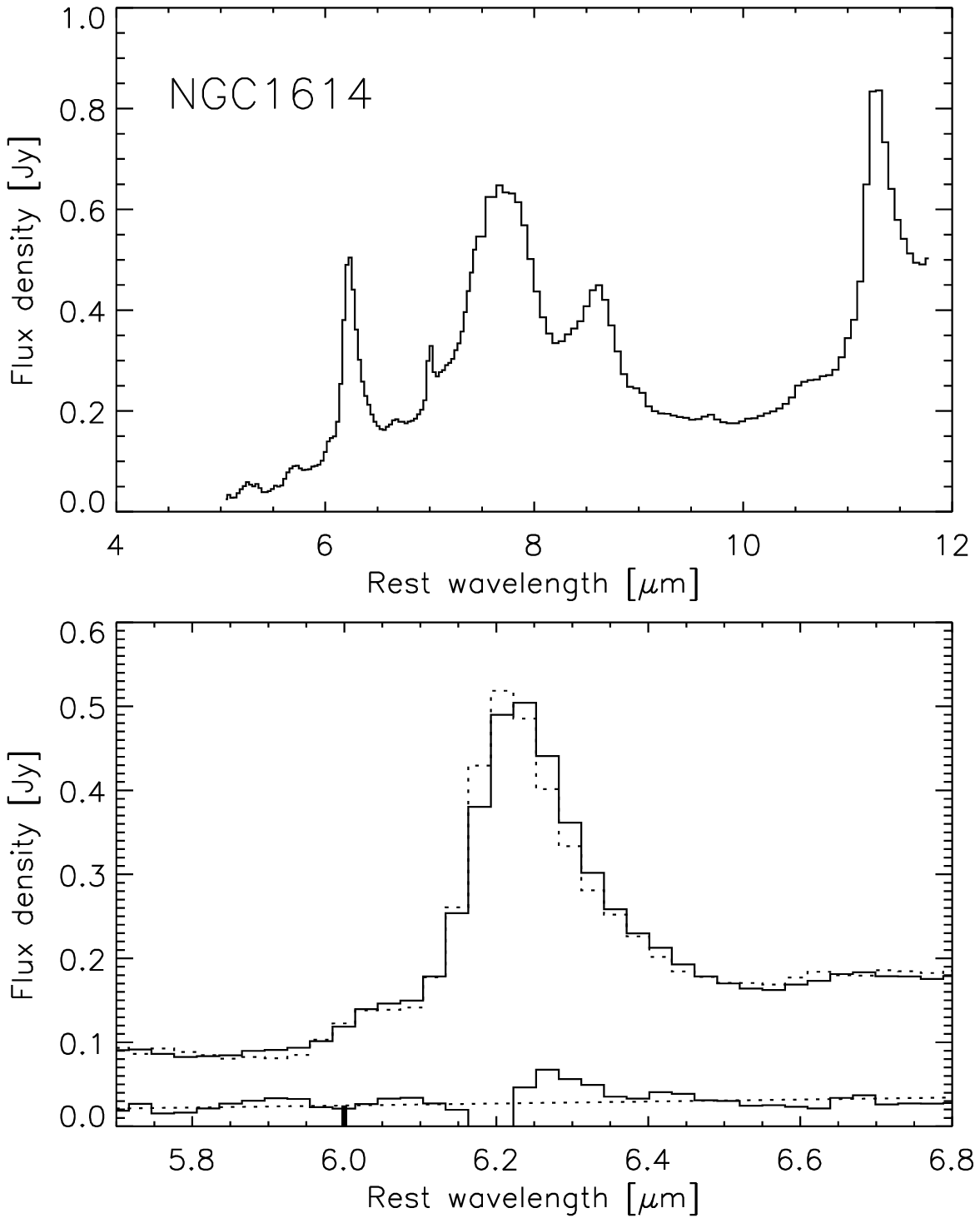}}
{\includegraphics[width=6.5cm,keepaspectratio]{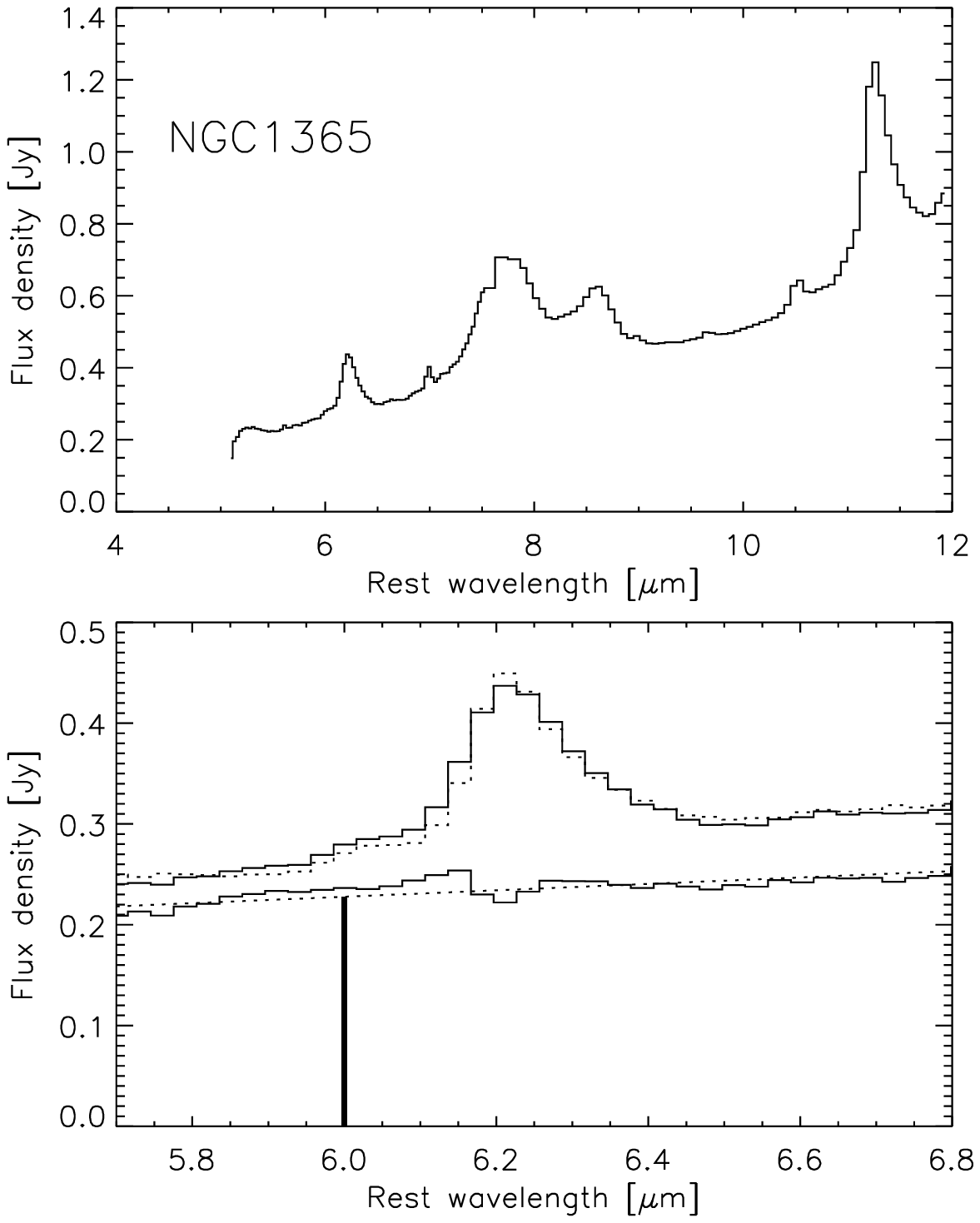}}\\[10mm]
{\includegraphics[width=6.5cm,keepaspectratio]{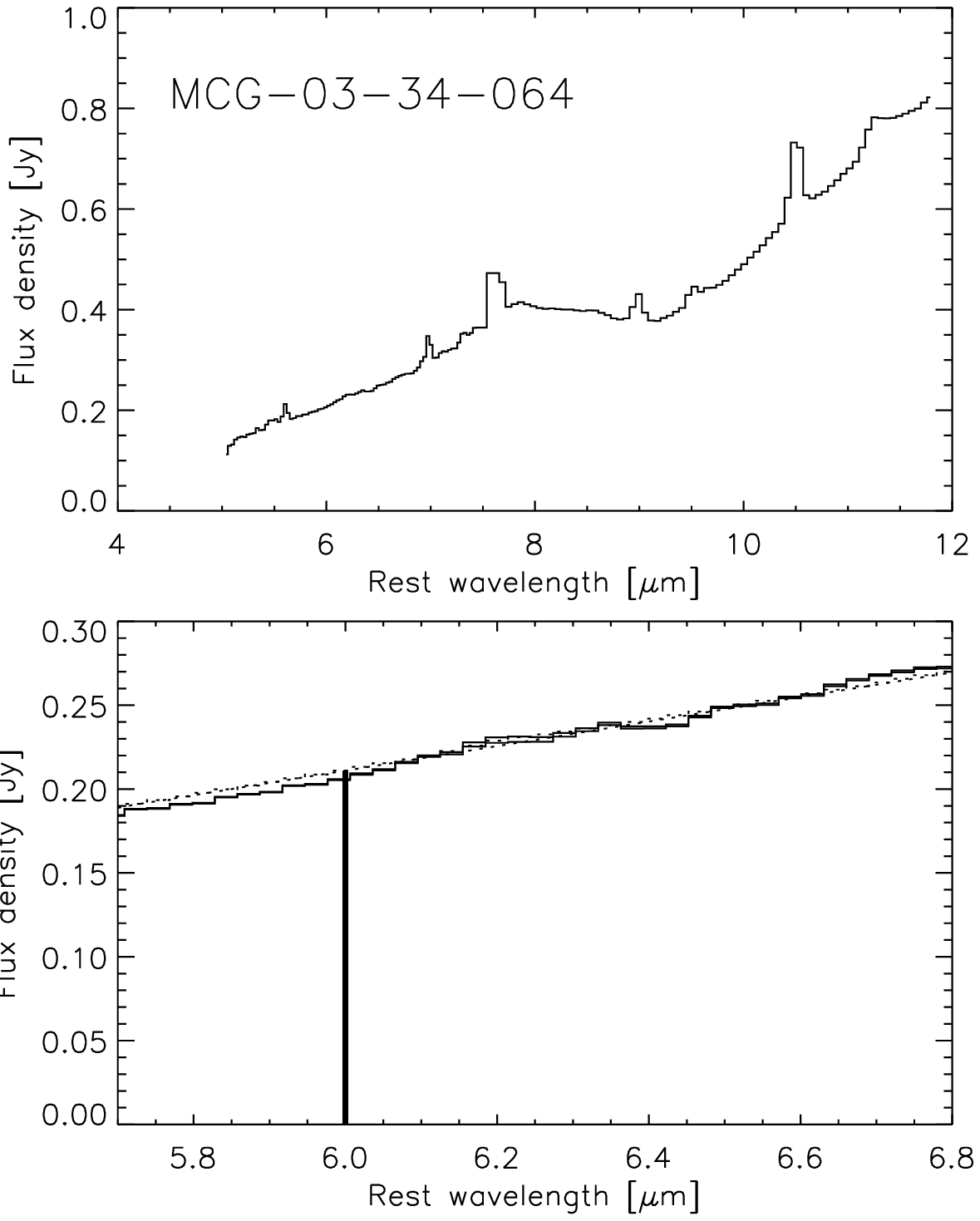}}
{\includegraphics[width=6.5cm,keepaspectratio]{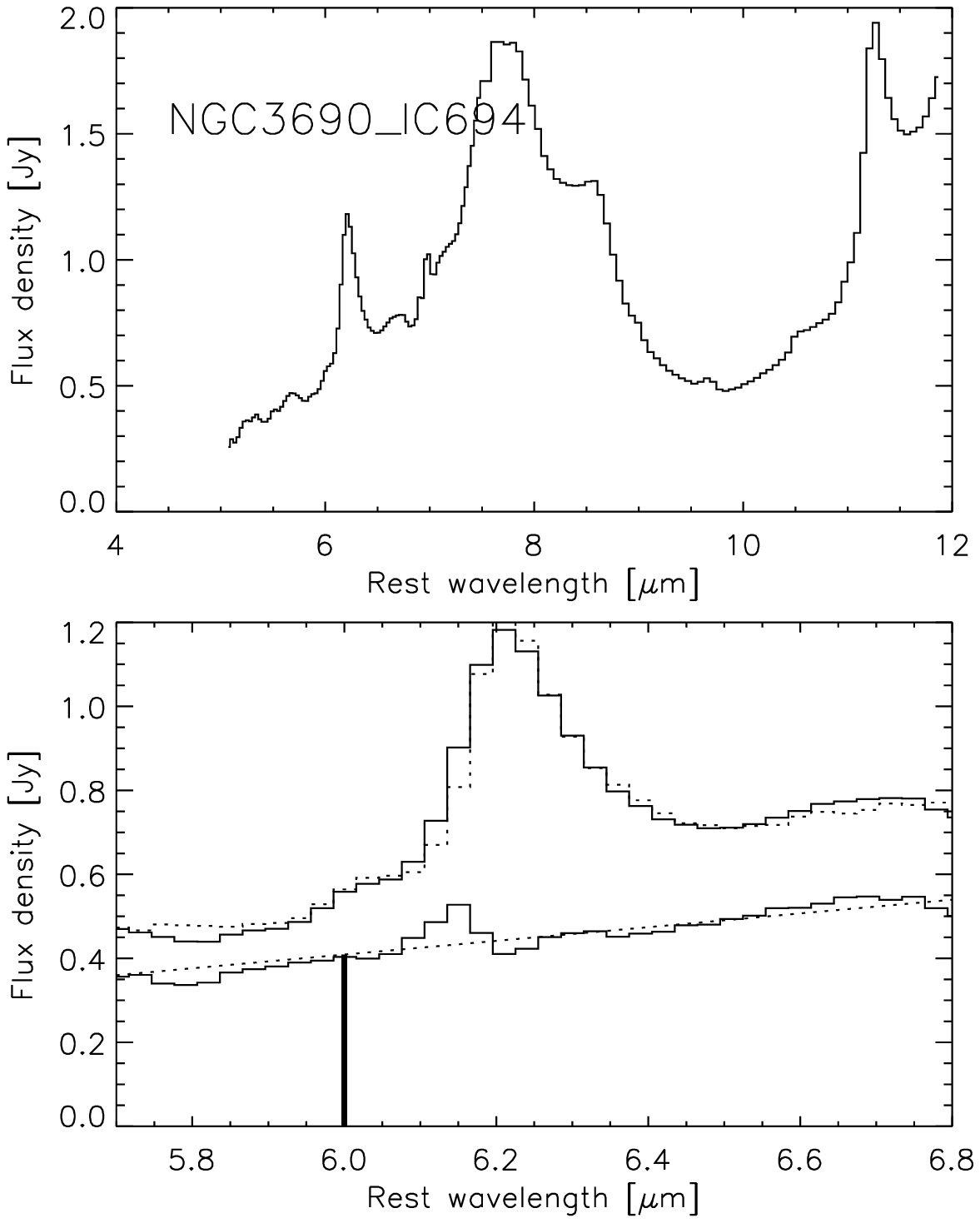}}
\caption{Examples for the decomposition used to isolate the AGN continuum. \textit{Top panels}: \textit{Spitzer} IRS low resolution spectra of four representative galaxies. \textit{Bottom panels}: Cutout of the region around the $6.2\,\mu{\rm m}$ PAH feature. Top continuous line~=~observed spectrum. Top dotted line~=~fit by the sum of the M82 spectrum and a linear AGN continuum. Bottom dotted line~=~fitted AGN continuum. Bottom continuous line~=~difference of observed spectrum and fitted PAH component. The thick vertical line indicates the $6\,\mu\rm{m}$ AGN continuum flux density $\tilde{f}_{6\rm{AGN}}$.}
\label{decomposition}
\end{figure}

We derive $f_{6\rm{AGN}}$ by a decomposition over a range from $5.5\,\mu{\rm m}$ to $6.85\,\mu{\rm m}$ rest wavelength. We fit the spectrum by the superposition of a star formation component dominated by the $6.2\,\mu{\rm m}$ PAH feature (M82 spectrum by \citealt{sturm00}) and a simple linear approximation of the AGN continuum. Fig.~\ref{decomposition} shows the decomposition of the spectrum for sources with different nature: NGC1614 is a pure starburst; NGC1365 is classified as a Seyfert 1.8 from its optical spectrum, has a Compton-thin AGN observed in the X-ray emission \citep{risaliti00} but shows also PAH features in the mid-IR spectrum; MCG-03-34-064 has a typical AGN-dominated mid-IR spectrum; the galaxy pair NGC3690/IC694 (Arp299) is classified as Compton-thick AGN from its X-ray spectrum \citep{dellaceca02} and its mid-IR spectrum reveals that the nuclear activity is quite high but coincident with strong star formation.

The results of the fitting procedure are $\tilde{f}_{6\rm{AGN}}$, represented in Fig.~\ref{decomposition} by the thick vertical line, and $\tilde{f}_{6.2{\rm peakSB}}$, an average flux density of the PAH component over the rest wavelength range $6.1$ to $6.35\,\mu{\rm m}$ (see Tab.~\ref{AGNdata}). A more comprehensive and in-depth study of the PAH features of this sample will be presented in future papers by the GOALS team. The spectrum of M82 is technically well suited for this decomposition but represents just a single object that may not be representative for star forming objects in general. We deal with this using the results and approach of \citet{lutz04} (their section 2.1). Specifically, we first use the M82 SWS spectrum for decomposition and then correct for systematic differences between M82 and a sample of 11 starbursts. To take into account the dispersion among star-forming objects in relative importance of all the components in the model of \citet{laurent00}, \citet{lutz04} have decomposed this small sample of 11 star-forming objects using the same approach and same M82 template. They find a mean residual of $0.096\times\tilde{f}_{6.2{\rm peakSB}}$ with a dispersion of 0.085. This corresponds to a a $6\,\mu\rm{m}$ continuum in M82 that is fainter than in other starbursts.  This finding agrees with the results of \citet{brandl06} who find more $6\,\mu\rm{m}$ continuum in their {\itshape Spitzer} starburst template composed of 13 sources than in the M82 SWS spectrum. Decomposition of the \citet{brandl06} template using our approach and correction for M82 systematics in the way described results in an insignificant residual continuum of less than 0.5\% of the 6.2$\mu$m PAH peak, showing that also quantitatively the agreement between their template and our approach is good. Tab.~\ref{AGNdata} implements our approach in the following way: the values $f_{6\rm{AGN}}$ used for all further analysis are obtained after subtracting $0.096\times\tilde{f}_{6.2{\rm peakSB}}$ from the direct fit result $\tilde{f}_{6\rm{AGN}}$. They are thus corrected for the slight systematic difference between M82 and other starbursts. Error estimates for $f_{6\rm{AGN}}$ are the quadratic sum of two components. The first is a measurement error based on individual pixel noise derived from the dispersion in the difference of observation and fit. The second is $0.085\times\tilde{f}_{6.2{\rm peakSB}}$, thus considering the dispersion in the properties of the comparison star forming galaxies.

In summary, AGN $6\,\mu{\rm m}$ continua obtained by our decomposition, are thus consistent with those obtained using \citet{brandl06} starburst template, and in addition include an estimate of the uncertainty arising from the spread of starburst properties.

\subsection{Distribution of the AGN contribution}

Next we study the AGN contribution to $L_{\rm TIR}$ in ULIRGs and LIRGs as a function of luminosity. Our results will be used in the modeling of \S~\ref{section_model} to build composite SEDs taking into account the two main energy sources (starbursts and AGN) of infrared galaxies.

\begin{figure}[p!]
\centering

{\includegraphics[width=6.5cm,height=5.3cm]{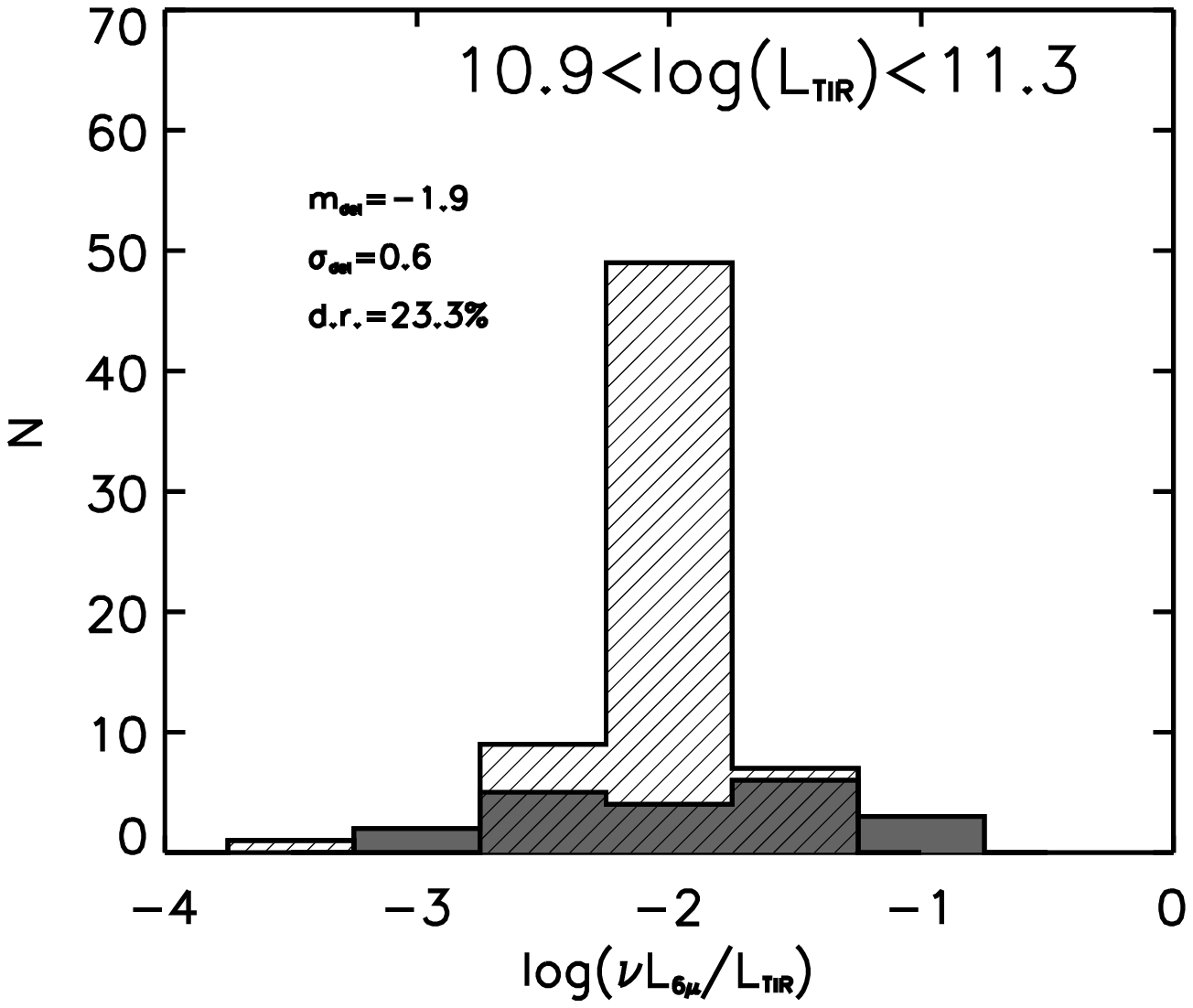}}
{\includegraphics[width=6.5cm,height=5.3cm]{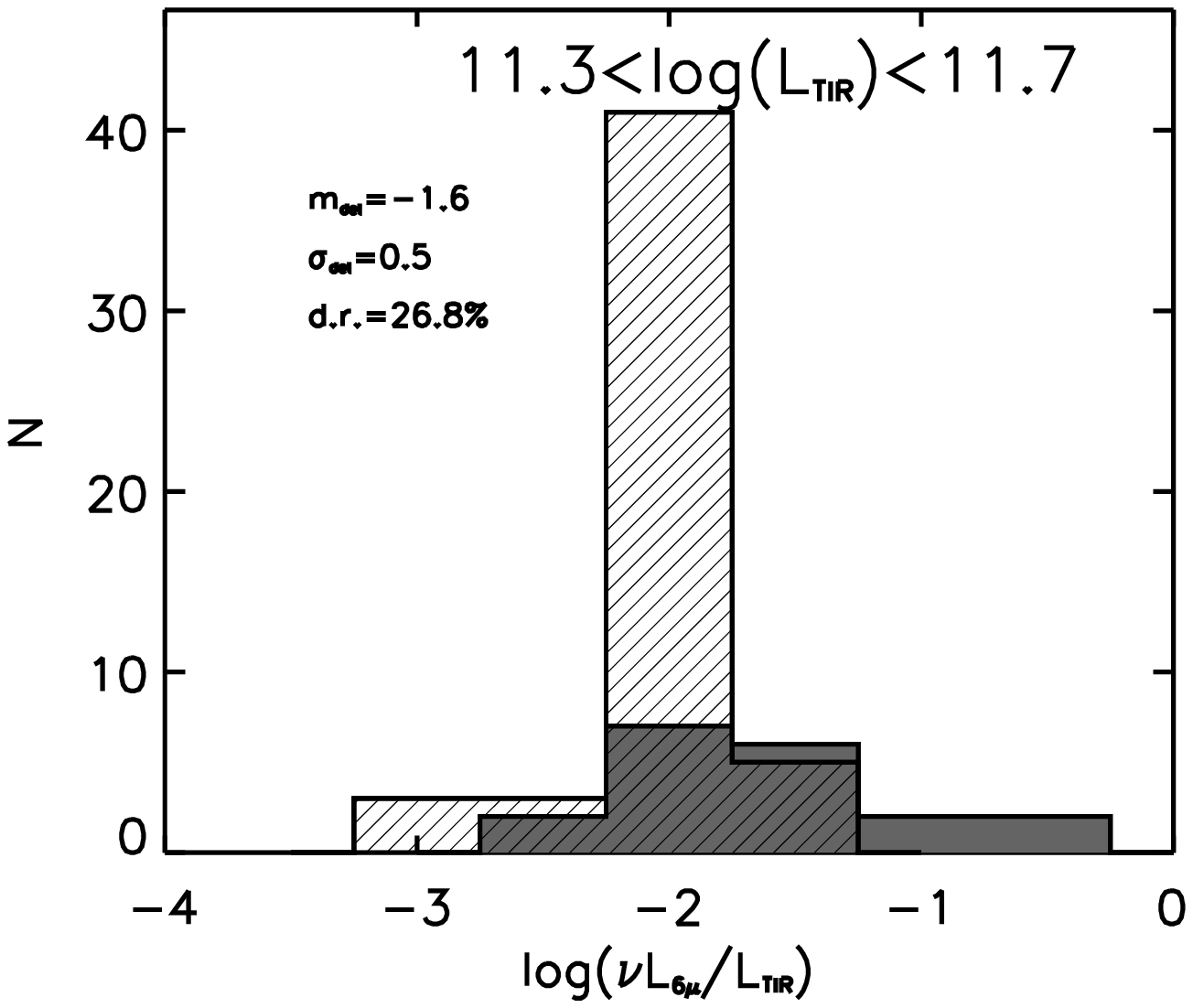}}\\
{\includegraphics[width=6.5cm,height=5.3cm]{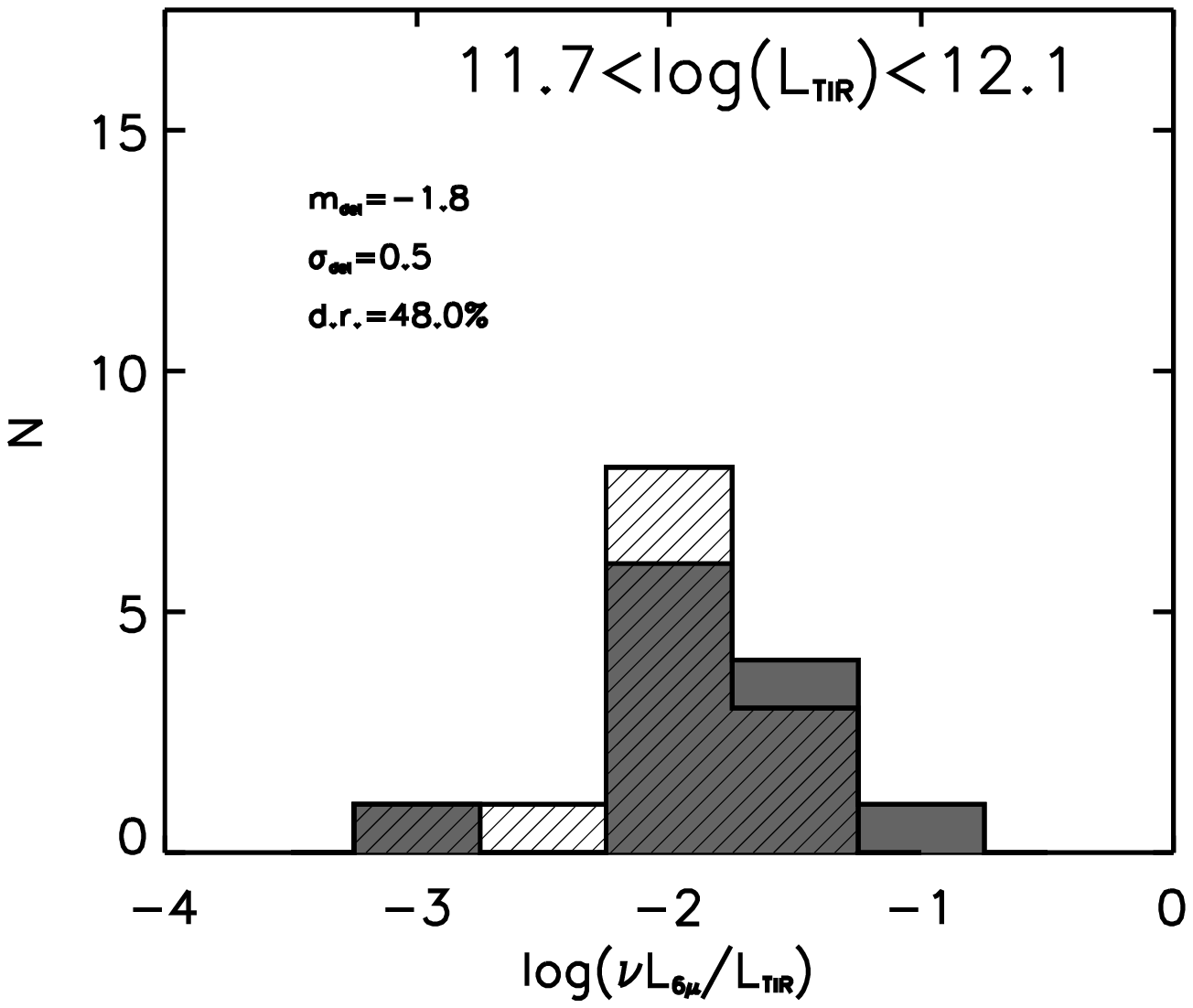}}
{\includegraphics[width=6.5cm,height=5.3cm]{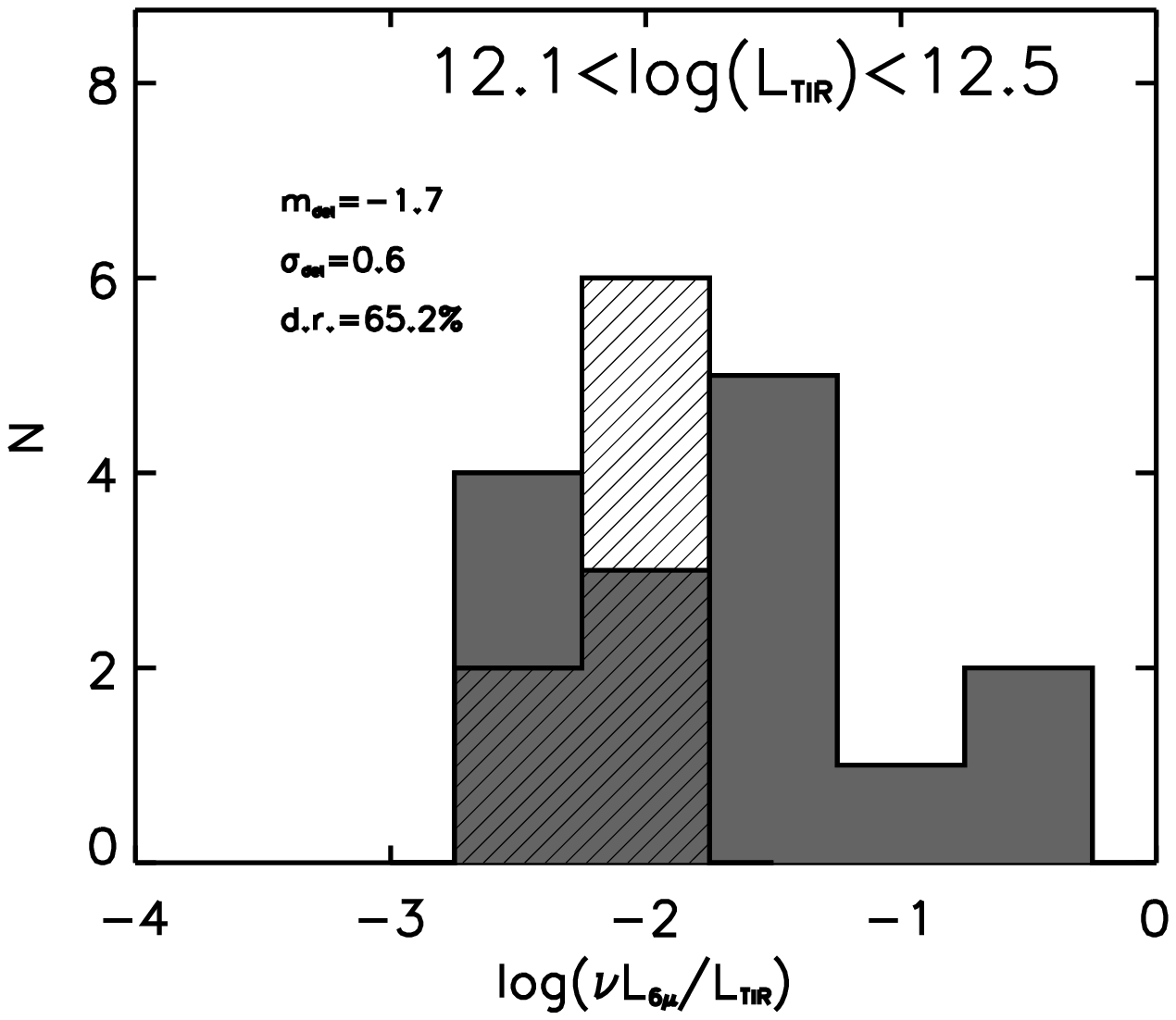}}\\
{\includegraphics[width=6.5cm,height=5.3cm]{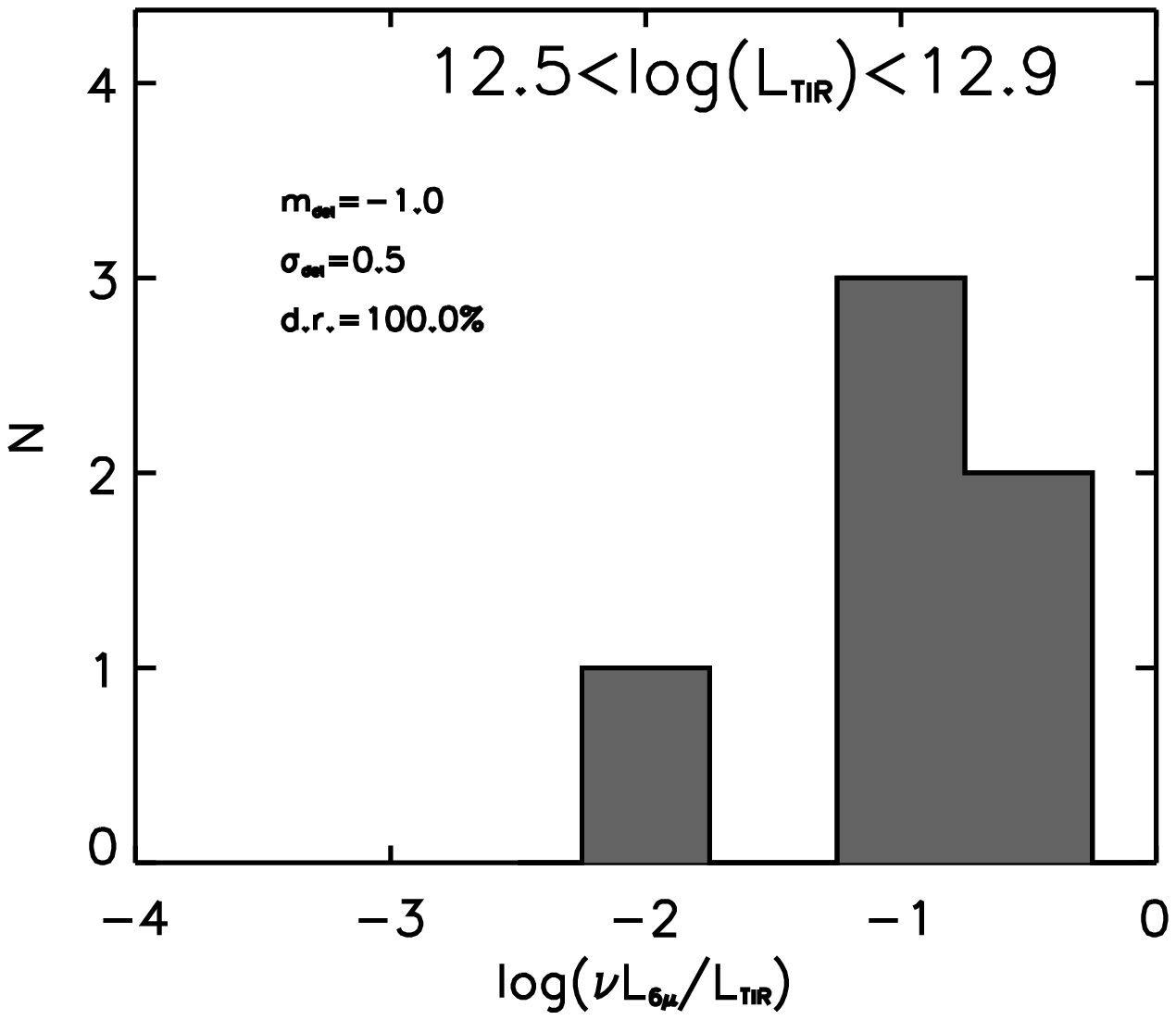}}
\caption{Distribution of $\nu L_{6\rm{AGN}}/L_{\rm TIR}$, for different luminosity bins. Solid histograms show the measurements, while the cross-hatched histograms indicate objects with upper limits. As indicated in each diagram, the mean values of the detections do not vary significantly between the different bins. What noticeably varies is the detection rate, spanning values from 23.3\% to 100\%.
}
\label{measurements}
\end{figure}

Fig.~\ref{measurements} shows the distributions of the ratio between the $6\,\mu{\rm m}$ AGN continuum luminosity and the total infrared luminosity ($\nu L_{6\rm{AGN}}/L_{\rm TIR}$) for five different luminosity bins, spanning $10^{10.9}$ to $10^{12.8}L_{\odot}$. These distributions do not show the intrinsic AGN luminosity, but the fraction of the IR emission due to the contribution of an AGN, because of the normalization used.



As shown from the distributions of Fig.~\ref{measurements} (solid and cross-hatched histograms) and also in the diagram of Fig.~\ref{L6um_vs_LIR} (open squares), there is no evident trend with luminosity in the values of the means of the {\itshape detections}. In the bins at lower $L_{\rm TIR}$ there are good statistics, but the percentage of upper limits to $L_{6\rm{AGN}}$ is rather high (upper limits are represented by cross-hatched histograms). In the bins at higher $L_{\rm TIR}$, where the $f_{6\rm{AGN}}$ detection rates increase in a significant way (solid histograms represent detections), the numbers of objects are more modest. In both cases, the averages obtained from detections only (open squares in Fig.~\ref{L6um_vs_LIR}) have to be considered upper limits to the real mean values of the distributions of $\nu L_{6{\rm AGN}}/L_{\rm TIR}$.

The clearest trend in our measurements is the detection rate as a function of $L_{\rm TIR}$. The decreasing fraction of upper limits clearly demonstrates that the distribution is changing across the luminosity bins and that there must be a trend also in the intrinsic distribution of $\nu L_{6{\rm AGN}}/L_{\rm TIR}$. With the assumption of a normal distribution, it is possible to estimate the main parameters, mean and standard deviation, of the intrinsic distribution of $\nu L_{6{\rm AGN}}/L_{\rm TIR}$ by Monte-Carlo simulations, taking into account the flux limit defining the RBGS sample as well as the {\itshape Spitzer}-IRS detection limit.

\subsection{Simulating the distribution of the AGN contribution}\label{simsec}



In order to reproduce our measurements, in terms of detection rate, mean and standard deviation of the detections (hereafter d.r., $m_{\rm det}$, $\sigma_{\rm det}$), using Monte-Carlo simulations, we need to make assumptions about the intrinsic distribution of $\nu L_{6{\rm AGN}}/L_{\rm TIR}$ we are looking for and the quantities that affect our upper limits. $\nu L_{6{\rm AGN}}/L_{\rm TIR}$ is tightly correlated with two different quantities. $L_{6\rm{AGN}}$ depends in a trivial way on $f_{6\rm{AGN}}$, i.e. on the sensitivity of the instrument. $L_{\rm{TIR}}$ is related to the $60\,\mu{\rm m}$ flux by relations well known and tested on IRAS data \citep{helou85,dale01}.

Our RBGS-based sample is selected at $60\,\mu{\rm m}$. Therefore, the first step of the simulation consists of generating a population of sources having a $60\,\mu{\rm m}$ flux density spanning a range from $5.24$ to $200\,{\rm Jy}$ and a distribution $N(f_{\nu}) \propto f_{\nu}^{-1.5}$ \citep{sanders03}, expected for a complete sample of objects in a non-evolving Euclidean universe that is a reasonable approximation for the relatively small redshift range covered by our sample. We derive $L_{\rm TIR}$ for each source generated starting from its $60\,\mu{\rm m}$ flux density, adopting $L_{\rm TIR}\sim 2\times \nu L_{60\,\mu\rm{m}}$ as appropriate for our sample. Assuming that the distribution of $\nu L_{6{\rm AGN}}/L_{\rm TIR}$ is gaussian with a certain mean and standard deviation ($m$, $\sigma$), we calculate the expected value of $f_{6{\rm AGN}}$. At this point we apply a detection limit to $f_{6{\rm AGN}}$ and calculate d.r., $m_{\rm det}$ and $\sigma_{\rm det}$. The values of $m$ and $\sigma$ that best reproduce the measured d.r., $m_{\rm det}$ and $\sigma_{\rm det}$ for each luminosity bin are assumed to be the parameters of the $\nu L_{6{\rm AGN}}/L_{\rm TIR}$ intrinsic distribution.

Our IRS data come from the public \textit{Spitzer} archive, consisting of several observing campaigns to varying depths. To assign a detection limit to $f_{6{\rm AGN}}$, we consider the values of $3\sigma$ for the detections and the values of the upper limits for the non-detections in each luminosity bin. We assume the median values of the distributions of these quantities as the detection limits of our measurements. These values span $\sim 18-1\,{\rm mJy}$, depending on the luminosity bin (see Tab.~\ref{AGNresults}).

\begin{figure}[p!]
\centering

{\includegraphics[width=7.cm,height=5.3cm]{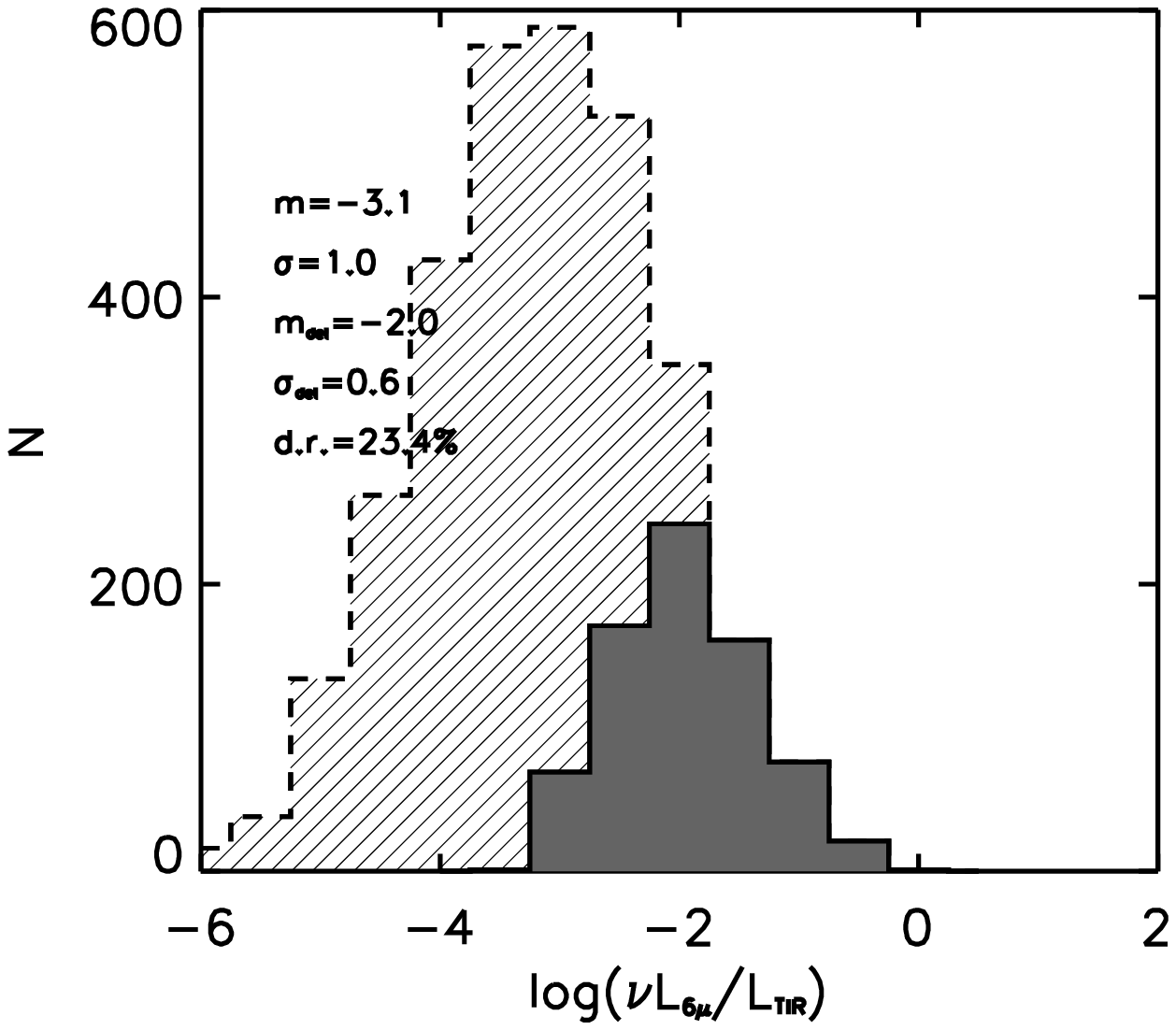}}
{\includegraphics[width=7.cm,height=5.3cm]{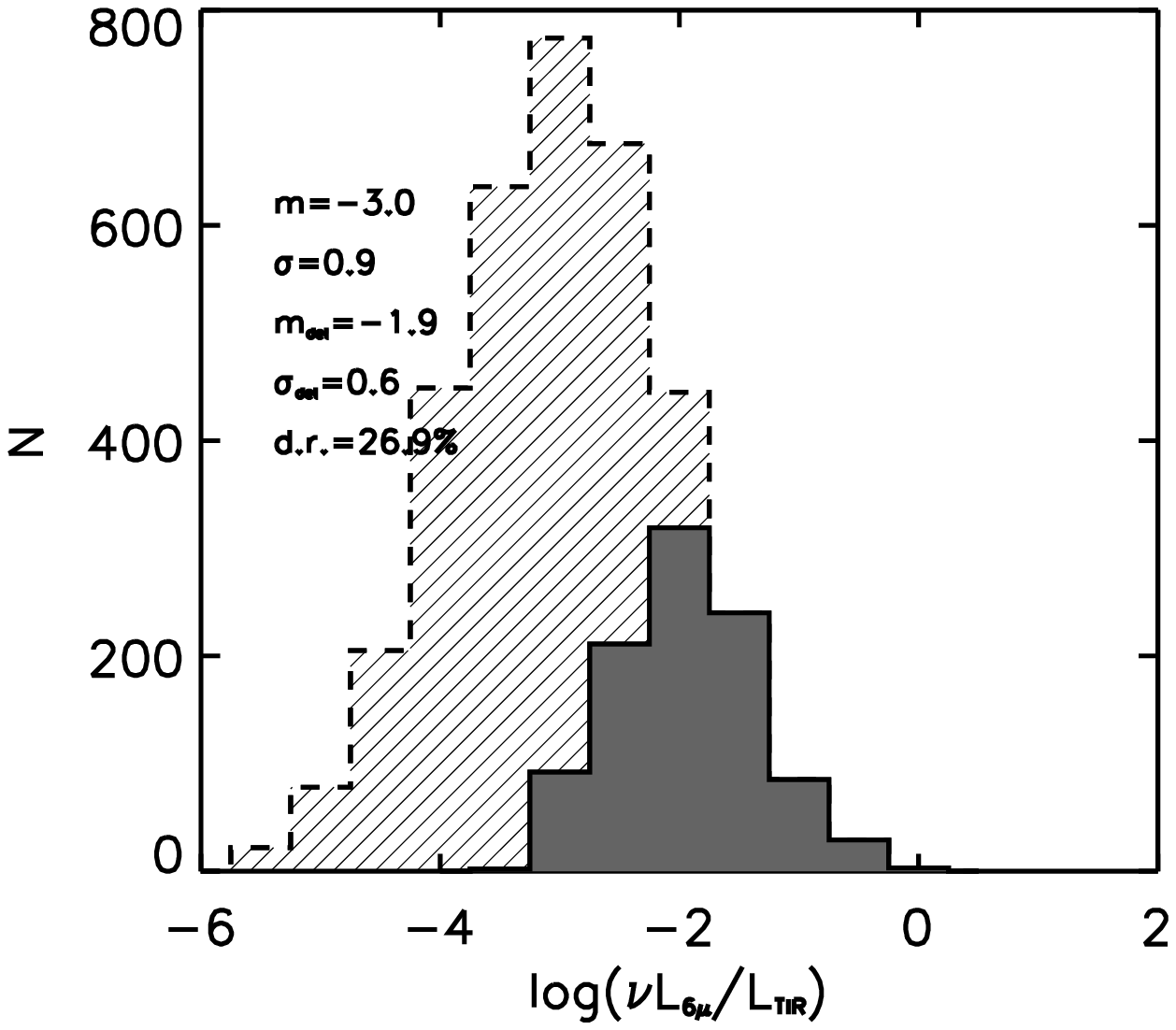}}\\
{\includegraphics[width=7.cm,height=5.3cm]{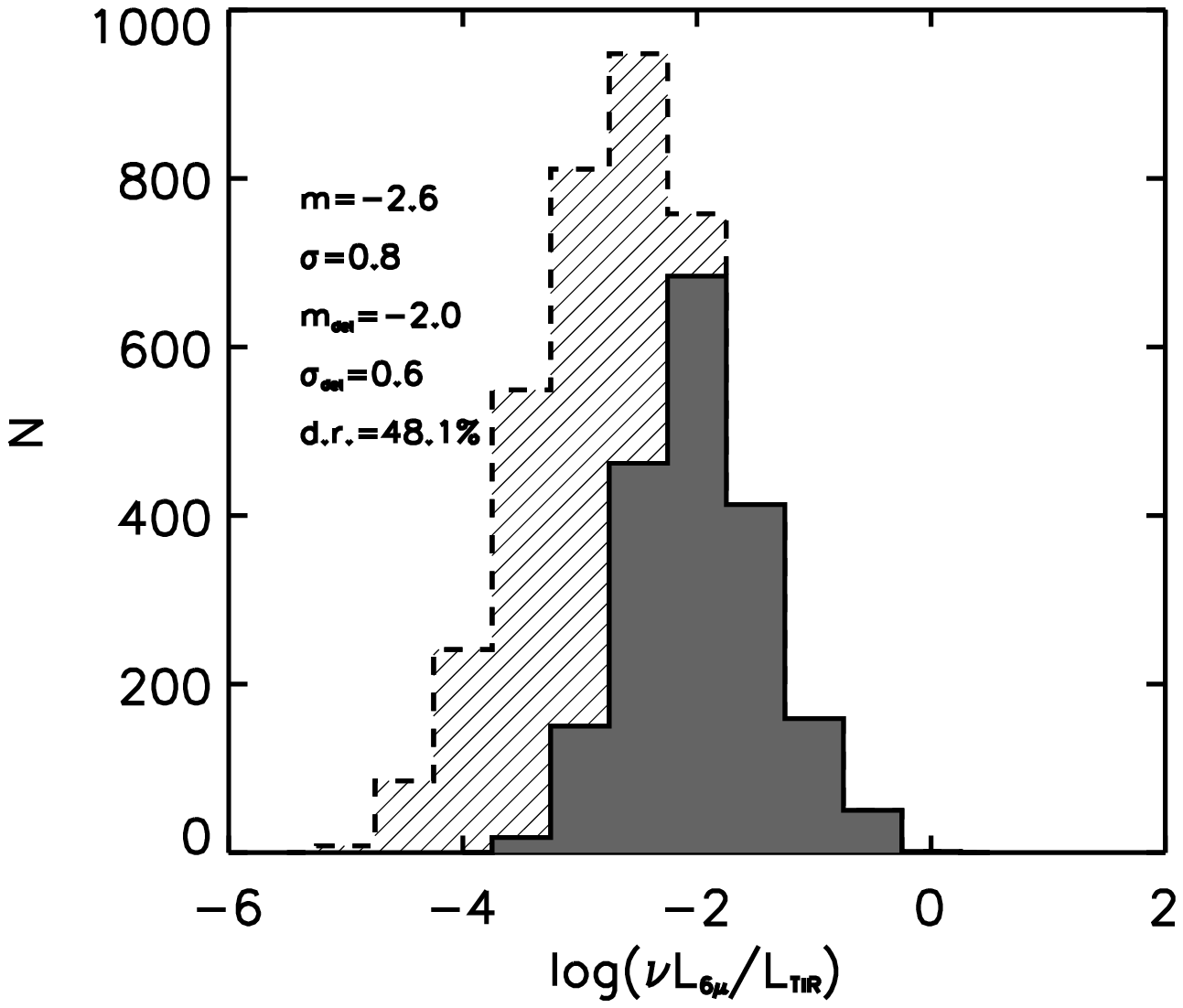}}
{\includegraphics[width=7.cm,height=5.3cm]{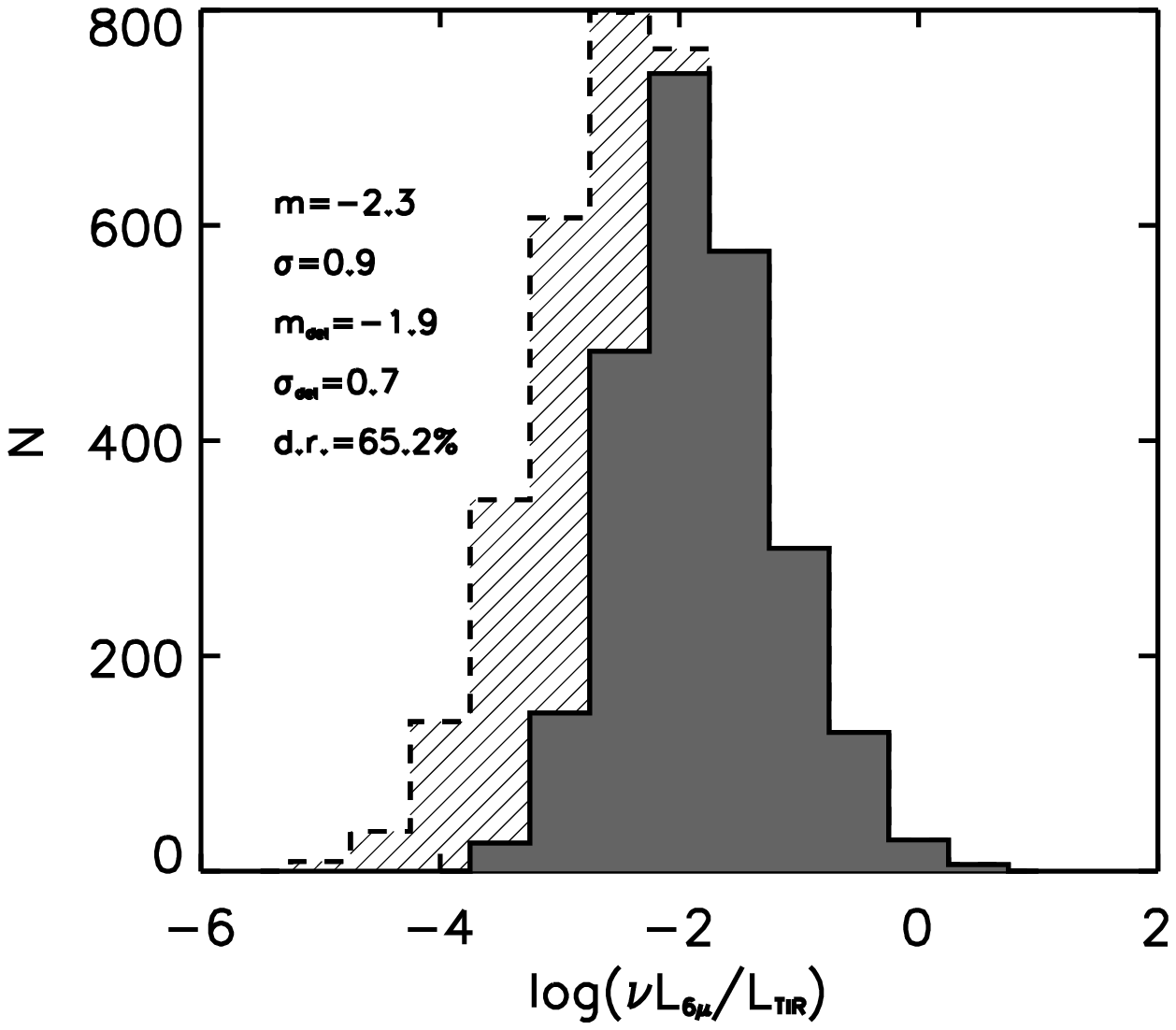}}\\
{\includegraphics[width=7.cm,height=5.3cm]{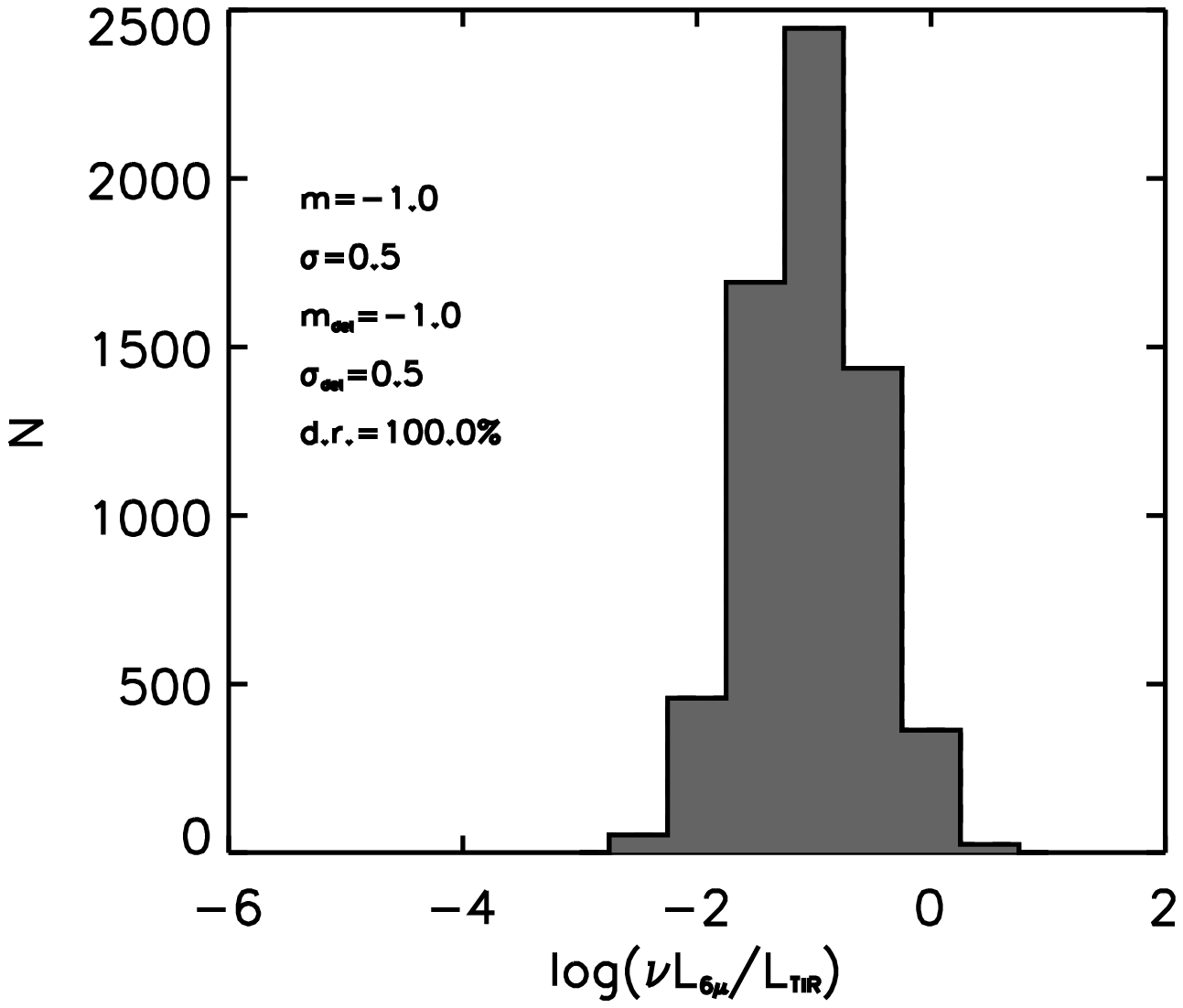}}
\caption{Simulations of $\nu L_{6\rm{AGN}}/L_{\rm TIR}$, assumed to be normally distributed, reproducing the same detection rates (d.r.), means ($m_{\rm det}$) and sigma ($\sigma_{\rm det}$) of the detections as the data. The detection limits assumed for $f_{6\rm{AGN}}$ in each bin is reported in Tab.~\ref{AGNresults}. Symbols have the same meaning as in Fig.~\ref{measurements}. Each simulation is made for a population of $2\times 10^4$ objects. Each diagram reproduces the values observed in the corresponding panel of Fig.~\ref{measurements} and shows also mean ($m$) and sigma ($\sigma$) of the adopted distribution.
} 
\label{simulations}
\end{figure}

Fig.~\ref{simulations} shows the results of the simulation for a population of $2\times 10^4$ objects. This number was chosen such that the sampled distributions converge. Each diagram shows the best gaussians that, given the detection limit, reproduce the observed values of d.r., $m_{\rm det}$ and $\sigma_{\rm det}$. A summary of the distribution parameters of measurements and simulations is presented in Tab.~\ref{AGNresults}. The values found for $m$ and $\sigma$ are also plotted in Fig.~\ref{L6um_vs_LIR} (green stars). Here, a correlation between the contribution to the infrared luminosity due to the AGN and the total infrared luminosity is clearly visible, with a dispersion decreasing with the luminosity. The  best fit, $\nu L_{6{\rm AGN}}/L_{{\rm TIR}} \propto L_{{\rm TIR}}^{\alpha}$, gives $\alpha=1.4 \pm 0.6$.

Our results assume the $\nu L_{6{\rm AGN}}/L_{\rm TIR}$ distribution to be gaussian for each luminosity bin. This is a reasonable first hypothesis and it reproduces the observations in terms of detection rate, mean and dispersion of the detections with a small numbers of parameters. More attention is needed when applying it to our Monte Carlo simulations of the infrared sky. The low AGN activity end of the distribution, corresponding mostly to IRS non-detections, is certainly poorly constrained, but such differences between already weak and totally insignificant AGN will not have a major effect on the predictions of our model. Of course, using these distributions to study physics of low luminosity AGN in these systems would be misleading. Differences at the high AGN activity end can be more important by introducing or missing very luminous mid-IR sources dominated by AGN. Comparing for example the $\log(\nu L_{6\rm{AGN}}/L_{\rm TIR})$ distribution for the third luminosity bin in Fig.~\ref{measurements} and Fig.~\ref{simulations} ($11.7<\log(L_{\rm TIR})<12.1$), we find the modelled distribution extending up to $\log(\nu L_{6\rm{AGN}}/L_{\rm TIR})\sim 0$ which is not reached by the observations. This could reflect limits in the statistics of our IRS sample as well as a true overprediction of our simple gaussian hypothesis at the high $\log(\nu L_{6\rm{AGN}}/ L_{\rm TIR})$ end. We will return to this issue in \S~\ref{section:addAGN} in the context of our backward evolution model. In the following we assume that the increase of the fraction of AGN occurs only up to the last luminosity bin measured ($L_{\rm TIR}\sim 10^{12.7}\,L_{\odot}$) and then the relation between $\nu L_{6{\rm AGN}}/L_{\rm TIR}$ and $L_{\rm TIR}$ becomes flat (Eq.~\ref{eq_AGN}; see also Fig.~\ref{L6um_vs_LIR}, green dotted line).


\subsection{Evolution with redshift}

It is interesting to check if and how the relation just found between $\nu L_{6{\rm AGN}}/L_{{\rm TIR}}$ and $L_{{\rm TIR}}$ evolves with redshift. Previous studies already showed that, even if in the local universe ULIRGs tend to be AGN-dominated at $L_{{\rm TIR}}>10^{12.5}\,L_{\odot}$ \citep{lutz98,veilleux99,tran01}, this trend does not extend to higher redshifts. For example, submillimeter galaxies that, with their typical infrared luminosity of $\sim 10^{13}\,L_{\odot}$, are among the most luminous objects known, are mainly starburst dominated. Mid-IR spectroscopy in fact shows that the contribution to the total luminosity due to AGN continuum is small in most of the sources observed with {\itshape Spitzer} IRS at high redshifts \citep{lutz05,menendez07,valiante07,pope08}. These results are consistent with X-ray observations of the same population \citep{alexander05}.

A systematic study at high redshifts is difficult with existing data. Ideally, to be consistent with the locally derived IRAS luminosity function, a high-$z$ sample should be selected at $60\times(1+z)\,\mu\rm{m}$, at the peak of the dust emission, but this cannot be done with our pre-{\itshape Herschel} observations.





Our high redshift sample includes all the sources observed by \citet{menendez07}, \citet{valiante07}, \citet{pope08} and \citet{brand08} for which the observations cover $5.5-6.85\,\mu{\rm m}$ in the rest frame (see Tab.~\ref{highz_AGNdata}). Objects from the first three works are submillimeter galaxies, some of them radio pre-selected. These galaxies are mainly starburst-dominated, but can present continuum emission in their mid-IR spectra, even if AGN is not the dominant power source. Galaxies from \citet{brand08} are optically faint objects selected at $70\,\mu{\rm m}$: their mid-IR spectra show that they can be either PAH or absorption dominated. All the IRS data have been retrieved from the {\itshape Spitzer} archive and reduced using the same techniques described in \S~\ref{section:AGNdatared}, in order to derive $L_{6\rm{AGN}}$ in a consistent way for all the objects. The results of the fit process are listed in Tab.~\ref{highz_AGNdata}. The values of $L_{\rm TIR}$ have been taken from the works cited above.

\begin{figure}
\centering

{\includegraphics[width=7.cm,height=5.7cm]{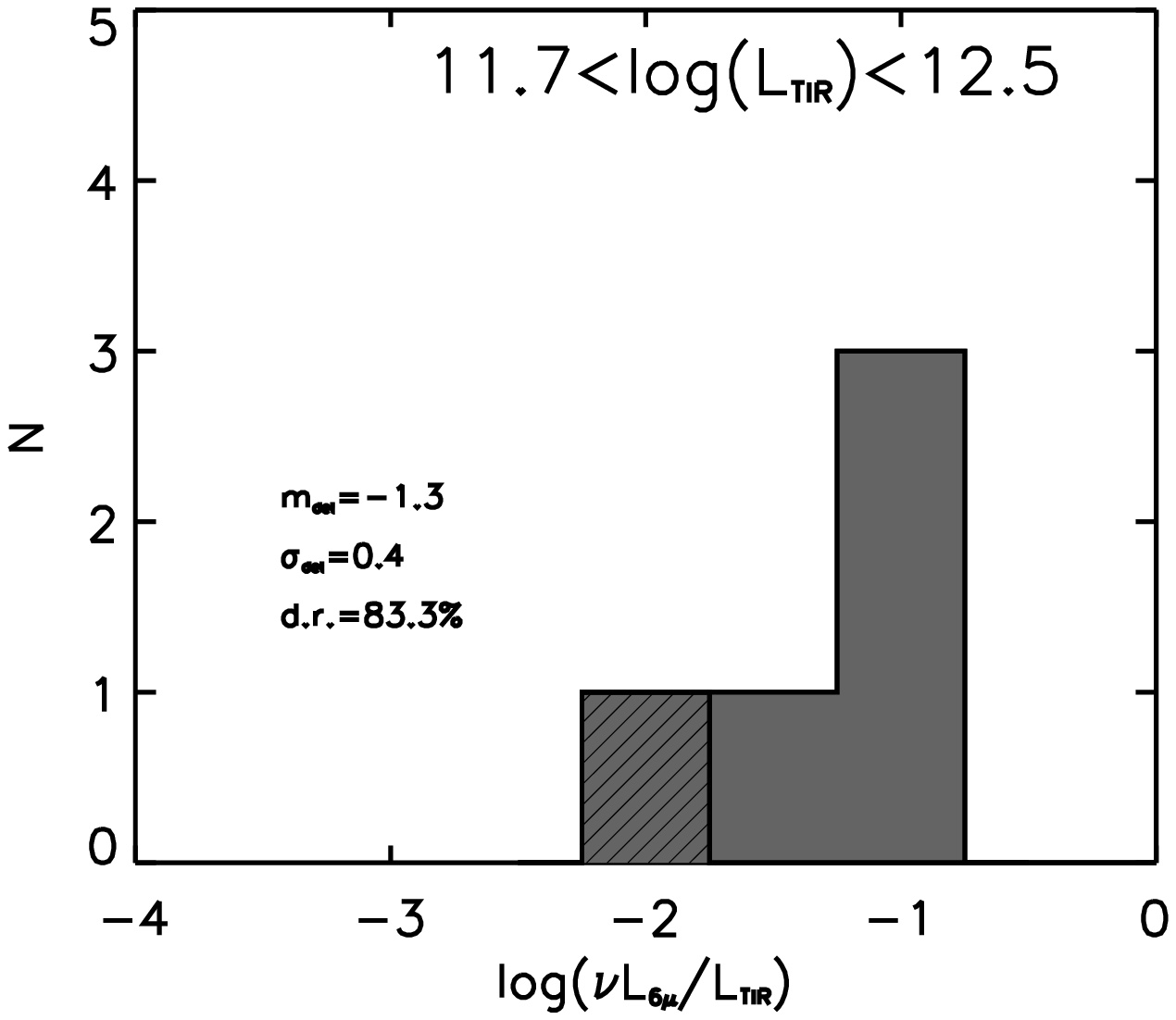}}\\
{\includegraphics[width=7.cm,height=5.7cm]{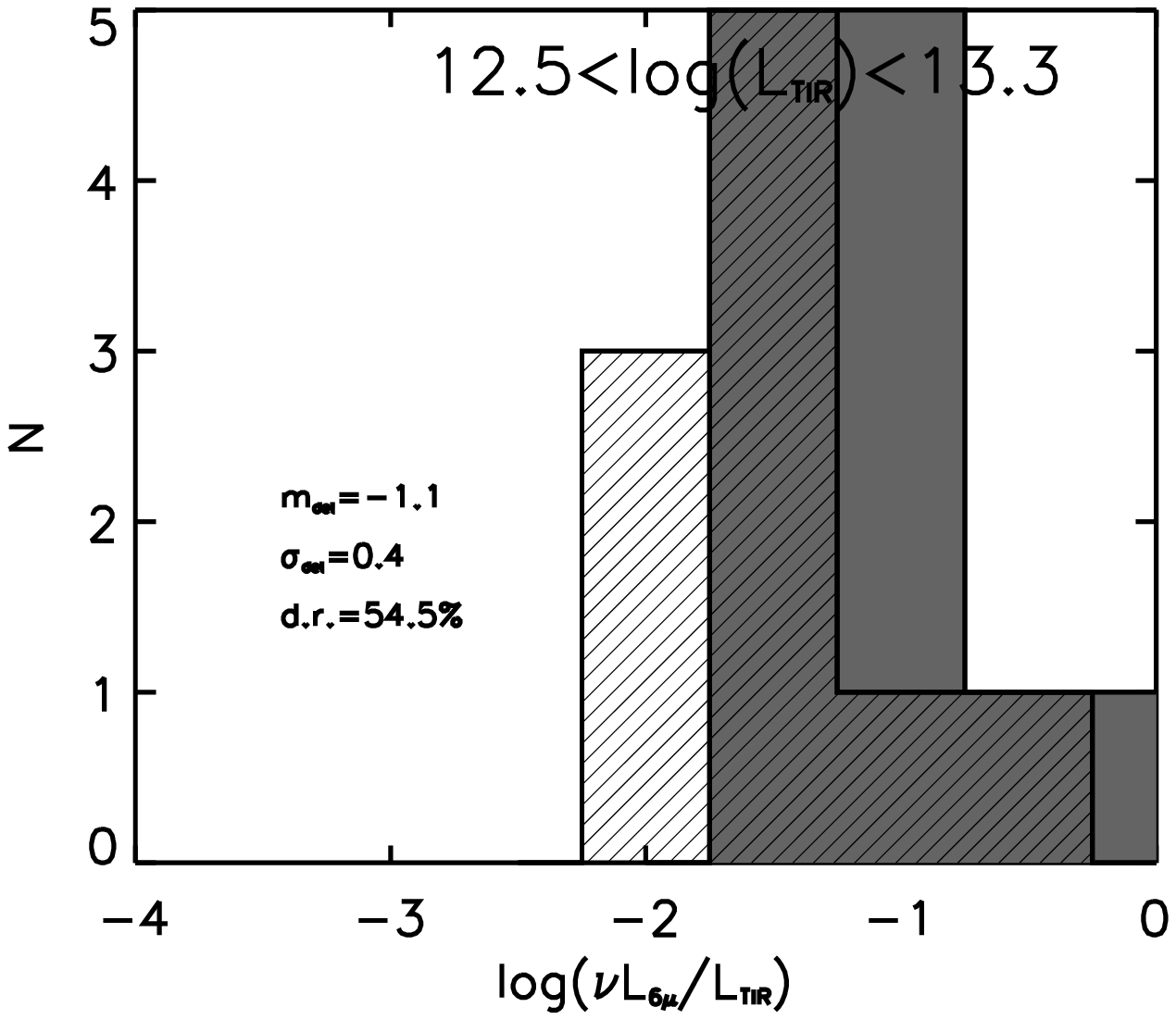}}\\
{\includegraphics[width=7.cm,height=5.7cm]{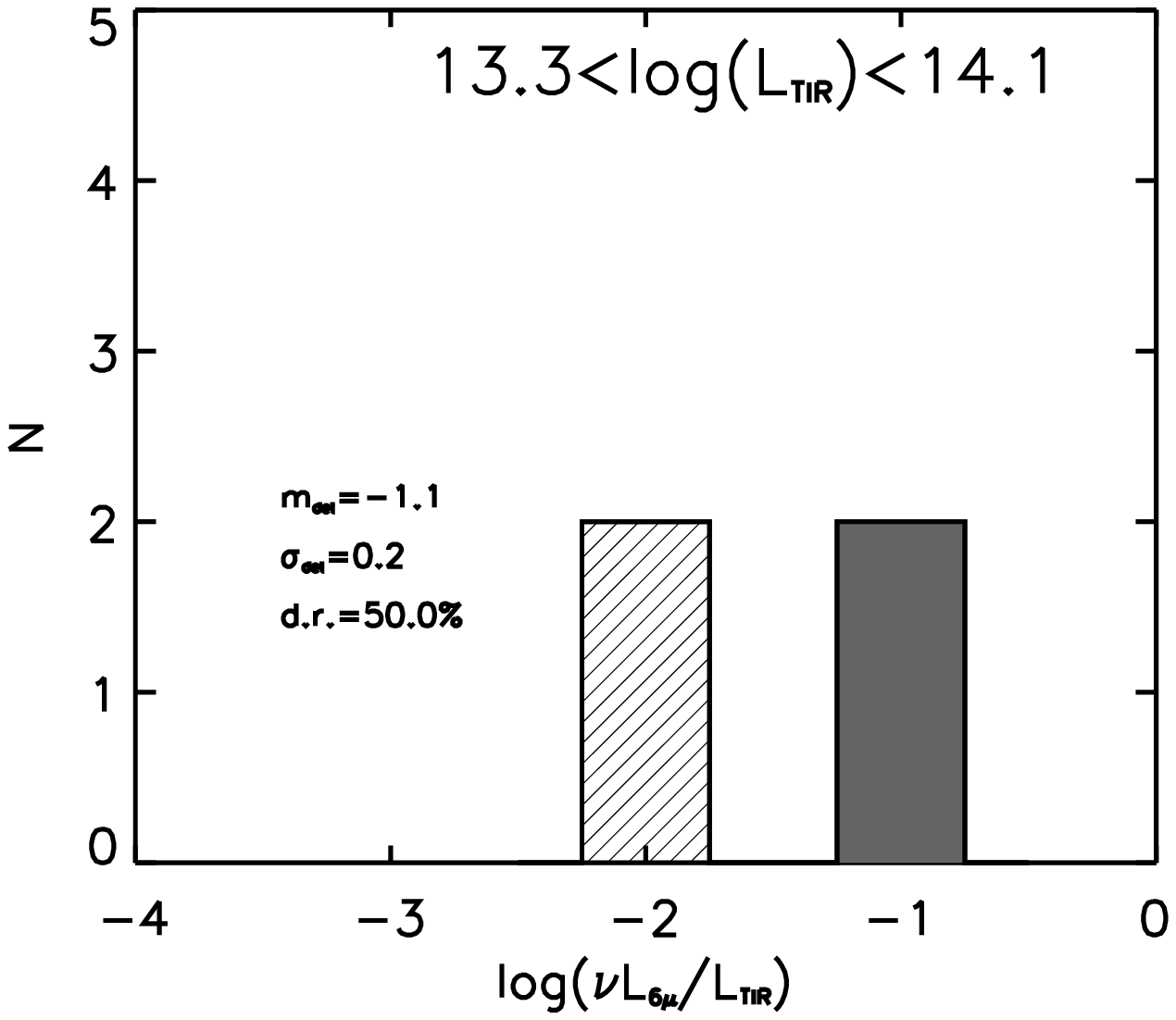}}
\caption{Distribution of $\nu L_{6{\rm AGN}}/L_{{\rm TIR}}$ for different luminosity bins. The sample lies at redshift $0.37<z<3.35$. Symbols have the same meaning as in Fig.~\ref{measurements}. Each diagram also shows the mean, the dispersion and the detection rate of the distribution.
}
\label{highzmeas}
\end{figure}

Fig.~\ref{highzmeas} shows the distributions of the $\nu L_{6{\rm AGN}}/L_{{\rm TIR}}$. Unfortunately, the statistics are rather modest for the bins at lowest and highest luminosities. The redshift range spanned by this sample is very broad ($0.37<z<3.35$). In particular, the bin at lowest luminosity is mainly populated by objects lying at $z\lesssim 1$, while the most luminous bins include the most distant sources. We treat the whole sample as a unique population, because the inclusion of the $z\lesssim 1$ objects does not significantly change the $\nu L_{6{\rm AGN}}/L_{{\rm TIR}}$ distribution in any of the luminosity bins. 

\begin{figure}
\centering

{\includegraphics[width=8.cm,height=5.7cm]{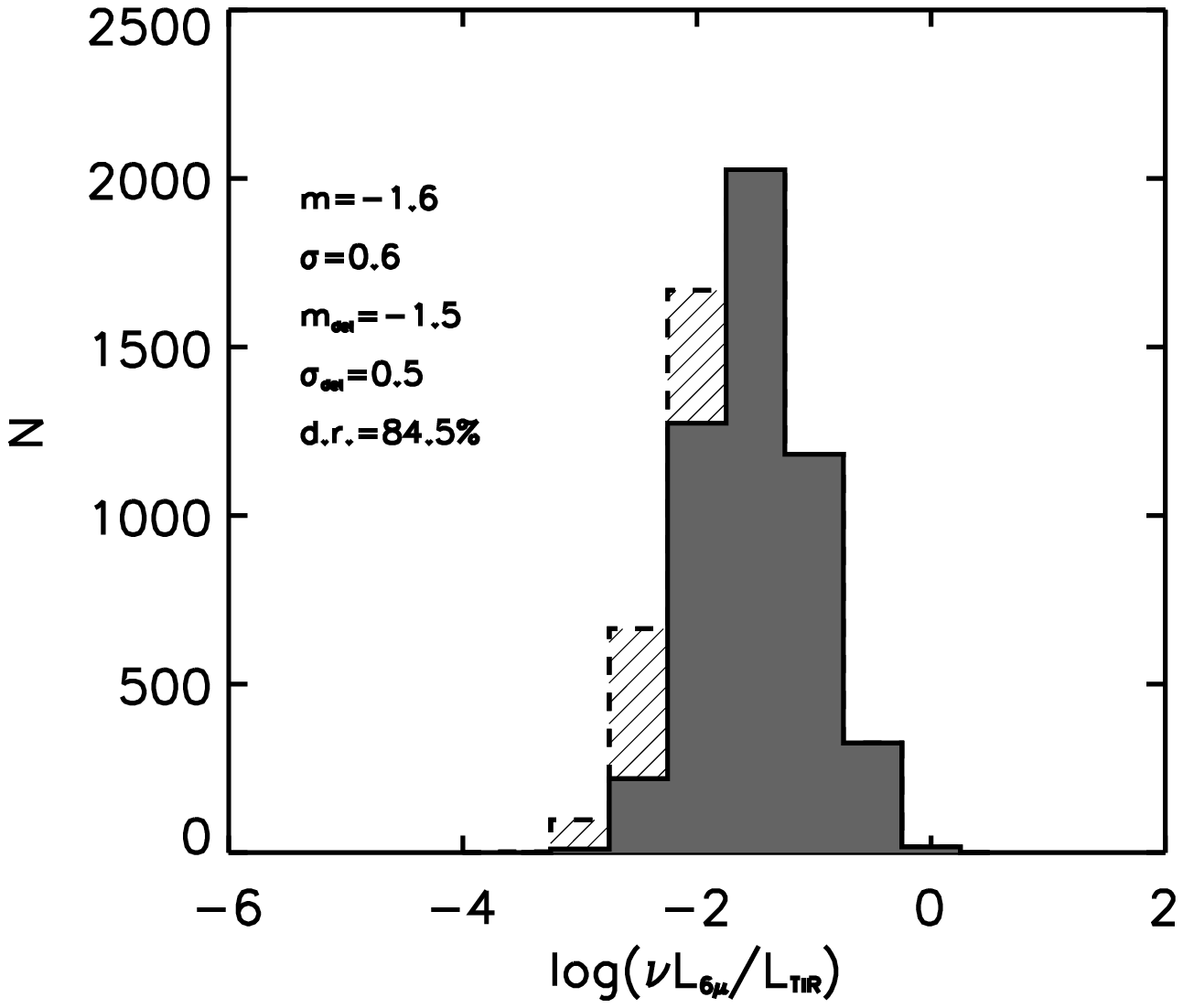}}\\
{\includegraphics[width=8.cm,height=5.7cm]{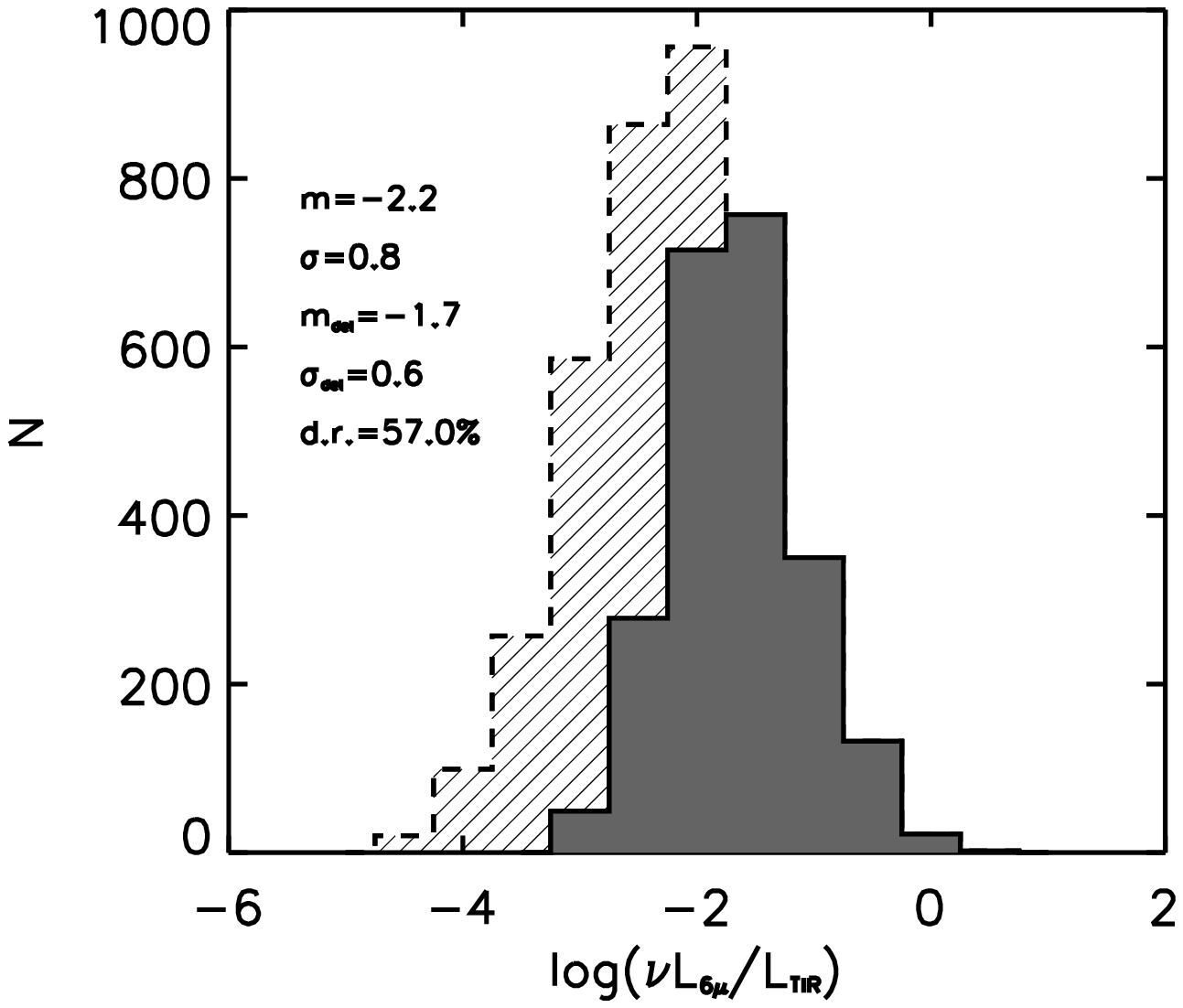}}
\caption{Simulations of $\nu L_{6\rm{AGN}}/L_{\rm TIR}$ normal distributions reproducing the same detection rates (d.r.), means ($m_{\rm det}$) and sigma ($\sigma_{\rm det}$) of the detections obtained from the data of the high redshift sample, assuming a detection limit of $0.13\,{\rm mJy}$. Each diagram reproduces the values observed in the corresponding panel of Fig.~\ref{highzmeas}, while the last luminosity bin has not been simulated because the statistic is too low for the fit to converge. Symbols have the same meaning as in Fig.~\ref{measurements}. Each diagram also shows mean ($m$) and sigma ($\sigma$) of the adopted distribution.
} 
\label{highzsim}
\end{figure}

\begin{figure}
\centering

\includegraphics[width=15.cm,height=10.7cm]{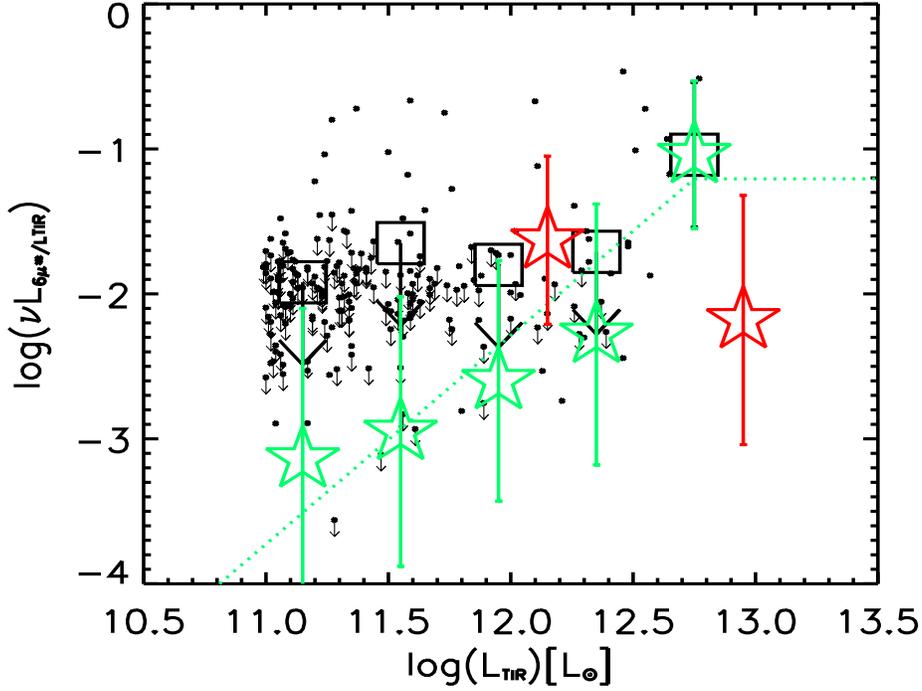}
\caption{$\nu L_{6{\rm AGN}}/L_{{\rm TIR}}$, vs. $L_{{\rm TIR}}$. Dots and small arrows represent measurements and upper limits of all the galaxies in our local sample. The large open squares are the means of the detections ($m_{\rm det}$) in bins of luminosity (see Tab.~\ref{AGNresults} and Fig.\ref{measurements}). Due to the large number of non-detections, they represent only an upper-limit on the true $\log{\nu L_{6\rm{AGN}}/L_{\rm TIR}}$ for almost all the bins. The green stars and the associated bars show means ($m$) and sigma ($\sigma$) of the gaussian distributions adopted in the simulations in order to reproduce detection rates (d.t.), means ($m_{\rm det}$) and sigma ($\sigma_{\rm det}$) of the detections from the measurements (see Tab.~\ref{AGNresults} and Fig.\ref{simulations}). The best fit, $\nu L_{6{\rm AGN}}/L_{{\rm TIR}} \propto L_{{\rm TIR}}^{\alpha}$, gives $\alpha=1.4 \pm 0.6$ and is shown as a dotted green line: the relation is assumed to be flat at high $L_{\rm TIR}$. The red stars and their bars show means and sigma of the gaussian distributions adopted to reproduce measurements of the high redshifts sample (see Fig.~\ref{highzmeas} and Fig.~\ref{highzsim}). The relation found for the local galaxies no longer holds.
}
\label{L6um_vs_LIR}
\end{figure}

In order to characterize the intrinsic distribution of $\nu L_{6{\rm AGN}}/L_{{\rm TIR}}$, we run Monte-Carlo simulations with the same techniques used for the local galaxies explained in \S~\ref{simsec}. We generate a population of sources having $60\,\mu{\rm m}$ flux density spanning $40$ to $1100\,{\rm mJy}$. We calculate the expected value of the $60\,\mu{\rm m}$ rest-frame flux density from the $850\,\mu{\rm m}$ and $70\,\mu{\rm m}$ fluxes, respectively for the submillimeter and the $70\,\mu{\rm m}$ selected sources, assuming a grey-body spectrum with emissivity index $\beta=1.5$ and $T_{\rm d}=32$K. The values obtained give us the range of the $60\,\mu{\rm m}$ population to simulate. Because of the low counts in the bin at highest luminosity, we are able to reproduce only the distributions of the first two bins, shown in Fig.~\ref{highzsim}. 

At high redshift, AGN contributions to the total infrared luminosity differ from local objects. For $L_{\rm TIR}\sim 10^{12}L_{\odot}$, the AGN contribution is consistent with the local universe distribution given the small number statistics. For the better populated $12.5<\log(L_{\rm TIR})<13.3$ bin however, which contains most of the SMGs, the AGN contribution is typically lower than for the few local objects reaching similar luminosities. All this evidence is tentative given the small sample size, the possible biases invoked by the selection and identification of the populations used, and the broad redshift range considered. It is hence too early to conclude a specific functional form of the change in the $\nu L_{6\rm{AGN}}/L_{\rm TIR}$ relation. For our subsequent simulations, we simply assume that the relation is the same as derived locally, with a change in the behaviour of the flat part: while in the local universe the increase of the AGN contribution stops at $L_{\rm TIR}\sim 10^{12.8}\,L_{\odot}$, for distant galaxies it occurs at lower luminosities ($L_{\rm TIR}\sim 10^{12}\,L_{\odot}$). It can be summarized as:
\begin{equation} \label{eq_AGN}
\nu L_{6\,\mu{\rm m}}/L_{{\rm TIR}} \propto 
\left\{ 
\begin{array}{ll}
L_{{\rm TIR}}^{1.4}   &      (L_{\rm TIR}< L_{\rm f})  \\
L_{\rm f}^{1.4}       &      (L_{\rm TIR}\geq L_{\rm f})
\end{array}
\right.;
\end{equation}
\begin{equation*}
L_{\rm f} =
\left\{
\begin{array}{ll}
10^{12.8}\,L_\odot  &    (z< 0.5) \\
10^{12}\,L_\odot    &    (z\geq 0.5)
\end{array}
\right.
\end{equation*}



\subsection{Discussion}

The different role of AGN at low and high redshift for objects in the same $L_{\rm TIR}$ class should not be a surprise. Fig.~\ref{L6um_vs_LIR} can be read in two (equivalent) ways: (1) the contribution of AGN to $L_{\rm TIR}$ is lower at high-$z$ with respect to the local universe for sources with $L_{\rm TIR}\sim 10^{13}L_{\odot}$, or (2) the AGN content of high-$z$ sources with $L_{\rm TIR}\sim 10^{13}L_{\odot}$ is similar to that of local galaxies with $L_{\rm TIR}\sim 10^{12}L_{\odot}$. This latter conclusion points out an analogy in the properties of SMGs and local ULIRGs. 

Such an analogy has been already discussed by \citet{tacconi06}: from CO observations of a sample of 14 SMGs, they conclude that the density of the molecular gas, the luminosity surface density and the temperature of the dust are the same in the two populations and SMGs are similar to local ULIRGs mergers, suitably scaled for their larger masses, luminosities and star formation rates, as well as their greater gas fractions. In spatially resolved observations, SMGs and ULIRGs show considerable similarities in their kinematics \citep{tacconi08}: the SMGs presenting multiple components are interacting systems and are similar to local double-nucleus ULIRGs, while SMGs showing characteristics of a rotational star-forming gas disk present the same surface and volume densities, for example, of the extremely compact ULIRG Arp220. All SMGs studied with sub-arcsecond millimeter interferometer so far, appear to be major mergers in different stages, similar to local ULIRGs.

Our analysis, nevertheless, is model-oriented: its purpose is not to find a theoretical explanation for the different populations observed, but to develop an evolutionary model based on observations capable of reproducing available data and to help plan forthcoming surveys with new instruments. 

\section{The model: number counts and redshift distributions of infrared sources} \label{section_model}

\subsection{The strong evolution of infrared galaxies: observational evidence} \label{section_evolution}

There has been, in the past years, strong observational evidence indicating high rates of evolution for infrared galaxies.

First, galaxy evolution can be observed through its imprint on the far-IR extragalactic background. Weakly constrained even as recently as the end of the 1990s, various observations now measure or give upper/lower limits on the background from the ultraviolet (UV) to the millimeter waveband (e.g.~\citealt{hauser01}). Data show the existence of a minimum between 3 and $10\,\mu\rm{m}$ separating direct stellar radiation from the infrared part due to radiation re-emitted by dust. This re-emitted dust radiation contains a comparable integrated power as the optical/near-IR: this amount is much larger than what is measured locally ($\sim 30$ per cent). The CIB is thus likely to be dominated by a population of strongly evolving redshifted infrared galaxies. Since the long-wavelength spectrum of the background is significantly flatter than the spectrum of local star-forming galaxies, it constrains the far-IR radiation production rate history \citep{gispert00}. The energy density must increase by a factor larger than 10 between the present time and redshift $z\sim 1-2$ and then stay rather constant at higher redshift (till $z\sim 3$).

Secondly, numerous deep cosmological surveys at 15, 24, 70, 90, 170, 850 and $1300\,\mu\rm{m}$ have resolved a fraction of the CIB into discrete sources. For all surveys, number counts indicate a very strong cosmological evolution of infrared galaxies, not only in the total power radiated but also in the shape of the LF. This is particularly obvious at submillimeter wavelengths, where the background is dominated by high-luminosity galaxies (SCUBA and MAMBO sources). The high rates of evolution exceed those measured in other wavelength regimes as well as those observed for quasars and active galactic nuclei (AGN).

Finally, high rates of evolution are suggested by the detection of Poissonian fluctuations of the CIB at a high level at 60 and $100\,\mu\rm{m}$ with IRAS \citep{mivilledeschenes02}, and $170\,\mu\rm{m}$ with ISOPHOT \citep{lagache00,matsuhara00}. For example, the constraints given by \citet{matsuhara00} on the galaxy number counts indicate the existence of a strong evolution in the counts. 

\subsection{SEDs} \label{section_sed}



Most existing backward evolution models (\S~\ref{section:intro}) separate the infrared sources into different populations. For example, \citet{franceschini88} consider three different populations: (1) normal late-type galaxies, (2) interacting/starburst galaxies and (3) galaxies with AGN. IRAS studies showed that galaxies of different nature in the local Universe have different spectral energy distributions (SEDs), e.g~different values of the $f_{60}/f_{25}$ ratio, so usually each population is associated with a particular SED family. The different populations of the model can evolve as a single population \citep{xu98,xu00} or at different rates, assuming a local LF for each component \citep{xu01}. Nevertheless, it is known that infrared galaxies, in particular ULIRGs, can host both starburst activity and an AGN \citep{genzel98,lutz98} and that galaxies classified as Seyfert show starburst tracers at infrared wavelengths \citep{lutz04}, while most of the far-IR emission in local QSOs is powered by starbursts \citep{schweitzer06}.



The SEDs can either be constructed empirically, considering the individual components responsible for the overall emission from the galaxies (e.g.~\citealt{rowanrobinson92}), or using radiative transfer models (e.g.~\citealt{efstathiou00}). While a complicated array of dust properties contributes to the SED of each galaxy, studies of IRAS galaxies have typically reduced the description to a best-fit single dust temperature, $T_{\rm{d}}$, with a one-to-one mapping to the $f_{60}/f_{100}$ flux ratio (hereafter referred to as $R(60,100)$). It has been demonstrated, as well, that local IRAS galaxies exhibit correlations of $R(60,100)$ with luminosity (e.g.~\citealt{dale01}) and that $R(60,100)$ changes systematically over a large range of luminosities. A statistical relation exists between $R(60,100)$ and infrared luminosity, even if the distribution is broad. Up to date, studies of the distribution spread in temperature for each luminosity bin exist, but analyses of the evolution are only preliminary, due to shortage of information on the temperature of the dust of high redshift sources \citep{chapman03,chapin09}. Difficulties of models with simple SED families and evolution functions became obvious with {\itshape Spitzer} data (e.g.~\citealt{lagache04}). There is already evidence for a change in SED properties with redshift, with the intrinsic SED shapes of high $z$ SMGs  resembling lower luminosity local objects rather than similarly luminous local HYLIRGs, but a deep study of SEDs and temperatures of high-redshift galaxies is still missing. The main problems are the degeneracy of dust temperature with redshift \citep{blain02} , the limited sample of sub-millimeter galaxies with measurements of spectroscopic redshifts (e.g.~\citealt{chapman05,valiante07,pope08}), and the selection effects on dust temperature induced by current methods to select high redshift infrared galaxies.

In building the SEDs for our model, we will take into account the coexistence of starburst and AGN in the same galaxies, their varying contribution with $L_{\rm TIR}$, and the spread in the luminosity-temperature ($L-T$) relation. AGN are hosted by infrared galaxies of different luminosities and their contribution to the total infrared emission can be either negligible or predominant. We studied and discussed the problem of the AGN contribution in \S~\ref{section:AGN}. We found that, on average, the contribution to $L_{\rm{TIR}}$ due to an AGN in IRAS galaxies is proportional to $L_{\rm{TIR}}^{1.4}$ of the host. Even if the dispersion in the relation allows the presence of low luminosity AGN-dominated objects (e.g.~NGC1068, $\log(L/L_{\odot})=11.3$), as well as ULIRGs powered by pure starbursts (e.g.~IRASF19297-0406, $\log(L)=12.4\, L_{\odot}$), this correlation holds over two-orders-of-magnitude, $\log(L)\sim 11-13\, L_{\odot}$. In \S~\ref{section:AGN}, we estimated the AGN contribution measuring the emission at $6\,\mu\rm{m}$, after the subtraction of a starburst-like spectrum; this method does not require to spatially resolve the AGN within the host.

In the simulation, we go in the opposite direction. Each source is assigned a redshift and a total infrared luminosity, $\tilde{L}_{\rm TIR}$, consisting of both the starburst ($\tilde{L}_{\rm SB}$) and AGN ($\tilde{L}_{\rm AGN}$) contributions ($\tilde{L}_{\rm TIR}=\tilde{L}_{\rm SB}+\tilde{L}_{\rm AGN}$). Starting from $\tilde{L}_{\rm TIR}$, we calculate the expected value $\langle\nu L_{6\,\mu{\rm m}}/L_{{\rm TIR}}\rangle$ for that luminosity, using Eq.~\ref{eq_AGN} (see also Fig.~\ref{L6um_vs_LIR}).

Then, we generate a random value of $\nu L_{6\,\mu{\rm m}}/L_{{\rm TIR}}$ assuming its distribution is a gaussian with mean $\langle\nu L_{6\,\mu{\rm m}}/L_{{\rm TIR}}\rangle$ and sigma corresponding to the luminosity bin where $\tilde{L}_{\rm TIR}$ lies (see Tab.~\ref{AGNresults}). From this random value we derive $\tilde{f}_{6\rm{AGN}}$ for the redshift of the source generated. Last, we scale an AGN template in order to achieve the desired flux density $\tilde{f}_{6\rm{AGN}}$ at $6\,\mu\rm{m}$. The AGN template used is the model calculated by \citet{efstathiou95} for the nuclear infrared continuum spectrum of the Seyfert galaxy NGC1068. The luminosity of the scaled template, $\tilde{L}_{\rm AGN}$, is the contribution of the AGN to $\tilde{L}_{\rm TIR}$.

\begin{figure}
\centering

\includegraphics[width=13.cm,height=9.3cm]{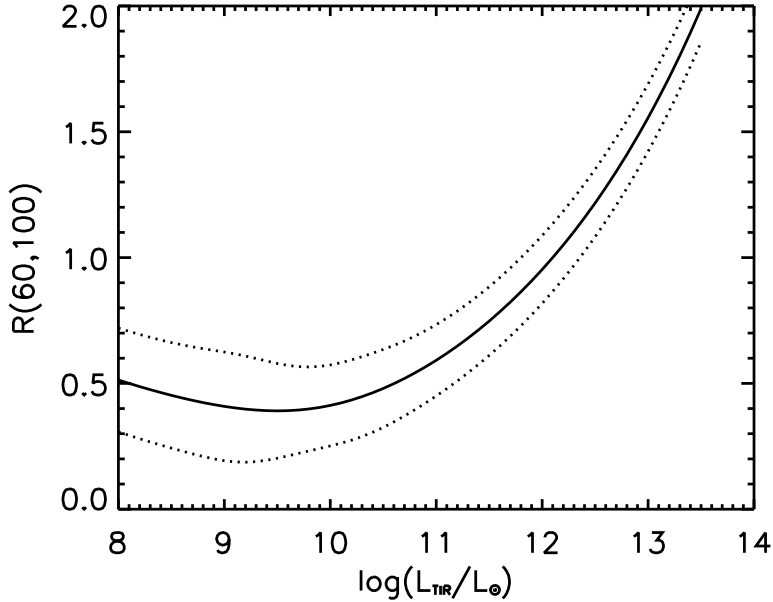}
\caption{Relation between the color ${R(60,100)}$ and $L_{\rm TIR}$ (solid and dotted lines give the mean and 1$-\sigma$ envelope of Eq.~\ref{eq:temp} respectively).
}
\label{fig:temp}
\end{figure}

Spectral templates for starbursts are taken from the \citet{dale01} (see also \citealt{dale02}) catalog, divided into 64 classes from $R(60,100)=0.29$ to 1.64, corresponding roughly to single-component dust-temperature models of $23-45\,$K. SEDs are normalized by integrating each spectrum over the $8-1000\,\mu\rm{m}$ range and scaling to the expected value of $L_{\rm{TIR}}$ for its $R(60,100)$, calculated from the relation found by \citet{chapin09} and adapted to the TIR luminosity (see Fig.~\ref{fig:temp}):
\begin{equation} \label{eq:temp}
R(60,100)=10^{C_*} \times\left(1+\frac{L^{\prime}}{L_{\rm SB}}\right)^{-\delta}\times\left(1+\frac{L_{\rm SB}}{L^{\prime}}\right)^{\gamma}
\end{equation}
with $\gamma=0.21$, $\delta=-0.12$, $C_*=-0.50$ and $L^{\prime}=5.6\times10^{9}\,L_{\odot}$. In order to fit this model we follow an identical methodology to \citet{chapin09}, calculating $L_{\rm TIR}$ by integrating the same \citet{dale02} SED catalogue, normalized to IRAS 60 and $100\,\mu\rm{m}$ data. In this way we obtain an SED family in the range $10^{9}-10^{15.7}\,\rm{Hz}$ spanning luminosities $\log(L)=8-13.5\,L_{\odot}$. During the simulation, in Eq.~\ref{eq:temp} we use the starburst contribution $\tilde{L}_{\rm SB}$ to the total infrared luminosity only, which we obtain by subtraction $\tilde{L}_{\rm AGN}$ from $\tilde{L}_{\rm TIR}$. From Eq.~\ref{eq:temp} we obtain the corresponding mean $\langle{R}(60,100)\rangle$ for the $\tilde{L}_{\rm SB}$ of the generated source. Then, we calculate the spread in the $L-T$ relation and generate a random value of $R(60,100)$, following the gaussian distribution calculated by \citet{chapin09} (see their Eq.~8 and Eq.~10) and centred on $\langle{R}(60,100)\rangle$ (see Fig.~\ref{fig:temp}). Finally, we rescale to the correct $\tilde{L}_{\rm SB}$ the \citet{dale02} library SED associated to the $R(60,100)$, which was obtained from the gaussian distribution implementing the spread in the $L-T$ relation.

\begin{figure}
\centering

\includegraphics[width=17.cm,height=15.cm]{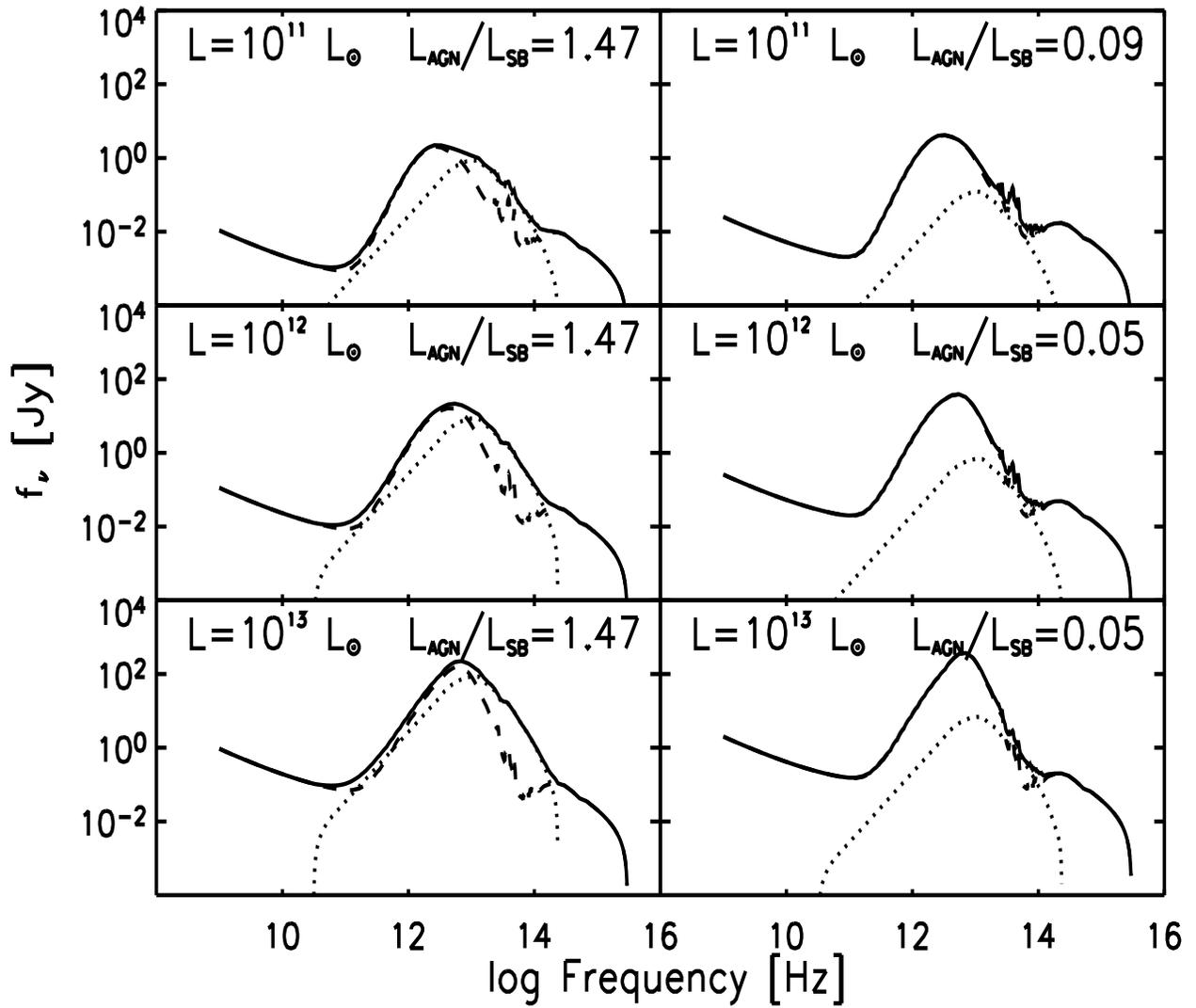}
\caption{Some examples of adopted SEDs (solid lines) obtained adding a star-forming galaxy spectrum (dashed lines) \citep{dale01,dale02} and an AGN template (dotted lines) \citep{efstathiou95}. The flux density is expressed in Jansky and is scaled for a distance of 100 Mpc. In each diagram we indicate $L_{\rm TIR}$ and the ratio between the luminosity of the AGN and the luminosity of the starburst spectra. For each luminosity class, two examples are shown: an AGN-dominated SED ($L_{\rm AGN}/L_{\rm SB}\geq 1$, left column) and a starburst-dominated SED ($L_{\rm AGN}/L_{\rm SB}<1$, right column).
}
\label{sedsAGN}
\end{figure}

To obtain the final SED, we add the two spectra derived for the AGN and starbursts parts. Fig.~\ref{sedsAGN} shows examples of SEDs with different $L_{\rm{TIR}}$ and different AGN contributions. Because the model takes into account both the average relation and its dispersion, it permits a wide range of combinations of $L_{\rm{TIR}}$ and AGN emission.

\subsection{Model parameters}

In this section we describe the algorithm to coherently model the number counts in different bands. A backward evolutionary model requires some basic ``ingredients'': 
\begin{itemize}
\item the spectral energy distributions, already described in \S~\ref{section_sed}, depending on $L_{\rm TIR}$, $R(60,100)$ and AGN content
\item a cosmological model describing the accessible volume in a survey solid angle as a function of redshift. This is now well constrained from WMAP results \citep{spergel07}. A concordance cosmology of $H_0 = 75\,\rm{km}\,\rm{s}^{-1}\,\rm{Mpc}^{-1}$,$\Omega_M=0.3$,$\Omega_\Lambda=0.7$ is assumed.
\item a luminosity function at $z=0$, e.g.~the most recent derivation of infrared luminosity function for IRAS galaxies by \citet{chapin09} 
\item the evolution functions for density and luminosity evolution.
\end{itemize}	

The solid angle observed (and therefore the number density of the sources) is geometrically defined by the assumed Universe model. The predicted number of sources in a given redshift interval $(z-0.5\,\mbox{d}z,z+0.5\,\mbox{d}z)$ and in a given infrared luminosity interval $(L-0.5\,\mbox{d}L,L+0.5\,\mbox{d}L)$ is given by
\begin{equation}\label{eq:number_sources}
\mbox{d}N(L,z) = g(z)\,\phi(L/f(z))\,\frac{\mbox{d}V}{\mbox{d}z}\,\mbox{d}L\,\mbox{d}z
\end{equation}
where $\phi$ is the local luminosity function and $f(z)$ and $g(z)$ are respectively the luminosity evolution function and the density evolution function and $\mbox{d}V/\mbox{d}z$ is the comoving volume element.

\begin{figure}
\centering

\includegraphics[width=14.cm,height=10.cm]{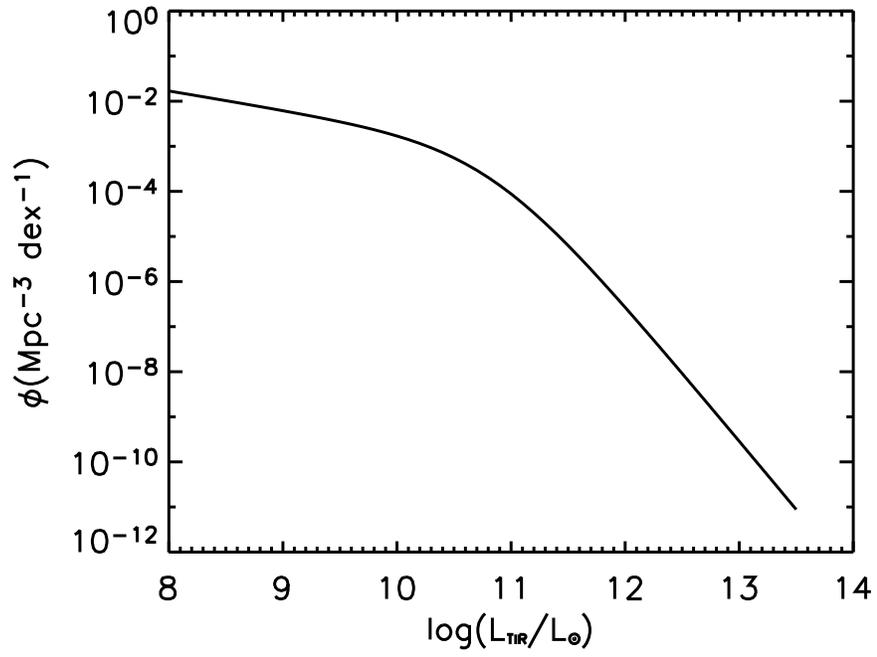}
\caption{The `double power-law''  parametrization of the TIR luminosity function.
}
\label{fig:lf}
\end{figure}

The luminosity function assumed at $z=0$ is a power law parametrization of IRAS galaxies. As explained by \citet{chapin09}, the LF calculated accounts for a temperature bias in the underlying $60\,\mu{\rm m}$ flux-limited sample, as well as luminosity evolution. The LF was computed using the $1/V_{\rm{max}}$ method \citep{schmidt68} with a modified formalism. The parametric form of the LF for the $L_{\rm{TIR}}$ is given by:
\begin{equation}\label{eq:chapinlf}
\phi(L) = \ln(10)L\rho_*\times\left(\frac{L}{L_*}\right)^{1-\alpha}\times\left(1+\frac{L}{L_*}\right)^{-\beta}
\end{equation}
with $\alpha=2.42$, $\beta=2.61$, $\rho_*=6.6\times10^{-15}\,{\rm Mpc}^{-3}\,L_{\odot}^{-1}$ and $L_*=6.9\times10^{10}\,L_{\odot}$ (see Fig.~\ref{fig:lf}). As in Eq.~\ref{eq:temp} we have re-calculated the fit for $L_{\rm TIR}$, since the results in \citet{chapin09} are for narrower-bandwidth FIR luminosities.

There exist observational evidences for strong evolution in the infrared galaxy population (see \S~\ref{section_evolution}). The amount of observed star formation shows that a large component of the infrared population has evolved to the present epoch. Possible scenarios are density evolution, where, due to merging/interactions, galaxies were more numerous in the past, or luminosity evolution, where, due to enhanced star formation, galaxies were more luminous in the past. These two schemes of evolution capture in a way that is practical for computational aspects what in principle is a more general evolution of the luminosity function with redshift. In general, some evolution is needed for all populations of active sources (e.g.~QSOs, radio galaxies and starburst galaxies) in order to model extragalactic source counts successfully. One of the ways to model luminosity and/or density evolution is using a simple power law. This assumption is motivated by similar evolutionary models at radio and X-ray wavelengths \citep{boyle88,benn93,condon94}. Similar evolution has also been observed in optically selected starburst galaxies \citep{lilly96} and in the submillimeter emission from radio loud galaxies observed at $850\,\mu\rm{m}$ by SCUBA \citep{archibald01}.

Our model assumes power law luminosity, $f(z)$, and density, $g(z)$, evolutions. The following functional forms are adopted:
\begin{equation}
f(z)=
\left\{
\begin{array}{ll}
(1+z)^{n_1}            &  (z<z_1) \\
(1+z)^{n_2}            &  (z\geq z_1)
\end{array}
\right.
\end{equation}
\begin{equation}
g(z)=
\left\{
\begin{array}{ll}
(1+z)^{m_1}            &  (z<z_2) \\
(1+z)^{m_2}            & (z\geq z_2)
\end{array}
\right.
\end{equation}
Both functions are assumed to be continuous at $z_1$ and $z_2$.

The algorithm calculates the number of sources predicted in each redshift bin (see Eq.~\ref{eq:number_sources}), introducing a poissonian scattering to the expected value. Then, a value $\tilde{L}_{\rm{TIR}}$ is associated to each source, by Monte-Carlo extraction, according to the luminosity function calculated for the redshift of the source, $\phi(L/f(z))$. $\tilde{L}_{\rm{TIR}}$ is then split into two parts: the fraction due to AGN contribution ($\tilde{L}_{\rm{AGN}}$), calculated following Eq.~\ref{eq_AGN}, and the remaining part associated to starbursts ($\tilde{L}_{\rm{SB}}$). The SED is thus ``built'' (see \S~\ref{section_sed}) adding a scaled AGN template (NGC1068) of luminosity $\tilde{L}_{\rm{AGN}}$ with the starburst template corresponding to $\tilde{L}_{\rm{SB}}$, after the application of a scatter in the $L-T$ relation, and scaled to the appropriate luminosity distance. The sources' flux densities in different bands are then calculated by convolving the redshifted SED with the bandpasses of the filters. In this way, we effectively simulate a virtual sky for a given evolution model. For each source, we know not only the flux densities for several bands and the $L_{\rm{TIR}}$, but also the temperature of the dust and the $L_{\rm{SB}}/L_{\rm{AGN}}$ ratio. The latter information, in particular, allows us to follow the co-evolution of starbursts and accretion, exploring photometric indicators to select different populations of infrared sources.



\section{Results and comparisons with available surveys}

Backward evolution models for the infrared spectral range include a significant number of tunable parameters, and using the proper observational constraints to adjust these parameters is the key to the successful use of such models. Traditionally, the first quantities to fit are the total IR/submm background as measured by COBE (CIB), and the number counts at mid-IR to submillimeter wavelengths. With improving identification and follow up of SCUBA, ISO and {\itshape Spitzer} sources, redshift distributions of different IR-selected populations now provide further powerful constraints. We are now about to enter the next level of physical characterization including determination of full rest frame far-IR SEDs, a field that will be given a big boost with BLAST and {\itshape Herschel}, and better disentangling the role of star formation and AGN in high redshift infrared galaxies. The model presented here is designed to allow comparison with this new type of constraints on SED and AGN content. By implicitly coupling AGN evolution to infrared galaxy evolution, this model is also amenable to comparison with AGN surveys at other wavelengths, if making use of assumptions about AGN SEDs and obscuration.

\begin{figure}
\centering

{\includegraphics[width=7.cm,keepaspectratio]{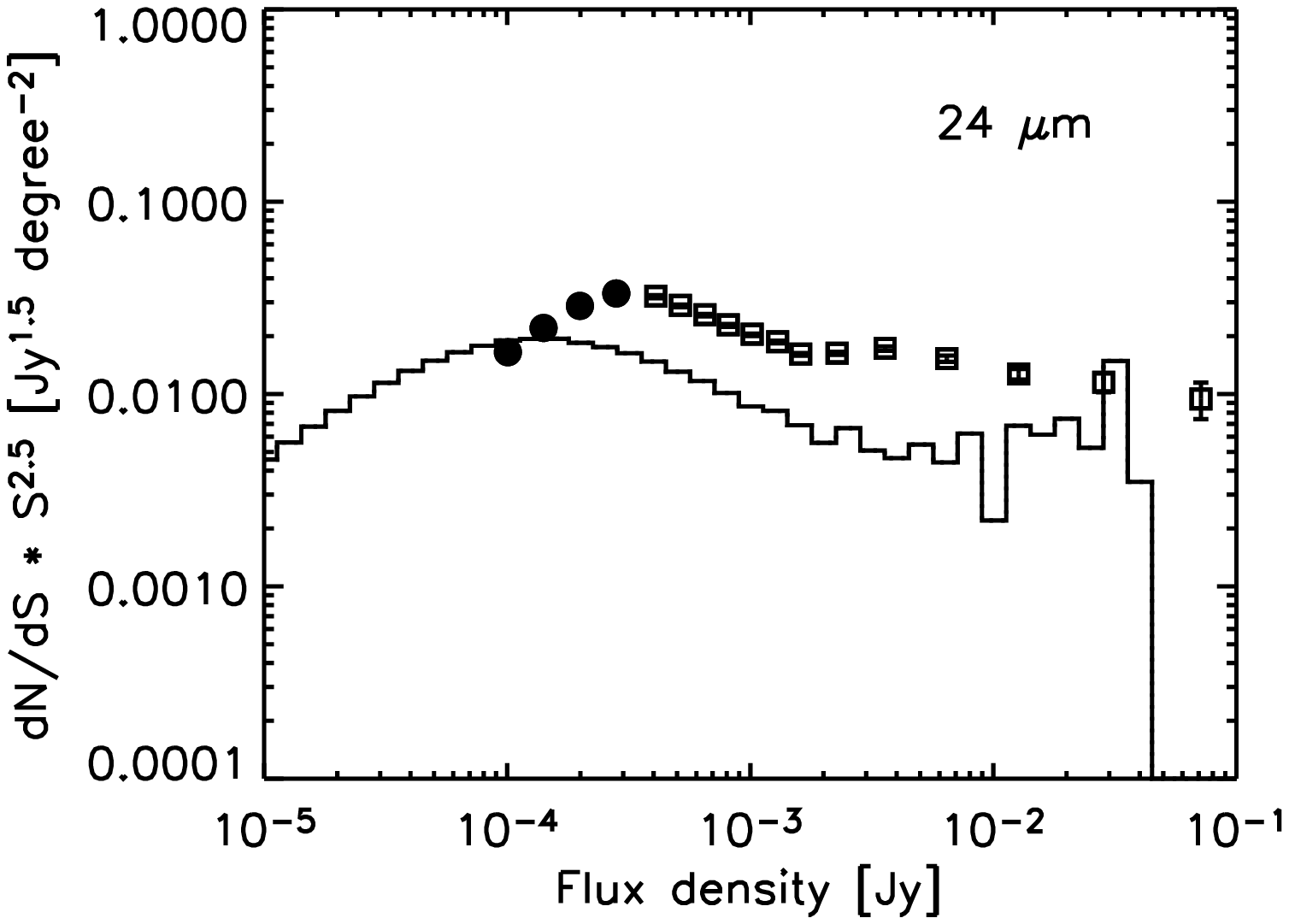}}
{\includegraphics[width=7.cm,keepaspectratio]{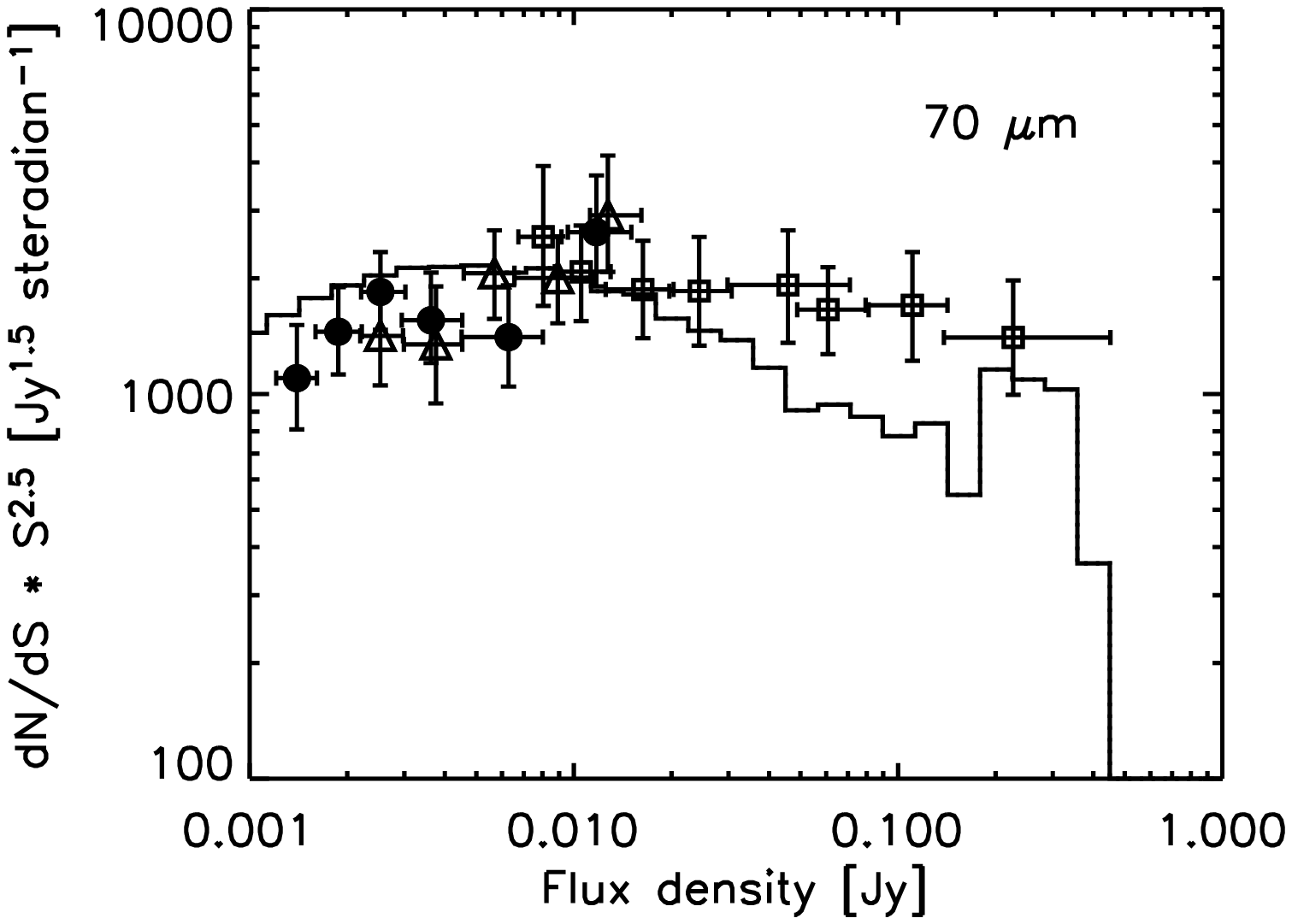}}\\
{\includegraphics[width=7.cm,keepaspectratio]{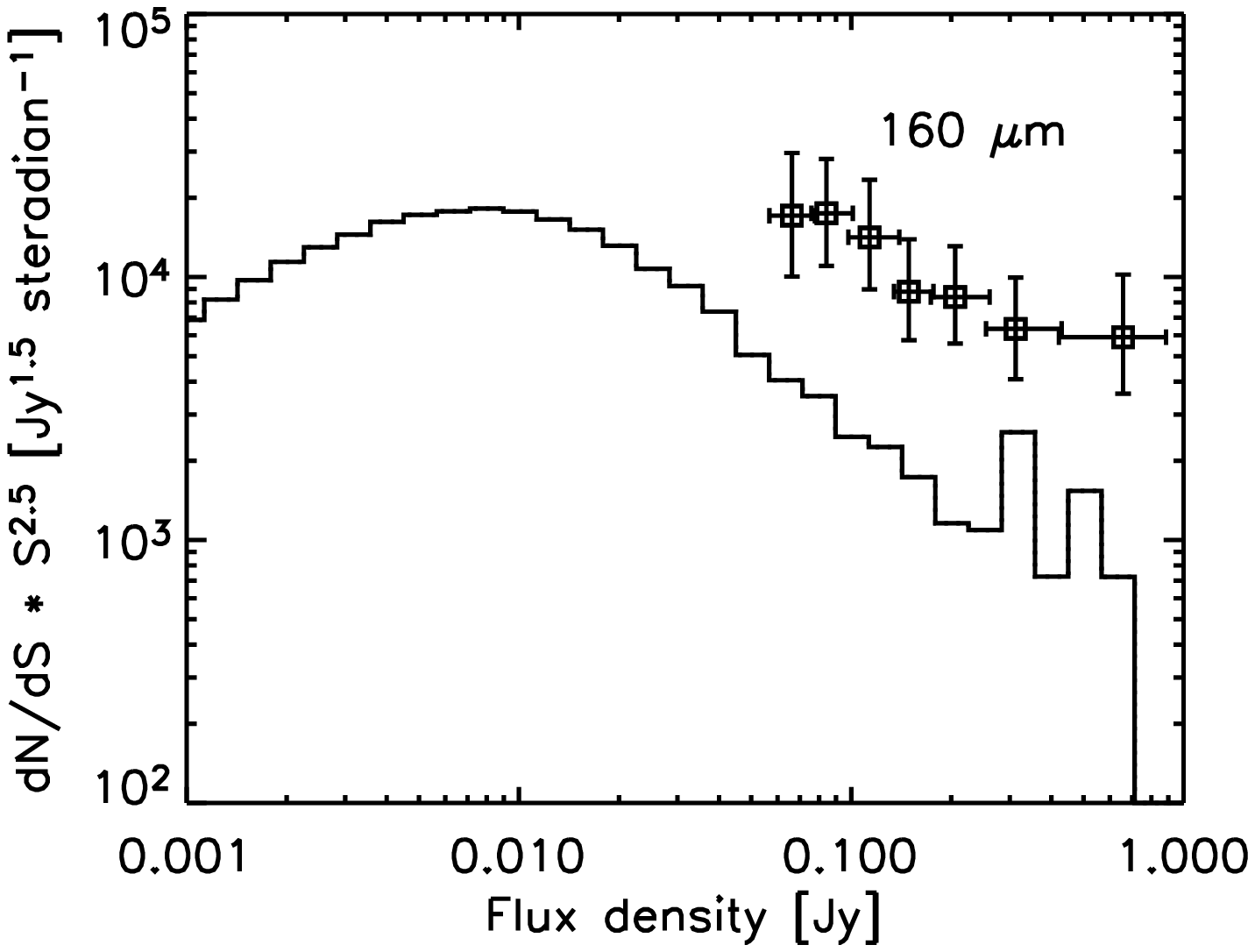}}
{\includegraphics[width=7.cm,keepaspectratio]{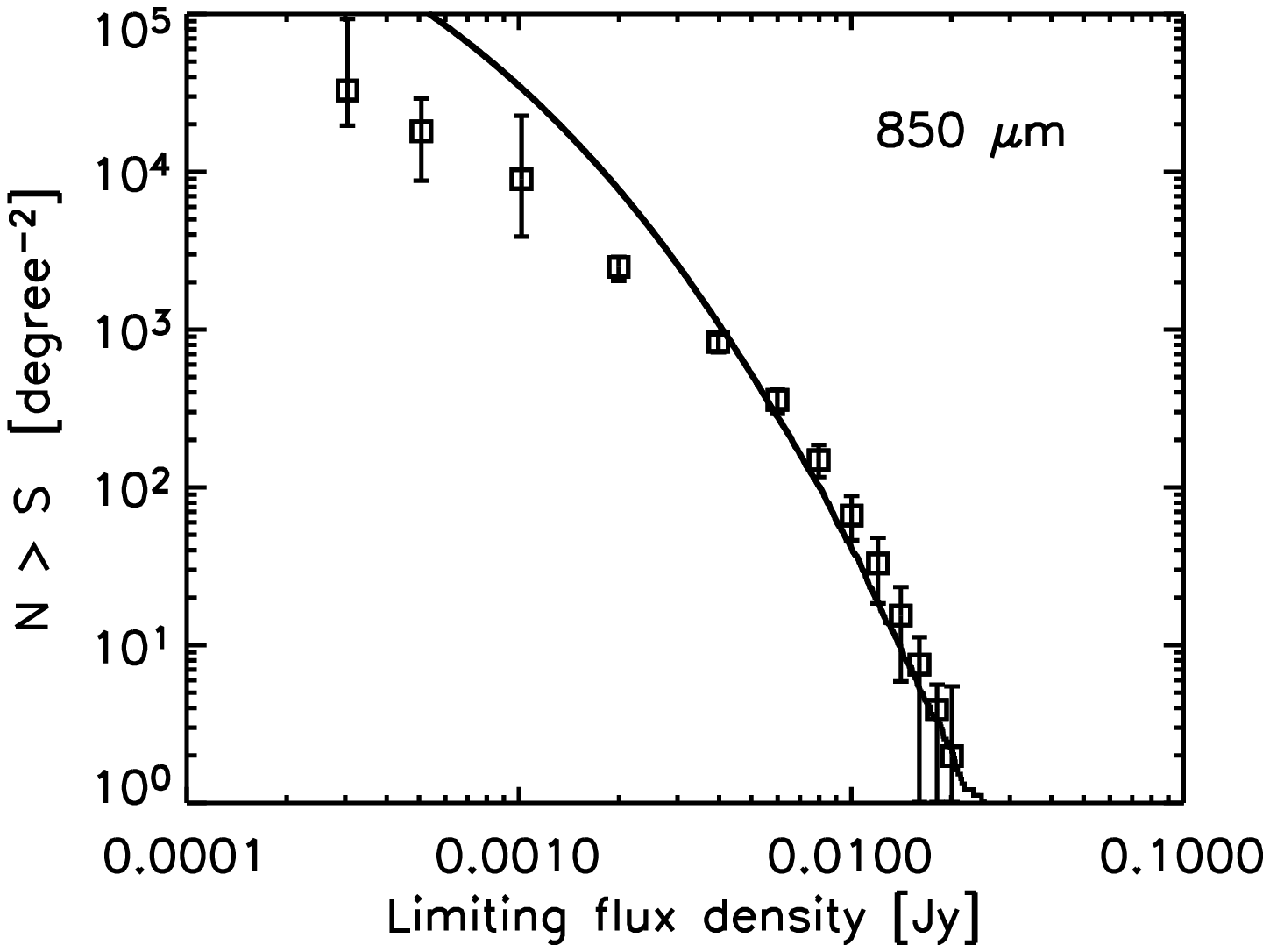}}\\
{\includegraphics[width=7.cm,keepaspectratio]{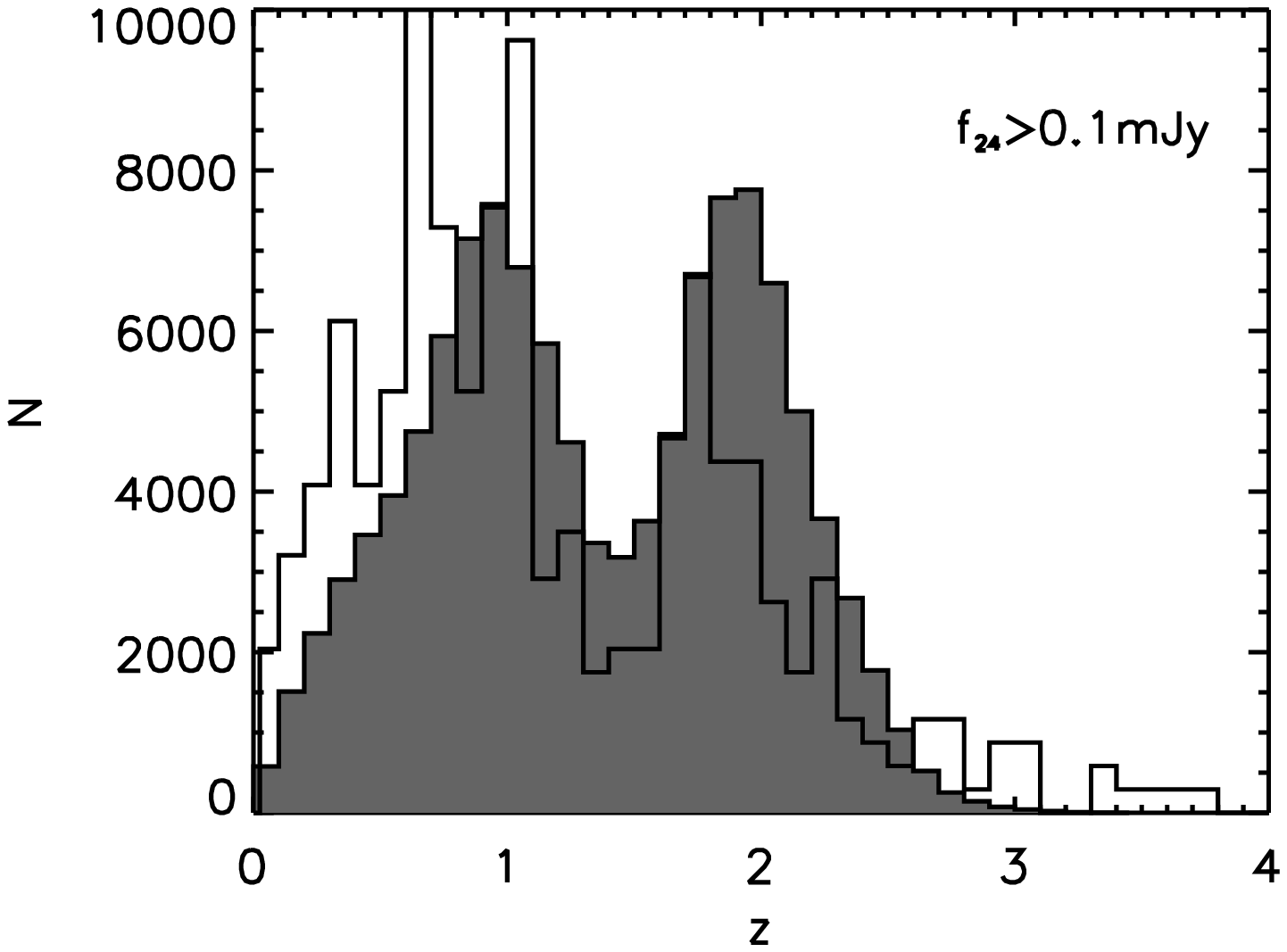}}
{\includegraphics[width=7.cm,keepaspectratio]{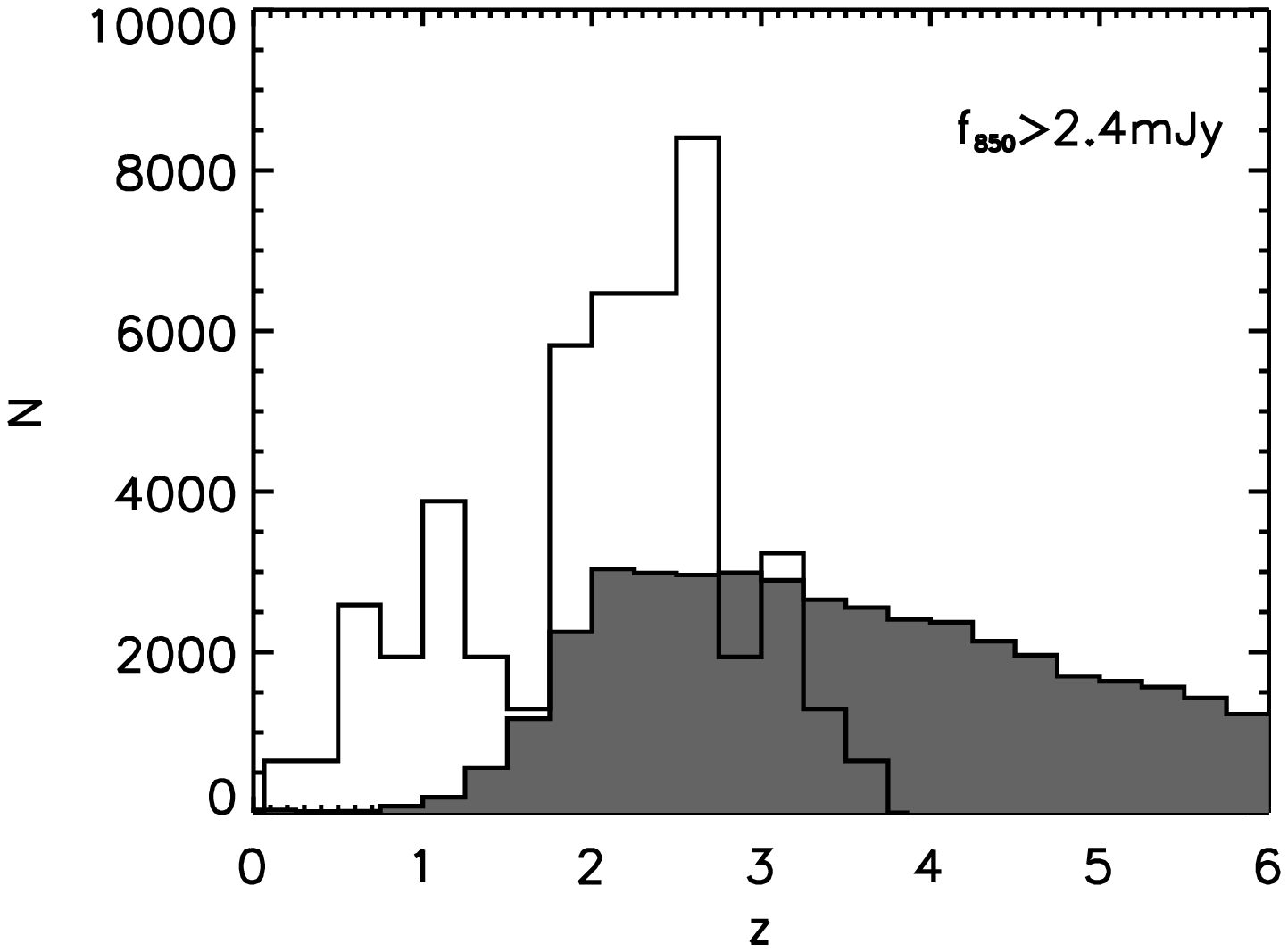}}\\
\caption{Number counts at 24, 70, 160, $850\,\mu\rm{m}$ and redshift distributions at 24 and $850\,\mu\rm{m}$, considering evolutions with $n_1={3.6}$, $n_2=0$, $m_1=1$, $m_2=0$, $z_1=z_2=2$, {\itshape without AGN contribution} (M1, solid lines). Data are from \citet{papovich04} ({\itshape filled circles}) and \citet{shupe08} ({\itshape open squares}) for the $24\,\mu\rm{m}$ number counts, \citet{frayer06,frayer06b} ($70\,\mu\rm{m}$ and $160\,\mu\rm{m}$ counts), \citet{coppin06} ($850\,\mu\rm{m}$ counts), \citet{wuyts08} ($24\,\mu\rm{m}$ redshifts) and \citet{chapman05} ($850\,\mu\rm{m}$ redshifts). Euclidean normalized differential counts are shown at 24, 70, $160\,\mu\rm{m}$, while integrated counts are shown at $850\,\mu\rm{m}$ following the style of the papers in which they were originally presented. Redshift data ({\itshape open histograms}) have been scaled for comparison with the simulations ({\itshape filled histograms}).
} 
\label{M1}
\end{figure}

\begin{figure}
\centering

{\includegraphics[width=7.cm,keepaspectratio]{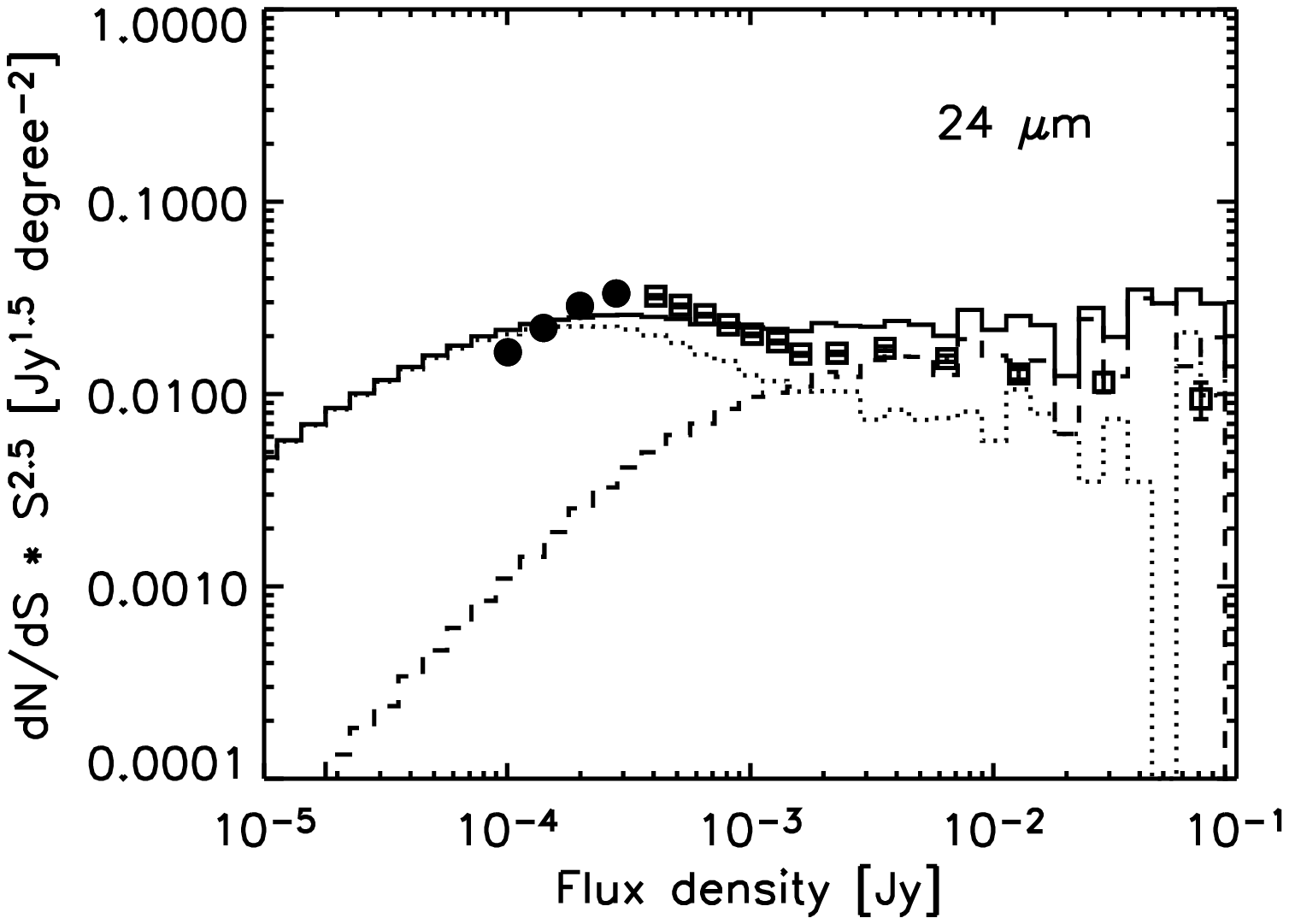}}
{\includegraphics[width=7.cm,keepaspectratio]{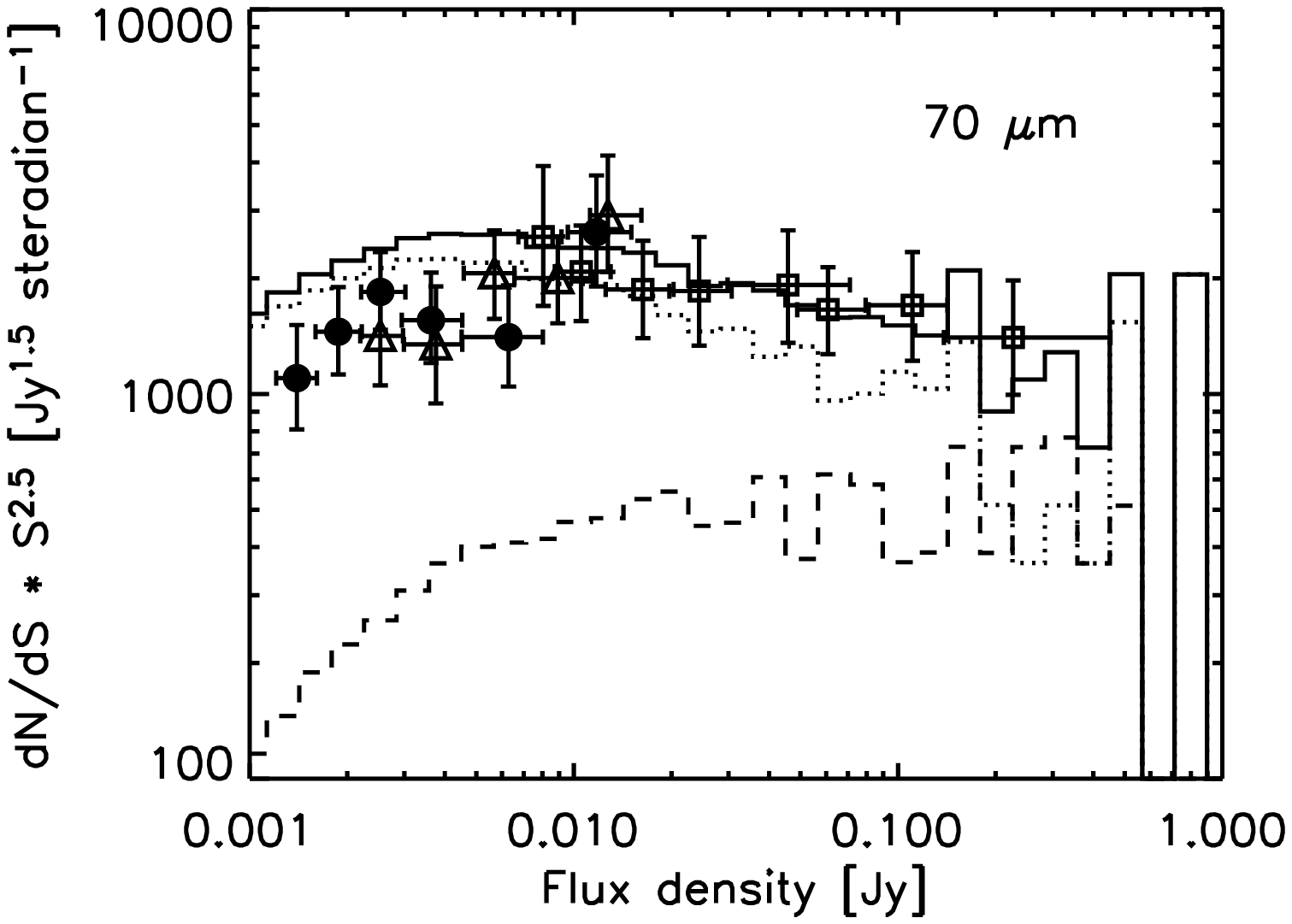}}\\
{\includegraphics[width=7.cm,keepaspectratio]{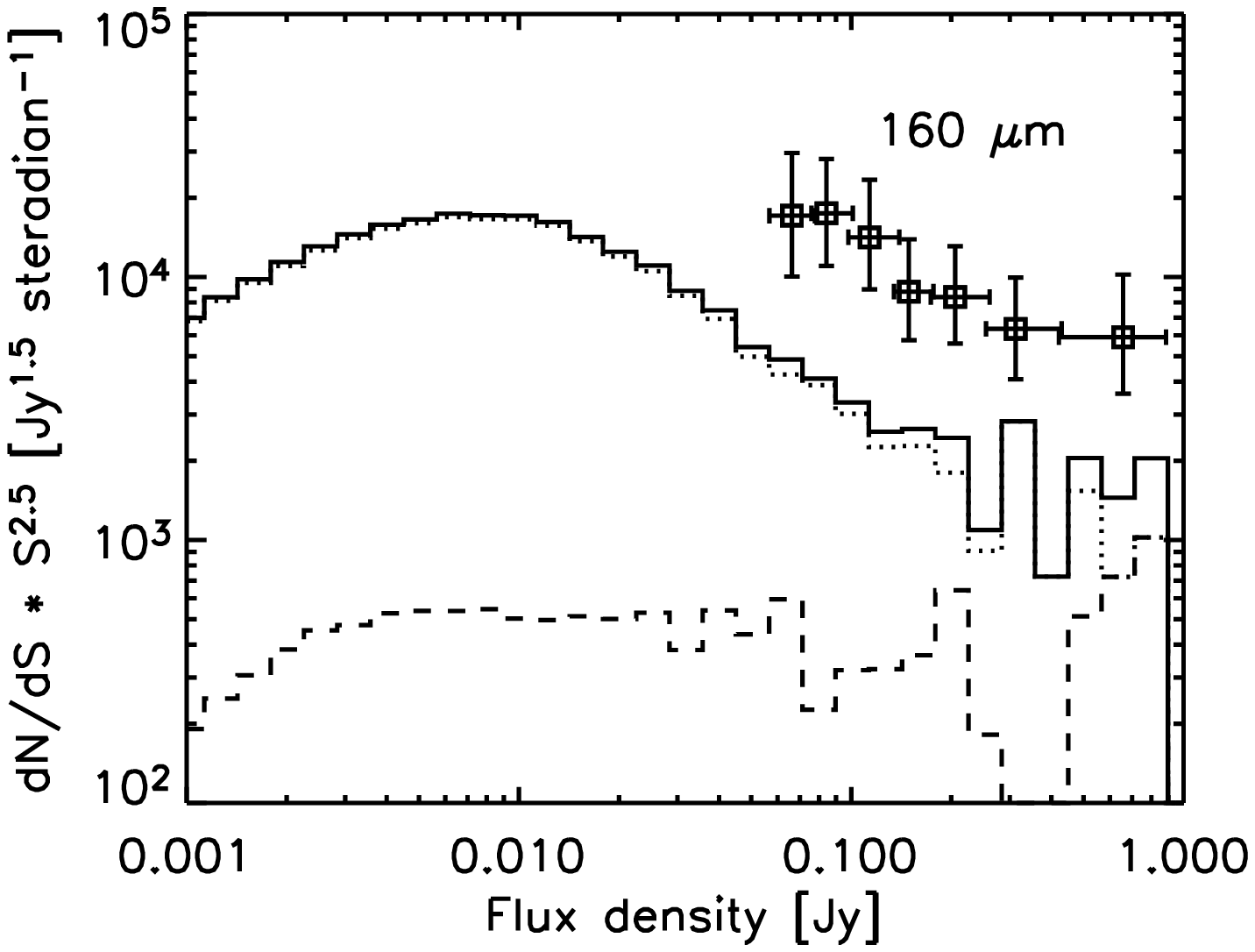}}
{\includegraphics[width=7.cm,keepaspectratio]{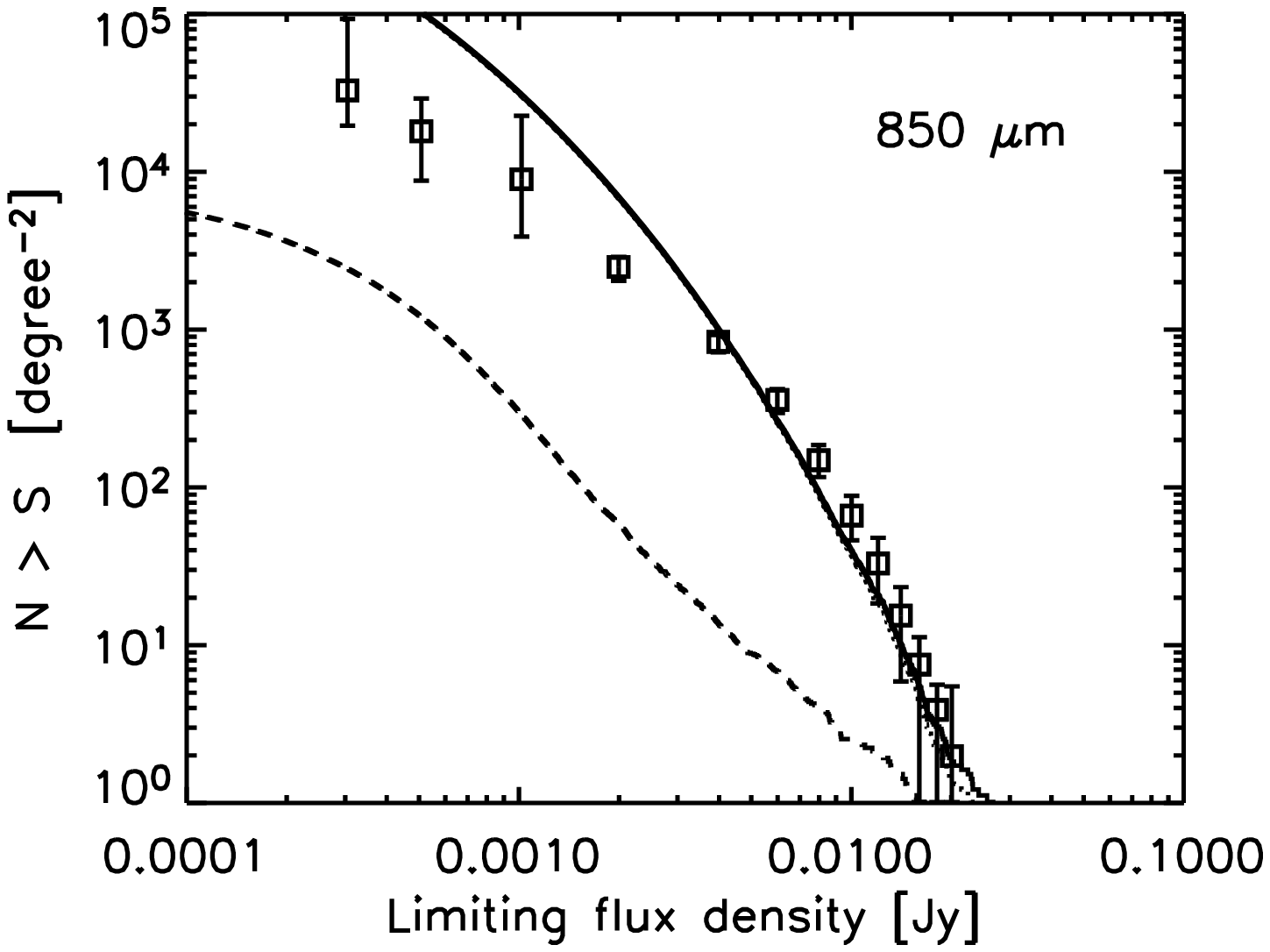}}\\
{\includegraphics[width=7.cm,keepaspectratio]{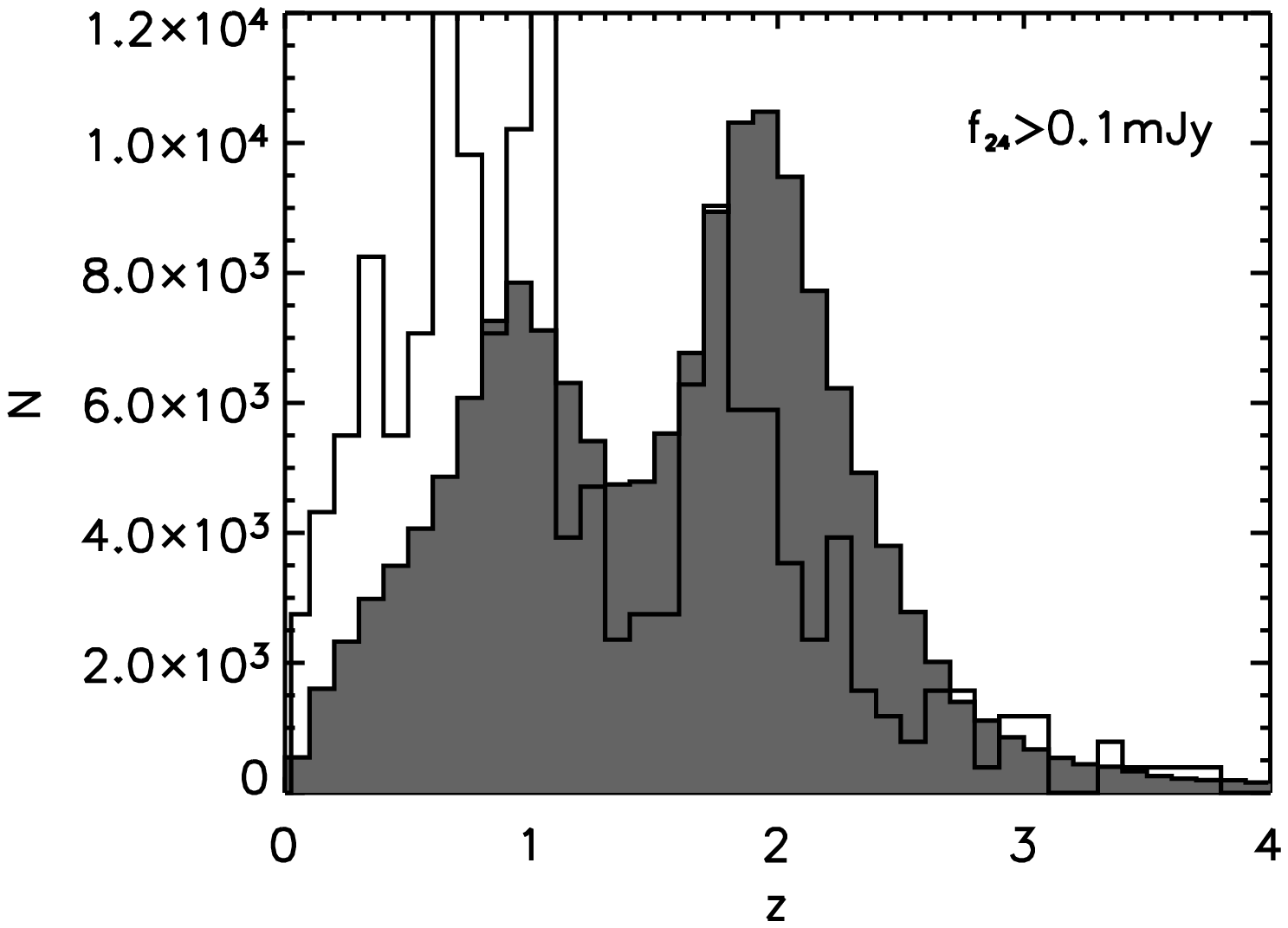}}
{\includegraphics[width=7.cm,keepaspectratio]{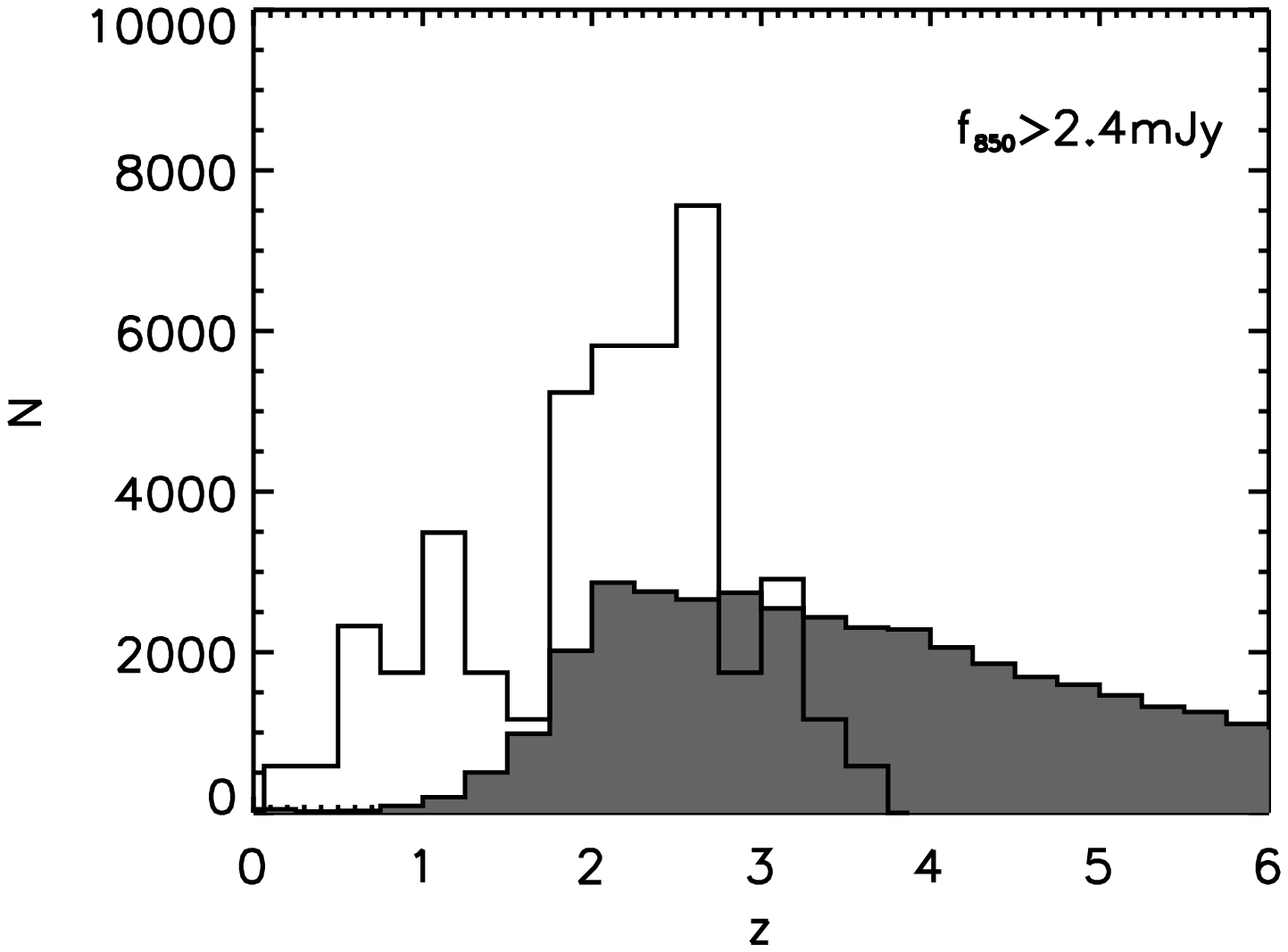}}\\
\caption{Same as Fig.~\ref{M1}, but now {\itshape with AGN contribution} (M2). AGN-dominated galaxies ($L_{\rm AGN}/L_{\rm SB}\geq 1$) are represented with dashed line, while starburst dominated galaxies ($L_{\rm AGN}/L_{\rm SB}< 1$) are in dotted line.
} 
\label{M2}
\end{figure}

\begin{figure}
\centering

{\includegraphics[width=7.cm,keepaspectratio]{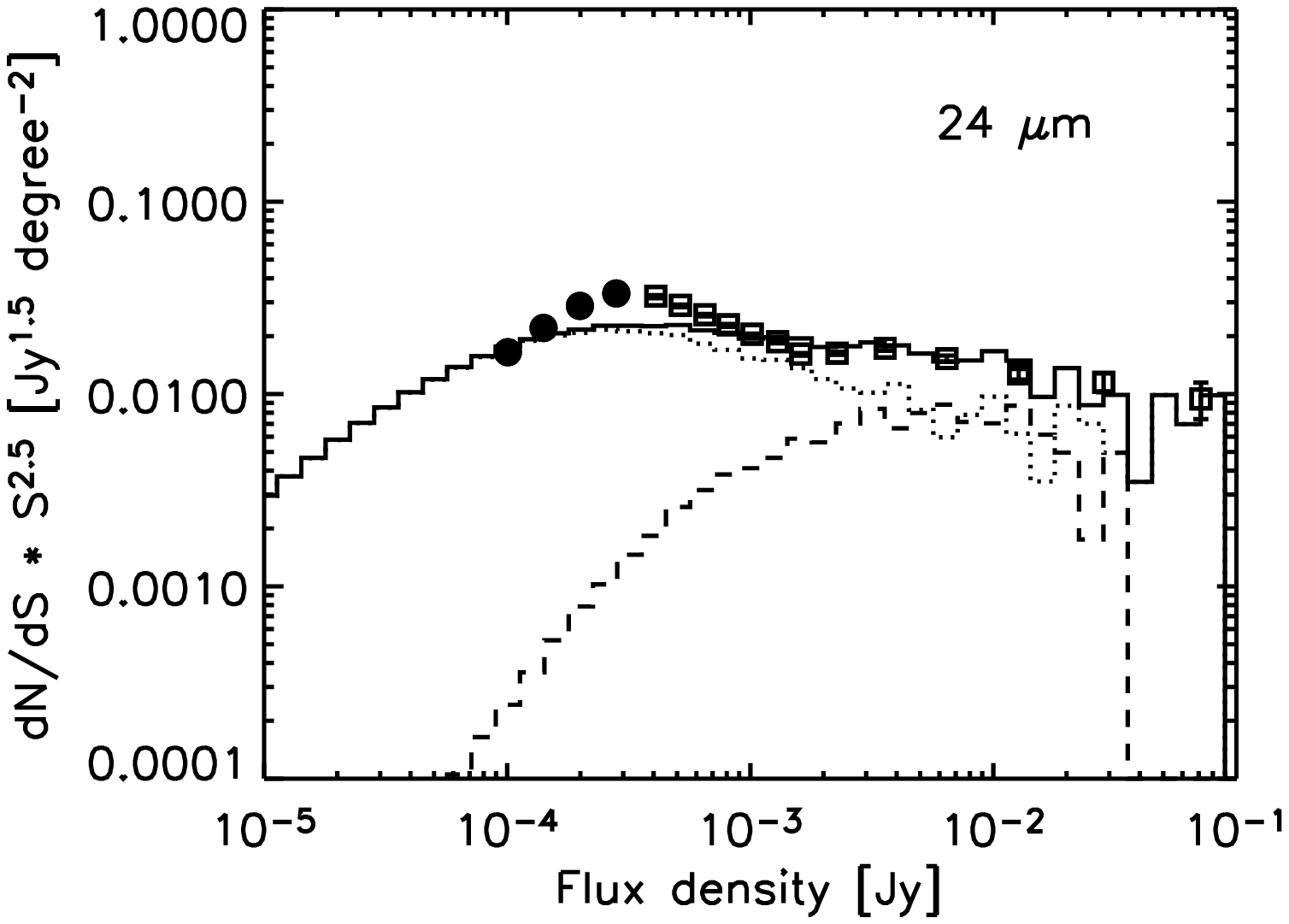}}
{\includegraphics[width=7.cm,keepaspectratio]{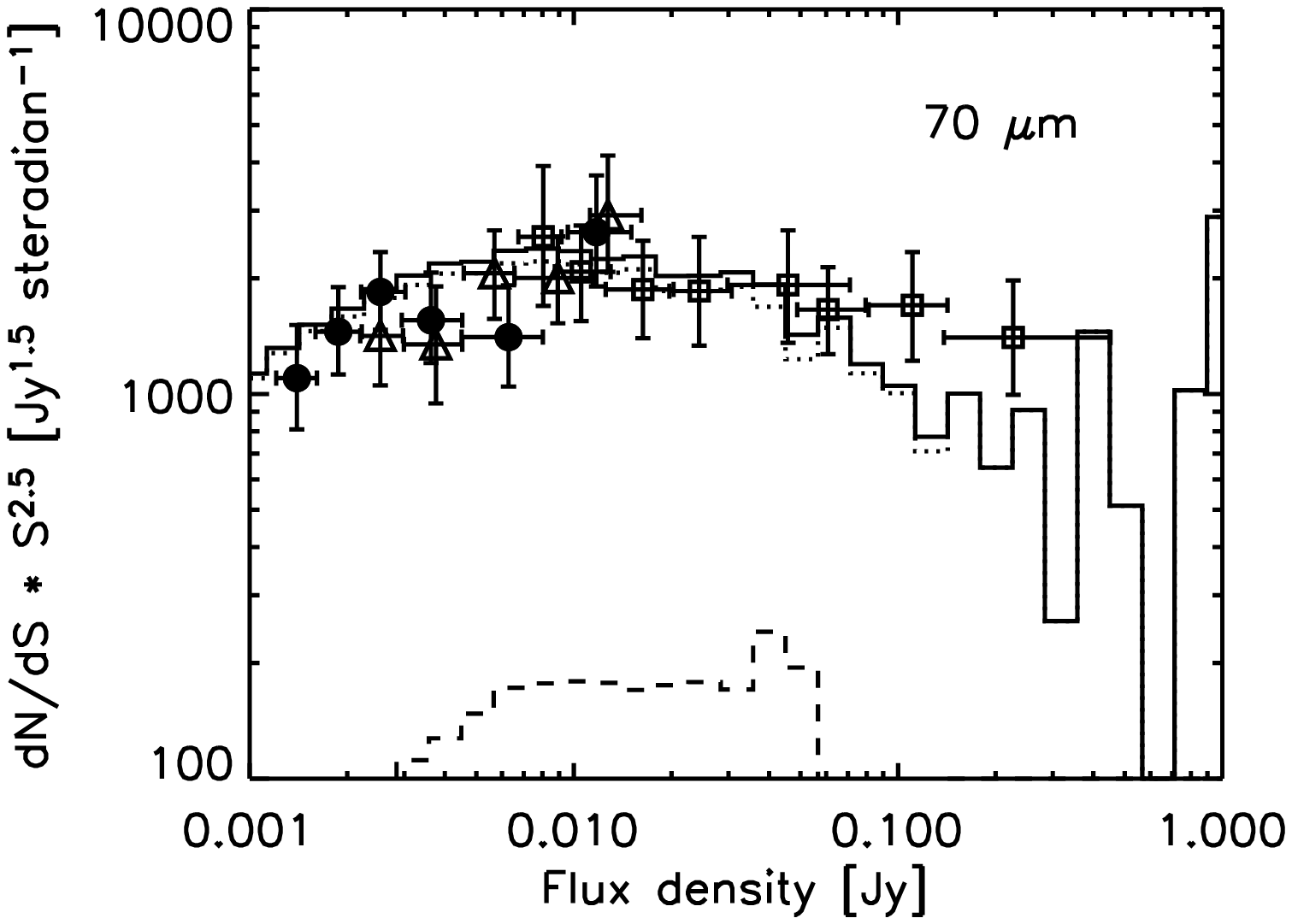}}\\
{\includegraphics[width=7.cm,keepaspectratio]{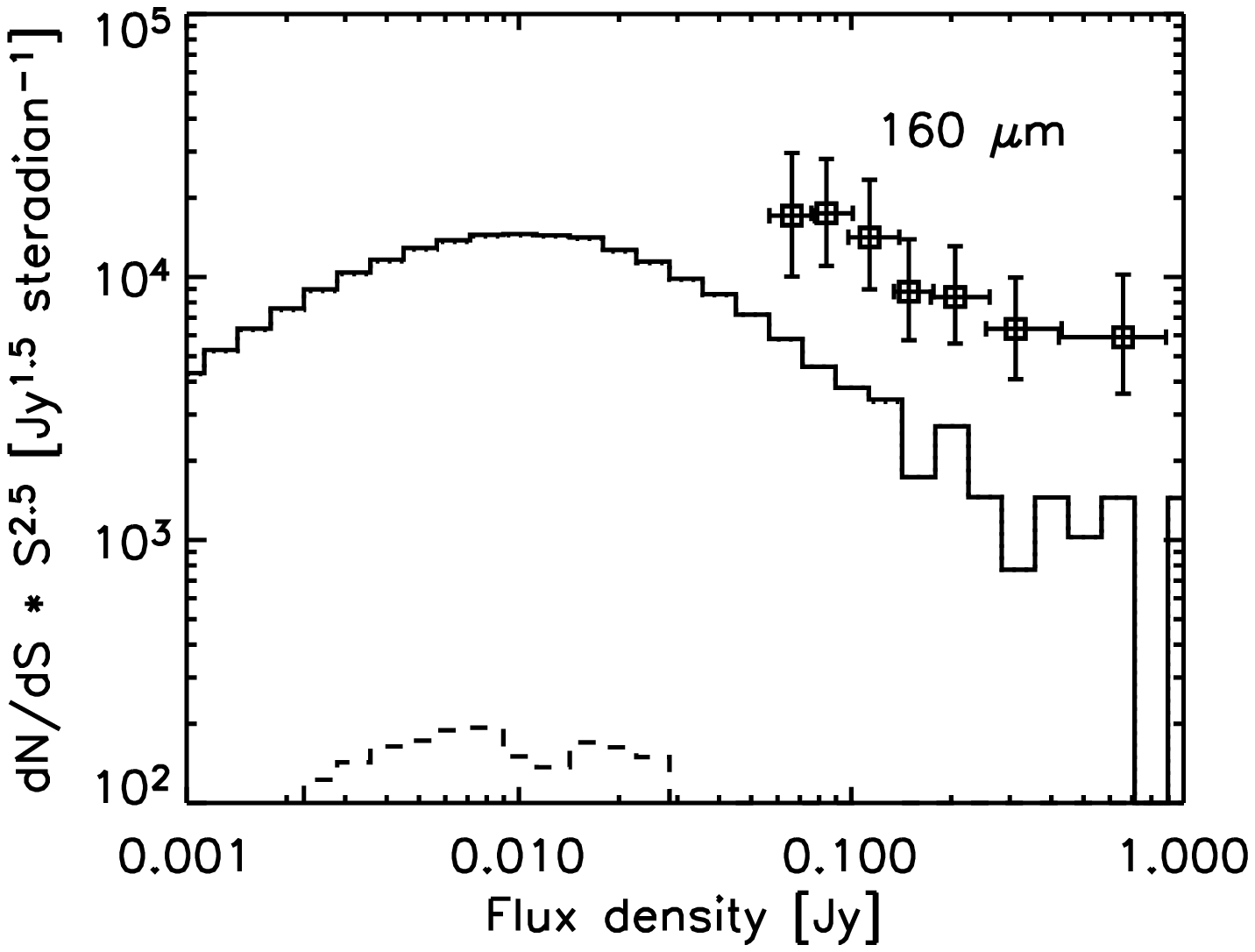}}
{\includegraphics[width=7.cm,keepaspectratio]{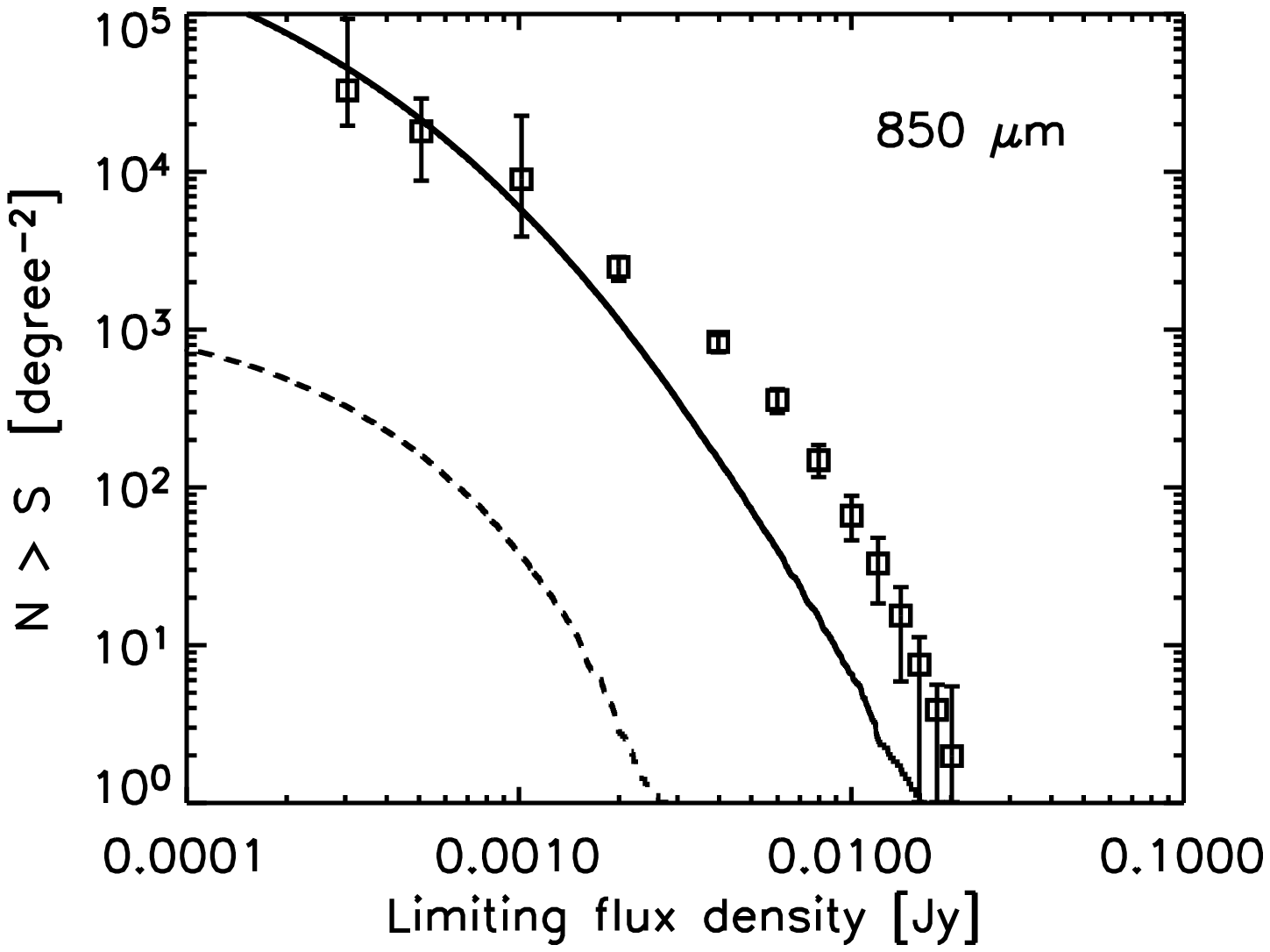}}\\
{\includegraphics[width=7.cm,keepaspectratio]{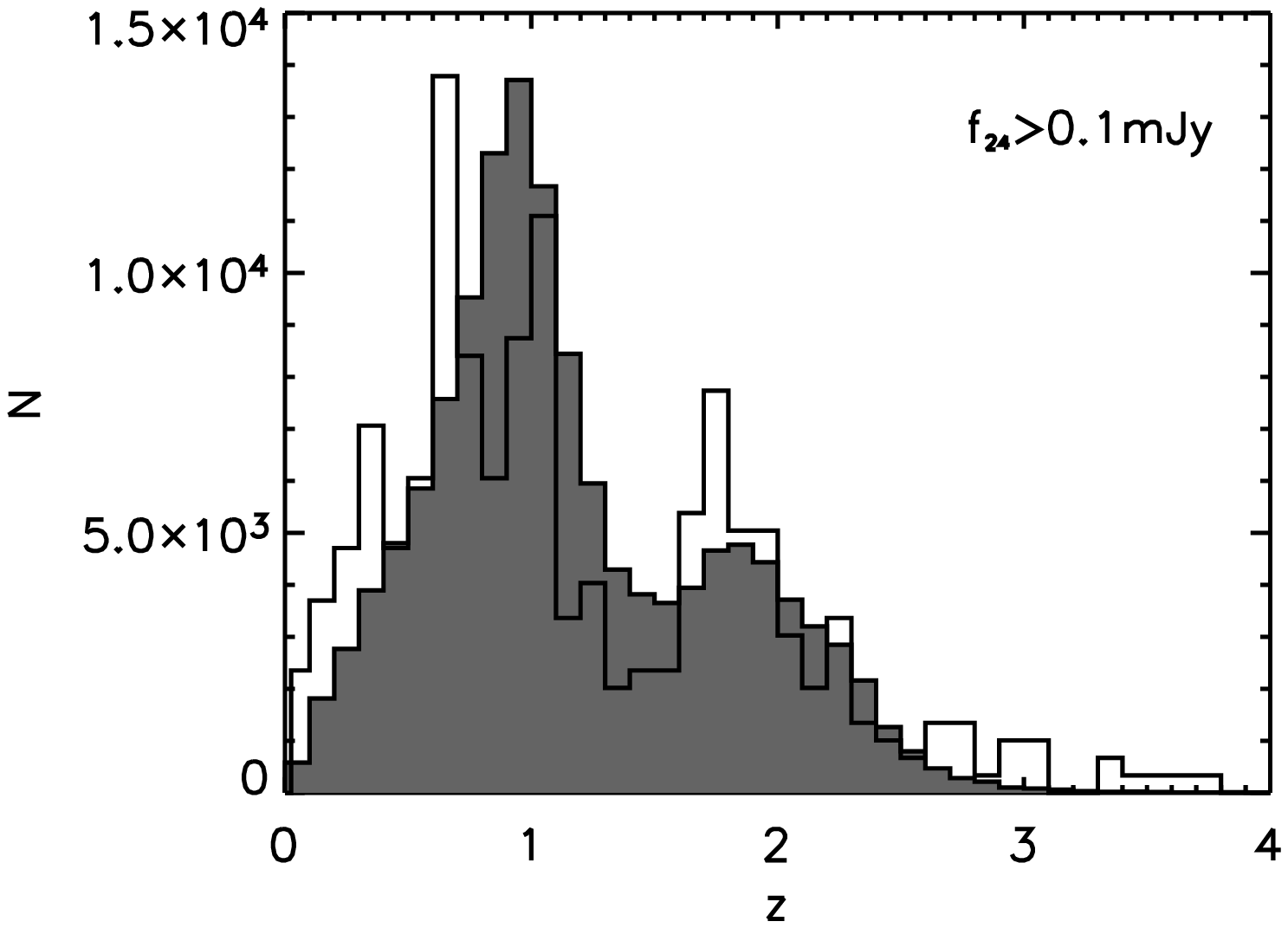}}
{\includegraphics[width=7.cm,keepaspectratio]{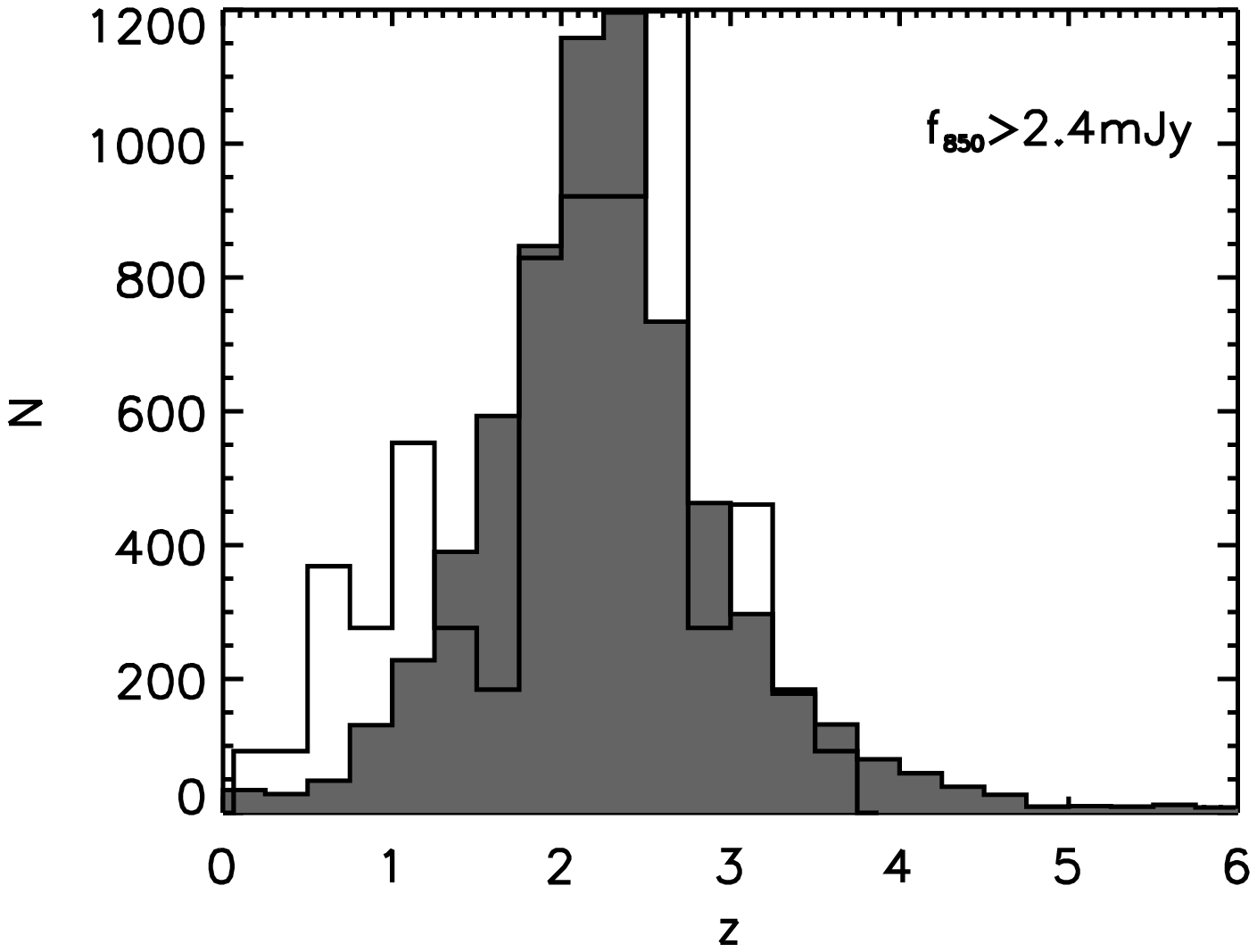}}\\
\caption{Number counts at 24, 70, 160, $850\,\mu\rm{m}$ and redshift distributions at 24 and $850\,\mu\rm{m}$, considering evolutions with $n_1={3.4}$, $n_2=-1$, $z_1=2.3$, $m_1=2$, $m_2=-1.5$ and $z_2=1$, {\itshape with AGN contribution} (M3). Symbols are like in Fig.~\ref{M2}. Data are like in Fig.~\ref{M1}.
} 
\label{M3}
\end{figure}

\begin{figure}
\centering

{\includegraphics[width=7.cm,height=5.cm]{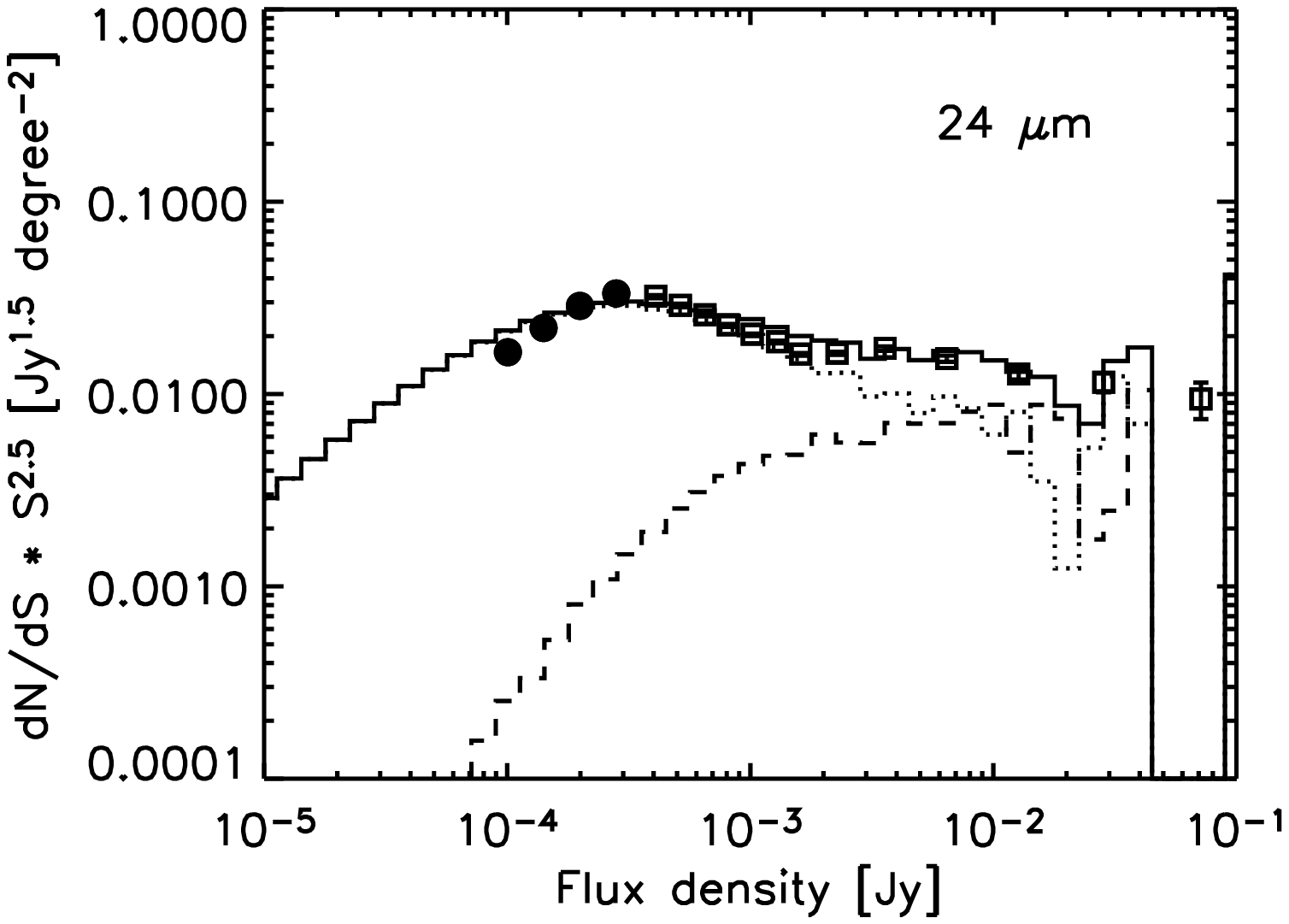}}
{\includegraphics[width=7.cm,height=5.cm]{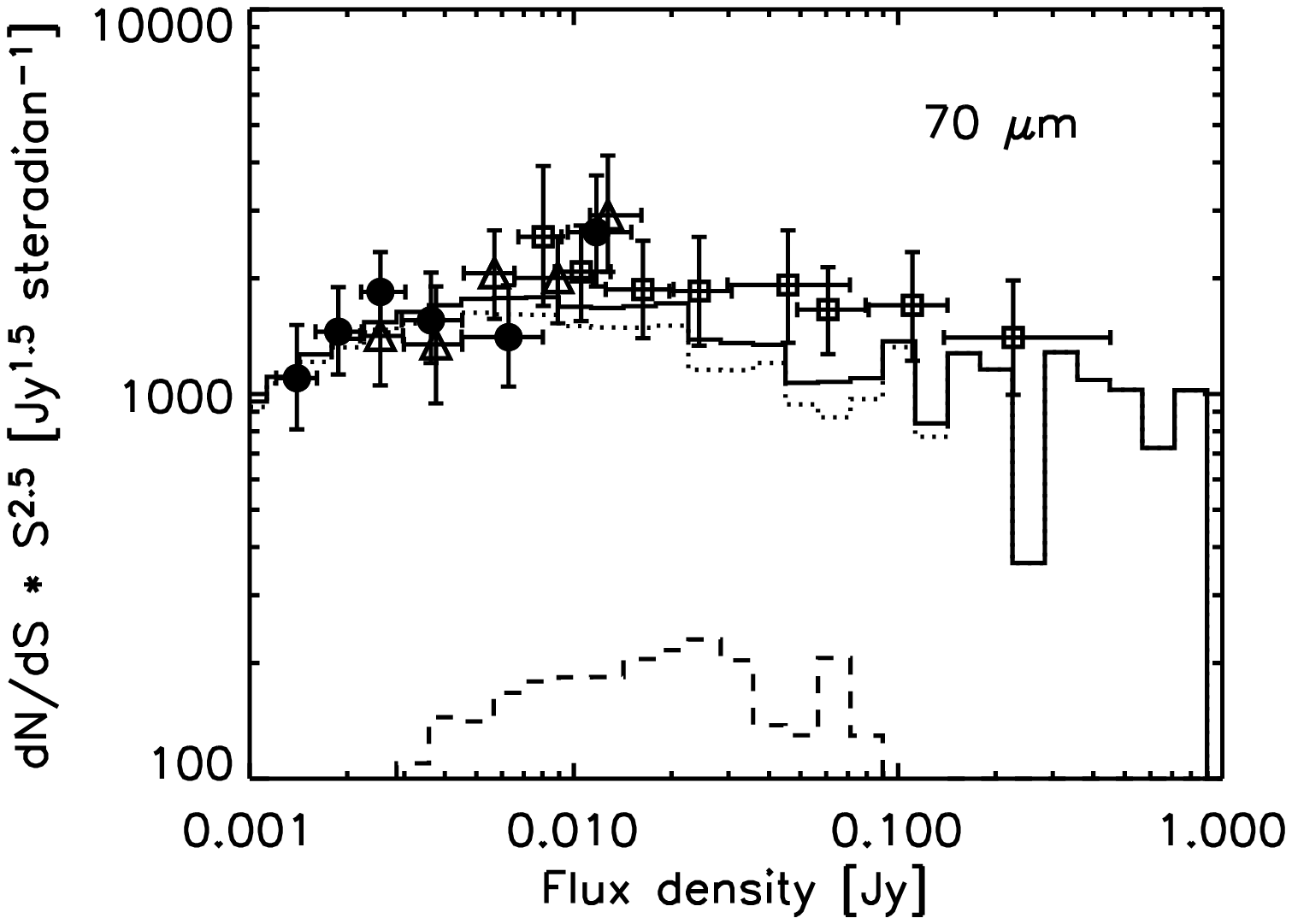}}\\
{\includegraphics[width=7.cm,height=5.cm]{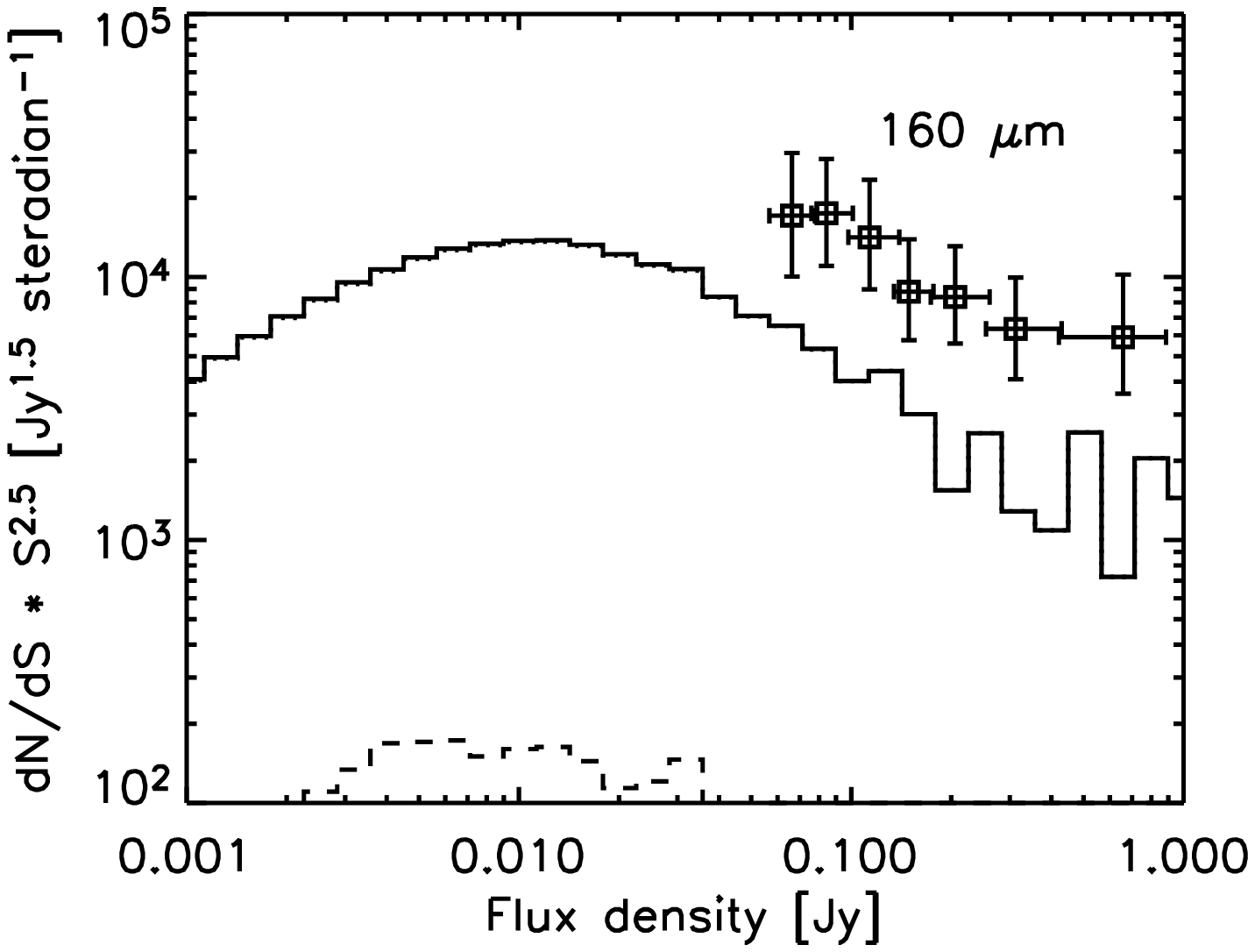}}
{\includegraphics[width=7.cm,height=5.cm]{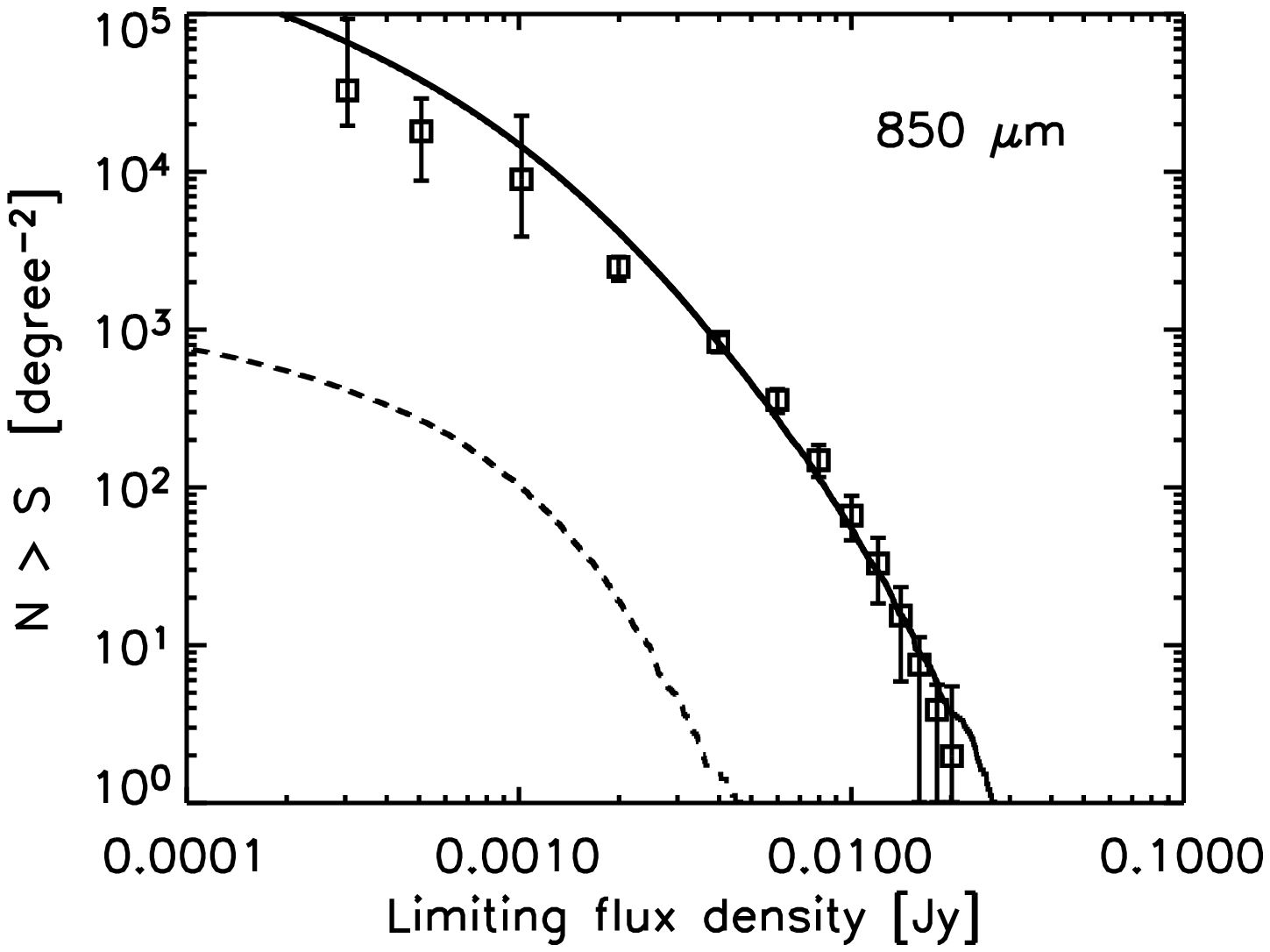}}\\
{\includegraphics[width=7.cm,height=5.cm]{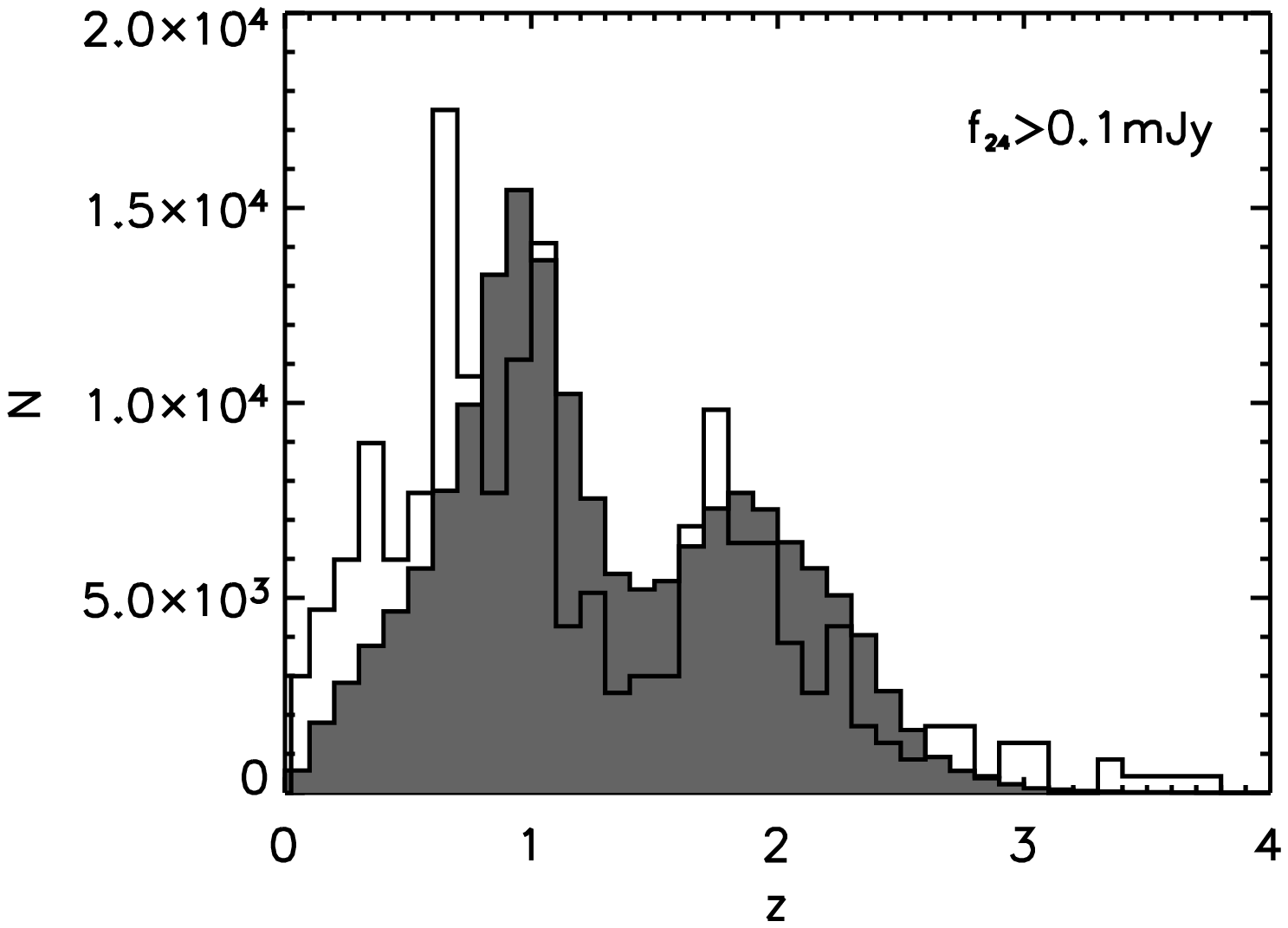}}
{\includegraphics[width=7.cm,height=5.cm]{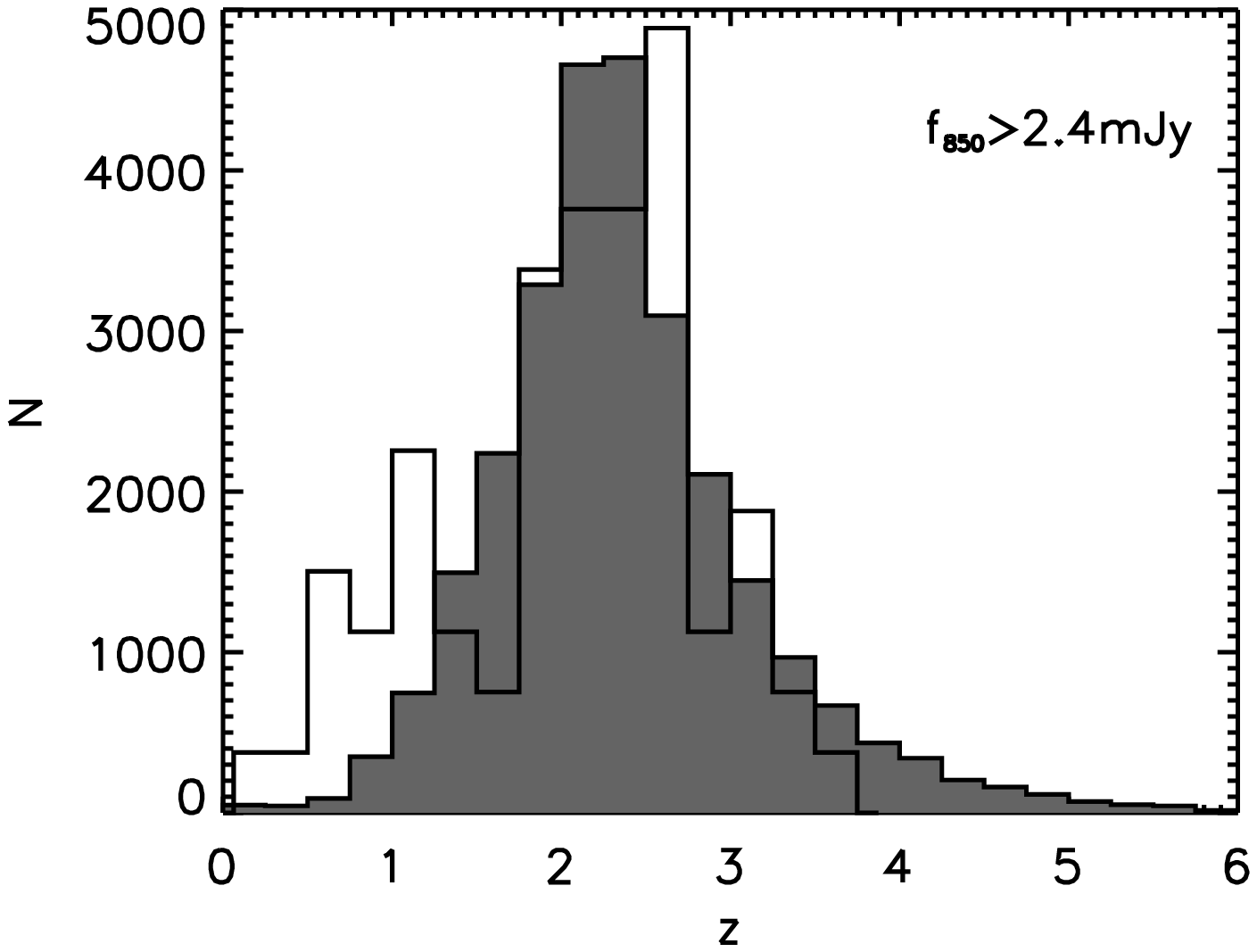}}\\
\caption{Same as Fig.~\ref{M3}, assuming an {\itshape evolution in the $L-T$ relation} (M4).
} 
\label{M4}
\end{figure}

The most relevant results of the last years are related to surveys carried out by SCUBA and {\itshape Spitzer}. Thus, we will compare our results with the data available for these instruments. Hereafter, we will use number counts from \citet{papovich04} and \citet{shupe08} for the $24\,\mu\rm{m}$ sources, \citet{frayer06,frayer06b} for the $70\,\mu\rm{m}$ sources, \citet{frayer06} for the $160\,\mu\rm{m}$ sources, \citet{coppin06} for the $850\,\mu\rm{m}$ sources. Redshift distributions are from \citet{wuyts08} ($24\,\mu\rm{m}$ sources) and \citet{chapman05} ($850\,\mu\rm{m}$ sources). Some of the models simulated are also compared with the CIB measurements \citep{fixsen98,lagache00,renault01,mivilledeschenes02,lagache03}. All the simulations are run on a field of view of $10\,\rm{deg}^2$, in order to be comparable with the main infrared surveys and sufficiently sample the high luminosity end. The redshift range covered is $0\leq z \leq 8$. In the plots of the redshift distributions (see Fig.~\ref{M1}$-$Fig.~\ref{M4}), the available data have been appropriately scaled in order to be comparable with the simulations.

In order to find the best set of parameters ($n_1,n_2,z_1,m_1,m_2,z_2$) fitting available data we have run simulations on small fields of view, changing slightly each parameter ``by hand''. Changing the parameters one by one allows us to understand how each parameter influences the final results: we have used this approach to have full control of the simulation and to understand the role and weight of each parameter. During the optimization process, we have considered differently the several constraints given from current surveys, depending on their reliability: first, we have worked to reproduce the most accurate measurements and only later we have tried to fit the less constrained results. For example, SHADES \citep{coppin06} has made accurate measurements of the number counts at $850\,\mu\rm{m}$ down to faint fluxes ($\sim 2\,\rm{mJy}$) and {\itshape Spitzer} has provided very sensitive measurements with minimum uncertainties of the number counts at $24\,\mu\rm{m}$, either with observations on large scales ($49\,\rm{deg}^2$, \citealt{shupe08}) or very deep, down to fluxes $\sim 0.1\,{\rm mJy}$ \citep{papovich04}. On the other hand, the redshift distribution of submillimeter galaxies is not well constrained at high $z$. In fact, the radio-selected sample of \citet{chapman05} is biased towards $z\lesssim 3$ because of the radio flux limit, and there are several hints of galaxies at $z\sim 4$ \citep{valiante07,knudsen08,daddi09}. The redshift distribution of $24\,\mu\rm{m}$ sources are obtained from photometric redshifts on a $K_S$-band selected galaxies in the GOODS-S field \citep{wuyts08}: cosmic variance can have an important role on the redshift distribution in a field so small, as well as the selection and the choice of the SEDs for photometric redshifts. All these effects could overestimate the number of low redshift objects. Finally, the surveys at $160\,\mu\rm{m}$ \citep{frayer06} are still rather shallow.

Below, we present a summary of the optimization process, showing the main steps followed to obtain the best parameter set ($n_1,n_2,z_1,m_1,m_2,z_2$). A summary of the parameters for all the models discussed are shown in Tab.~\ref{tab:models}.




\subsection{Pure starbursts SEDs}



As a starting point, we compare the observations with a model considering only the starburst part of the SED, including its scatter in the $L-T$ relation (model M1). The submillimeter galaxies are starburst dominated, so we look for the best parameter set reproducing at least the $850\,\mu\rm{m}$ number counts. We assume a strong evolution in luminosity ($n_1=3$, $n_2=0$) and a weaker evolution in density ($m_1=1$, $m_2=0$), with $z_1=z_2=2$. The results are shown in Fig.~\ref{M1}. The bright $850\,\mu\rm{m}$ counts are well reproduced. However, all the other counts are underestimated: using this evolutionary model, there is still room to introduce an SED taking into account the AGN contribution. The redshift distribution at $24\,\mu\rm{m}$ is quite well reproduced, while at $850\,\mu\rm{m}$ the modeled redshift distribution shows a strong excess at $z\gtrsim 3.5$. Adding the AGN part, we can be confident that the submillimeter distribution will not change much, but we should not expect the same trend in the mid-IR, since there the AGN contribution can be significant, in particular at high redshifts. 

\subsection{Adding the AGN contribution} \label{section:addAGN}

In this second step, we build SEDs as described in \S~\ref{section_sed}. In particular, we determine the AGN contribution following Eq.~\ref{eq_AGN}. As calculated in \S~\ref{section:AGN}, we assume a gaussian distribution in each infrared luminosity bin (see Tab.~\ref{AGNresults}). The results, assuming the same evolution applied previously, are shown in Fig.~\ref{M2} (M2). Compared to model M1, the $24\,\mu\rm{m}$ number counts are now better reproduced, even if the shape of the peak is slightly different. The redshift distribution at $24\,\mu\rm{m}$ has become worse: the AGN contribution has introduced a large excess at $z\sim 2$. The parameter set used up to now is no more acceptable and we have to explore the parameter space further, in order to reproduce all the observations.



Comparing different kinds of evolution, we learn that we need a strong luminosity evolution in order to reproduce the number counts at $24\,\mu\rm{m}$ and the $850\,\mu\rm{m}$, but this always introduces an overprediction of the $24\,\mu\rm{m}$ $z\sim 2$ sources. On the other side, with a strong density evolution we obtain a redshift distribution at $24\,\mu\rm{m}$ that is very similar to the observed one, but we completely lose the peak in the $24\,\mu\rm{m}$ number counts. We note, moreover, that a shift of the peak of the density evolution, $z_2$, towards lower redshifts, causes a change in the shape of the redshift distribution at $24\,\mu\rm{m}$, reducing the number of the $z\sim 2$ sources, without changing the $24\,\mu\rm{m}$ number counts. Last, if we want to reduce the number of high redshift $850\,\mu\rm{m}$ sources, we need a luminosity evolution peaking at $z_1$ and then dropping down ($n_2$ has to be negative). Fig.~\ref{M3} shows number counts and redshift distributions for a model with $n_1=3.4$, $n_2=-1$, $z_1=2.3$, $m_1=1$, $m_2=-1.5$ and $z_2=1$ (M3). Data at $24\,\mu\rm{m}$ are well reproduced, in particular the predicted peak of the number counts has now the same shape of the measurements even if it is slightly lower, while at $850\,\mu\rm{m}$ the simulation underpredicts the integral number counts but reproduces the redshift distribution.



\subsection{Evolving the $L-T$ relation}

In this simulation, the starburst part of the SED still does not vary from the local to the distant universe. This last point is in contradiction to recent works already mentioned in \S~\ref{section_sed}, showing that submillimeter galaxies, even being as luminous as local ``HYLIRGs'', have dust temperatures typical of less luminous local galaxies (ULIRGs) \citep{chapman05,pope06,valiante07}. \citet{pope06}, in particular, needed to modify the ``classical'' SED model of \citet{chary01} increasing the cold component of the dust, in order to fit the SEDs of the submillimeter galaxies in the GOODS-N field. 



We then include in our model a simple evolution in the $L-T$ relation, accounting for the cooler temperatures of high redshift infrared galaxies. We modify Eq.~\ref{eq:temp} in the following way: 
\begin{equation}
\begin{array}{l l} 
R(60,100)= & 10^{C_*} \times\left(1+\frac{L^{\prime}}{L_{\rm SB}/(1+z)^{1.5}}\right)^{-\delta} \\
& \times\left(1+\frac{L_{\rm SB}/(1+z)^{1.5}}{L^{\prime}}\right)^{\gamma}
\end{array}
\end{equation}
with all the parameters as in Eq.~\ref{eq:temp}. This evolution is consistent with the observations cited above. Fig.~\ref{M4} shows number counts and redshift distributions for this new model with $n_1=3.4$, $n_2=-1$, $z_1=2.3$, $m_1=1$, $m_2=-1.5$ and $z_2=1$ (M4). All the submillimeter measurements are now well reproduced, with perhaps some excess in the number counts at strong fluxes, together with the mid-IR data, whose number counts have improved and redshift distribution has not felt the effect of the new SEDs. The number counts in the far-IR, in particular at $160\,\mu\rm{m}$, are nevertheless underestimated.

Even if the peak of the $24\,\mu{\rm m}$ number counts is well reproduced by the model M4, there is still some excess in the number counts at strong fluxes, where AGN dominate, but the uncertainties are large given the small number statistics. We compare the number of AGN predicted by our model with the number of sources expected adopting QSO luminosity functions. We select only the brightest $24\,\mu\rm{m}$ sources ($f_{24}\geq 5\,\rm{mJy}$) and consider only the AGN-dominated sources ($L_{\rm AGN}/L_{\rm SB}\geq 1$). With the exception of some local and faint objects, all the sources selected in this way are highly luminous, distant AGN. Applying a simple bolometric correction, $L_{\rm BOL}\sim 10 \nu L_{\nu}$ at mid-IR rest wavelengths, we can calculate the expected number of QSOs for the same $L_{\rm BOL}$ and $z$ ranges of our subsample. We integrate the luminosity function of \citet{hopkins07} (``full'' model, with pure luminosity evolution and bright- and faint-end slopes evolving with redshift) for $10^{12.2}\leq L_{\rm BOL}\leq 10^{14.5}\,L_{\odot}$ and $0.1\leq z\leq 2.3$. The number of the sources selected from our simulation does not exceed the expected number given by the bolometric luminosity function. Nevertheless, we already pointed out that the high part of the $\nu L_6(\rm AGN)/L_{\rm TIR}$ distribution is not well constrained (see \S~\ref{simsec}). In particular, the hypothesis of the gaussian distribution could be wrong: it was assumed following the principle of the minimization of the free parameters, but a different kind of distribution, for example an asymmetric distribution with a cut-off at high luminosities, cannot be excluded (see \S~\ref{AGNcontrib}).

\subsection{Evidence for a local ``cold'' population?}



Our models were no able to reproduce the number counts at $160\,\mu\rm{m}$, underestimating them by a factor of $\sim 5$. The ISOPHOT Serendipity and FIRBACK surveys have revealed a population of nearby cold galaxies \citep{stickel98,stickel00,chapman02,patris03,dennefeld05,sajina06}, under-represented in the $60\,\mu{\rm m}$ IRAS sample. These objects are often associated with bright optical spiral galaxies, and their far-IR colours ($f_{170}/f_{100}\sim 1.3$) indicate a rising spectrum beyond $100\,\mu\rm{m}$, similar to that seen for example in the Milky Way galactic ridge \citep{serra78}. 

\citet{lagache03} implemented this class of objects (hereafter ``cold galaxies'') in their evolutionary model: cold galaxies dominate the $z=0$ luminosity function at low luminosities and become less important at higher redshifts, because their evolution is passive and short ($n_1=1,n_2=0,z_1=0.4$). Their contribution to the high luminosity part of the luminosity function is still substantial up to $L_{\rm TIR}\sim 10^{12}\,L_{\odot}$, while at larger luminosities it is quite small. \citet{lagache03} show that the fraction of cold galaxies can contribute to the total number counts up to $\sim 50\%$ at $170\,\mu\rm{m}$ in the flux range where a comparison with measurements is possible ($0.1-1.0\,{\rm mJy}$, see their Fig.~9). The modifications added later to the model have not changed the number counts at $160\,\mu\rm{m}$ (\citealt{lagache04}, their Fig.~4). 

In order to introduce a similar population in our model, we have to assume that probably the scatter in the $L-T$ relation calculated by \citet{chapin09} and implemented in our simulations needs some corrections in order to take into account the cold galaxies population. The lack of these corrections can explain why we underestimate the $160\,\mu\rm{m}$ number counts. Moreover, any changes of the adopted $L-T$ relation and its scatter has to be constrained by the $850\,\mu{\rm m}$ counts. The $160\,\mu\rm{m}$ counts need to be increased without creating excess $850\,\mu{\rm m}$ counts. This may support a mostly local change. Moreover, colder galaxies with a higher $160\,\mu\rm{m}$ flux have typically lower $70\,\mu\rm{m}$ flux: any change introduced cannot be too much strong. \citet{chapin09} discussed two SMGs that are outliers respect to their $L-T$ relation: they explain the exceptions assuming a mismatching in the optical counterparts. Nevertheless, they do not exclude that both the galaxies are really at low redshift and a population of cold galaxies do exist at $z<1$ and are largely missed in IRAS surveys. 

\begin{figure}
\centering

{\includegraphics[width=7.cm,keepaspectratio]{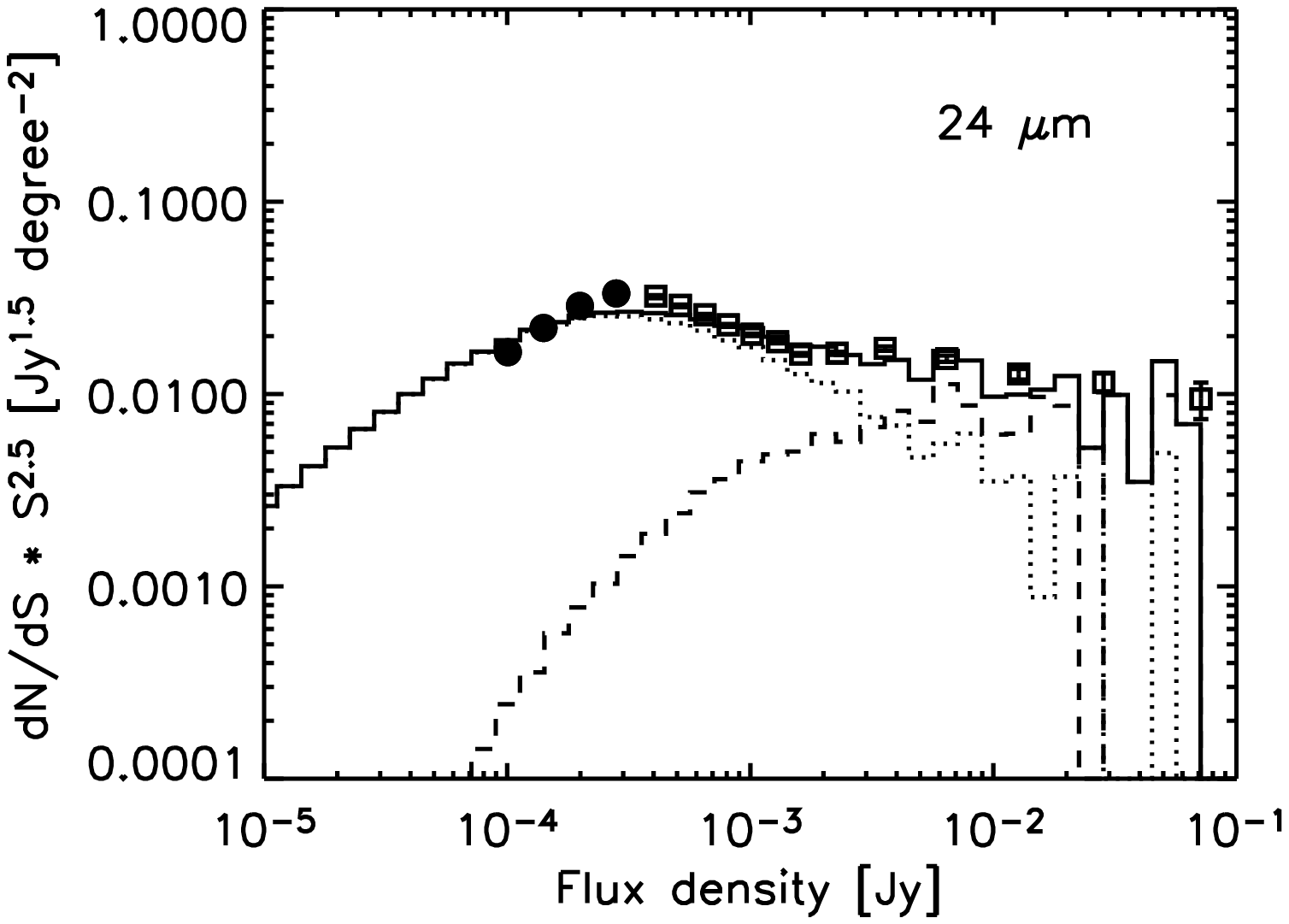}}
{\includegraphics[width=7.cm,keepaspectratio]{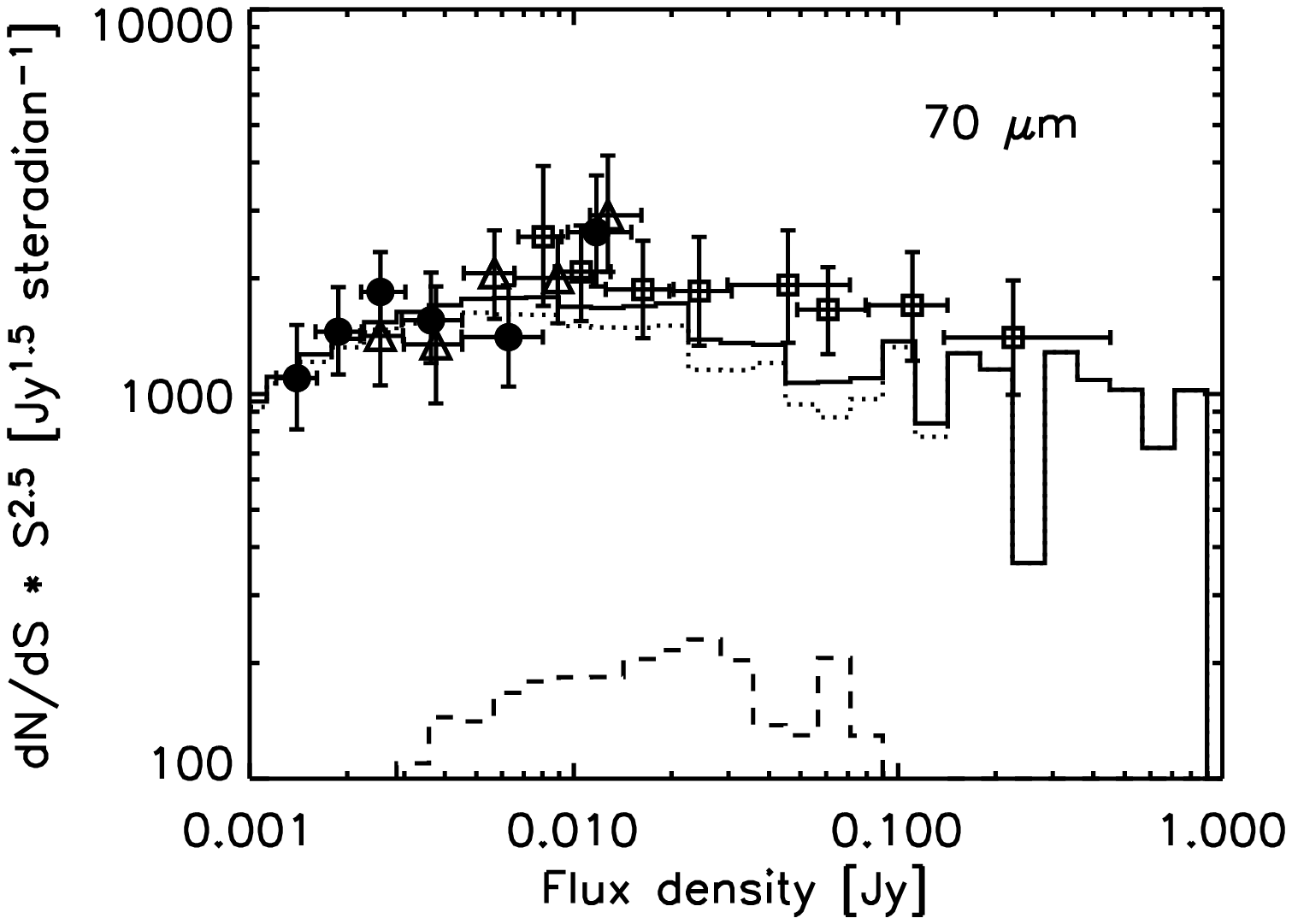}}\\
{\includegraphics[width=7.cm,keepaspectratio]{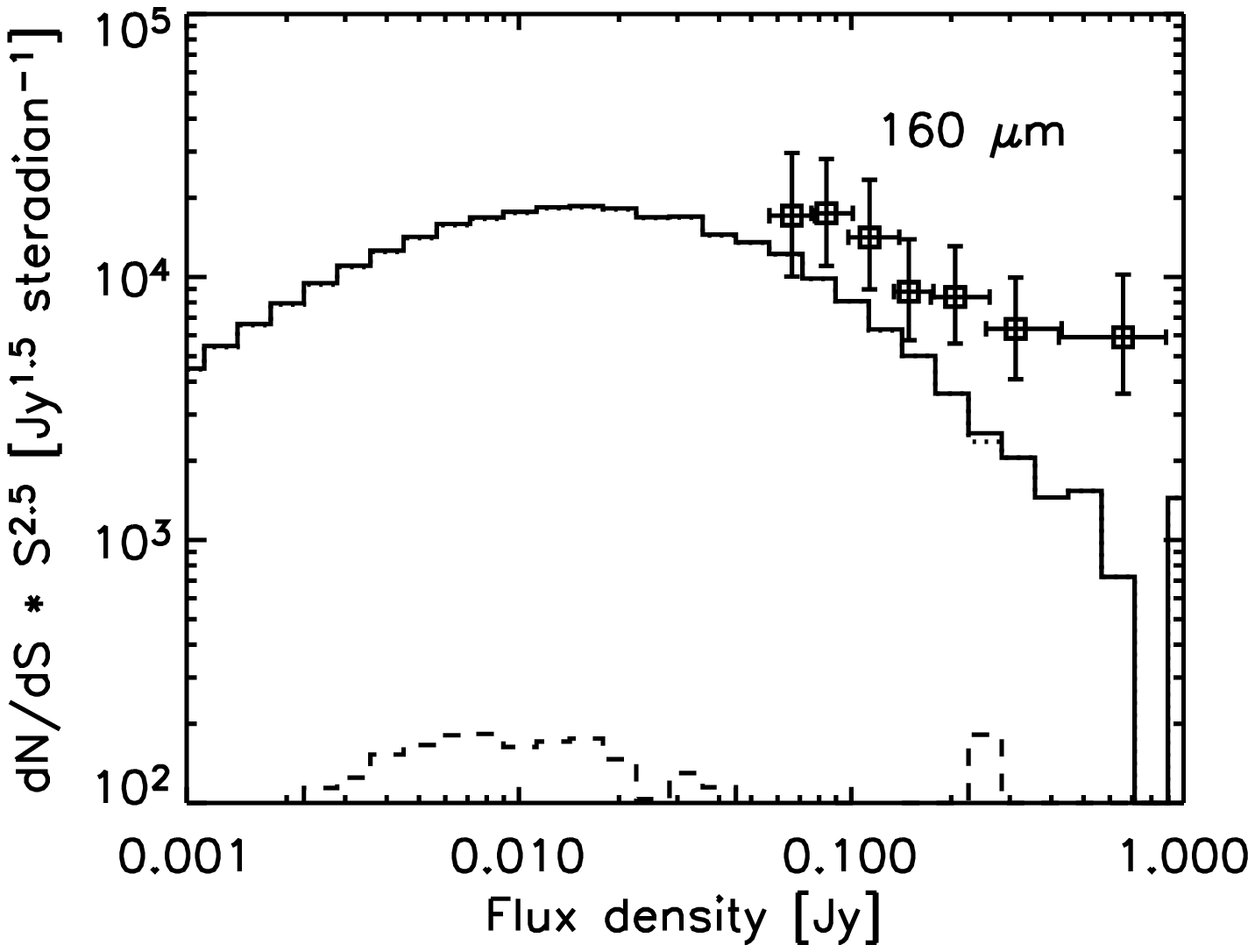}}
{\includegraphics[width=7.cm,keepaspectratio]{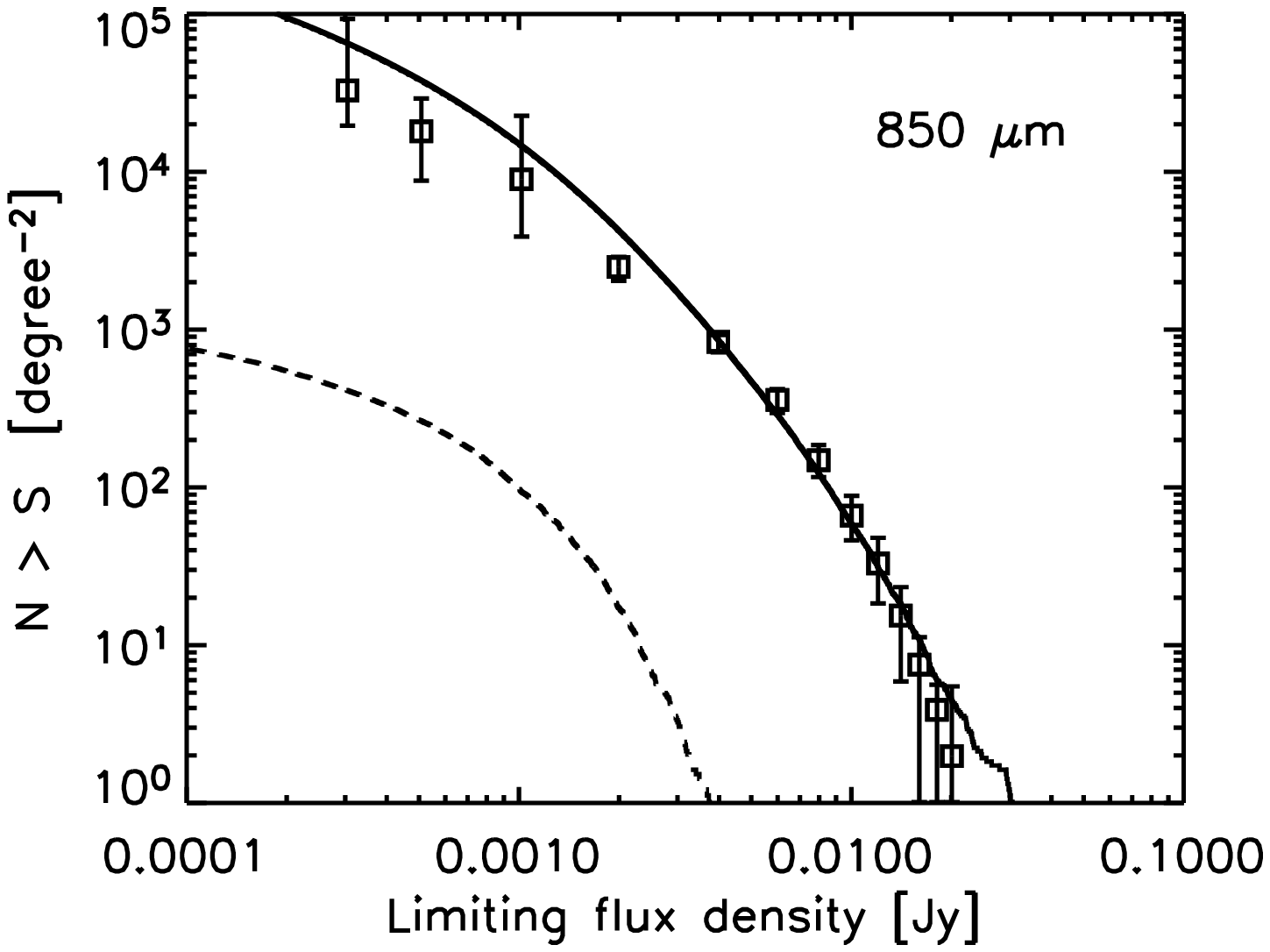}}\\
{\includegraphics[width=7.cm,keepaspectratio]{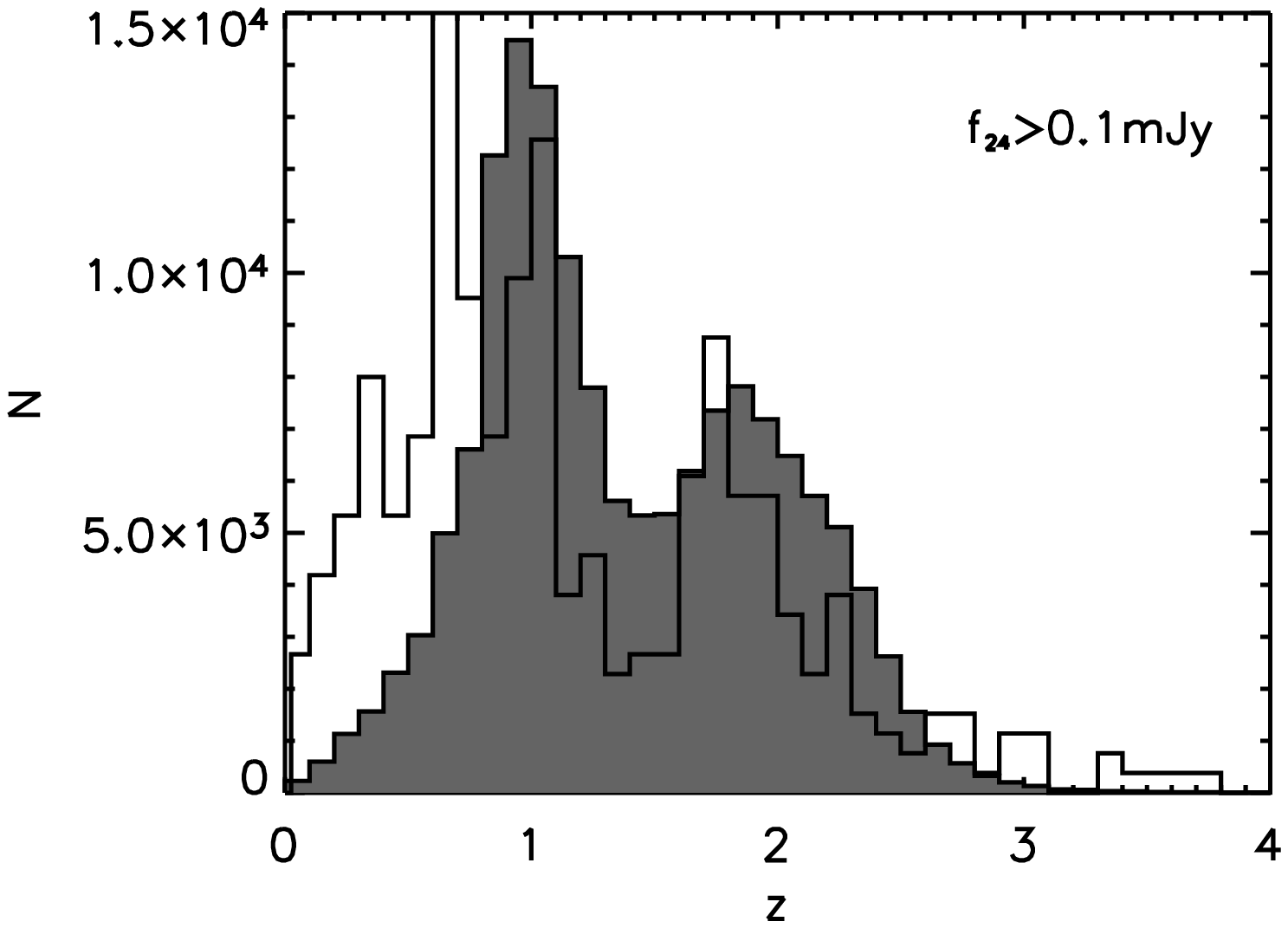}}
{\includegraphics[width=7.cm,keepaspectratio]{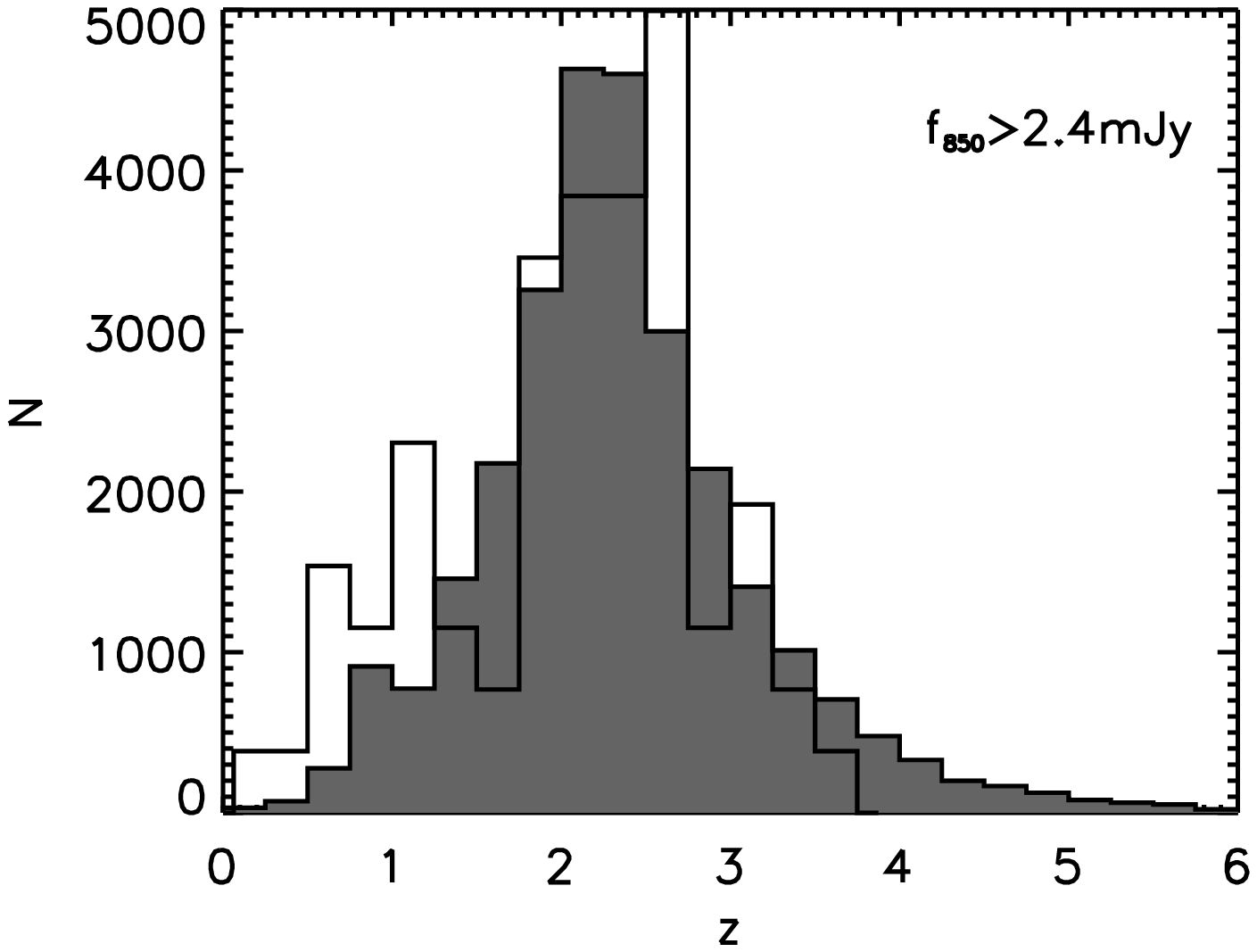}}\\
\caption{Same as Fig.~\ref{M3}, assuming an {\itshape evolution} and a {\itshape spread towards lower temperatures} in the $L-T$ relation (M5).
} 
\label{M5}
\end{figure}

We then modify the spread in the $L-T$ relation assuming an asymmetric gaussian distribution. The original distribution calculated by \citet{chapin09} is a standard gaussian in $\log(R(60,100))$. We assume a 7 times broader gaussian on the ``cold'' side of the peak respect to the ``warm'' side at low redshifts ($z<1$). In this way more cold galaxies are generated with respect to the IRAS population. The results for this model with $n_1=3.4$, $n_2=-1$, $z_1=2.3$, $m_1=2$, $m_2=-1.5$ and $z_2=1$ (M5) are shown in Fig.~\ref{M5}. Finally, both the far-IR ($160\,\mu{\rm m}$ and $70\,\mu\rm{m}$ wavebands) number counts are reproduced within the uncertainties as well as the submm measurements. We note that a qualitatively similar need to add local cold sources arose when using in our models both the earlier IRAS-based determination of temperature spread by \citet{chapman03} and the recent work of \citet{chapin09}. We can consider M5 our best-fit model: all the number counts are well reproduced (within a factor $\lesssim 2$), and also the redshift distributions are very similar to the observed ones, even if there are some discrepancies, due probably to the reasons explained above.

An example of how this population of cold galaxies could influence results of {\itshape Herschel} surveys is given in \S~\ref{predictions} (see Fig.~\ref{PEPpredict}{\itshape bottom}).





\begin{figure}
\centering

\includegraphics[width=13.cm,height=9.3cm]{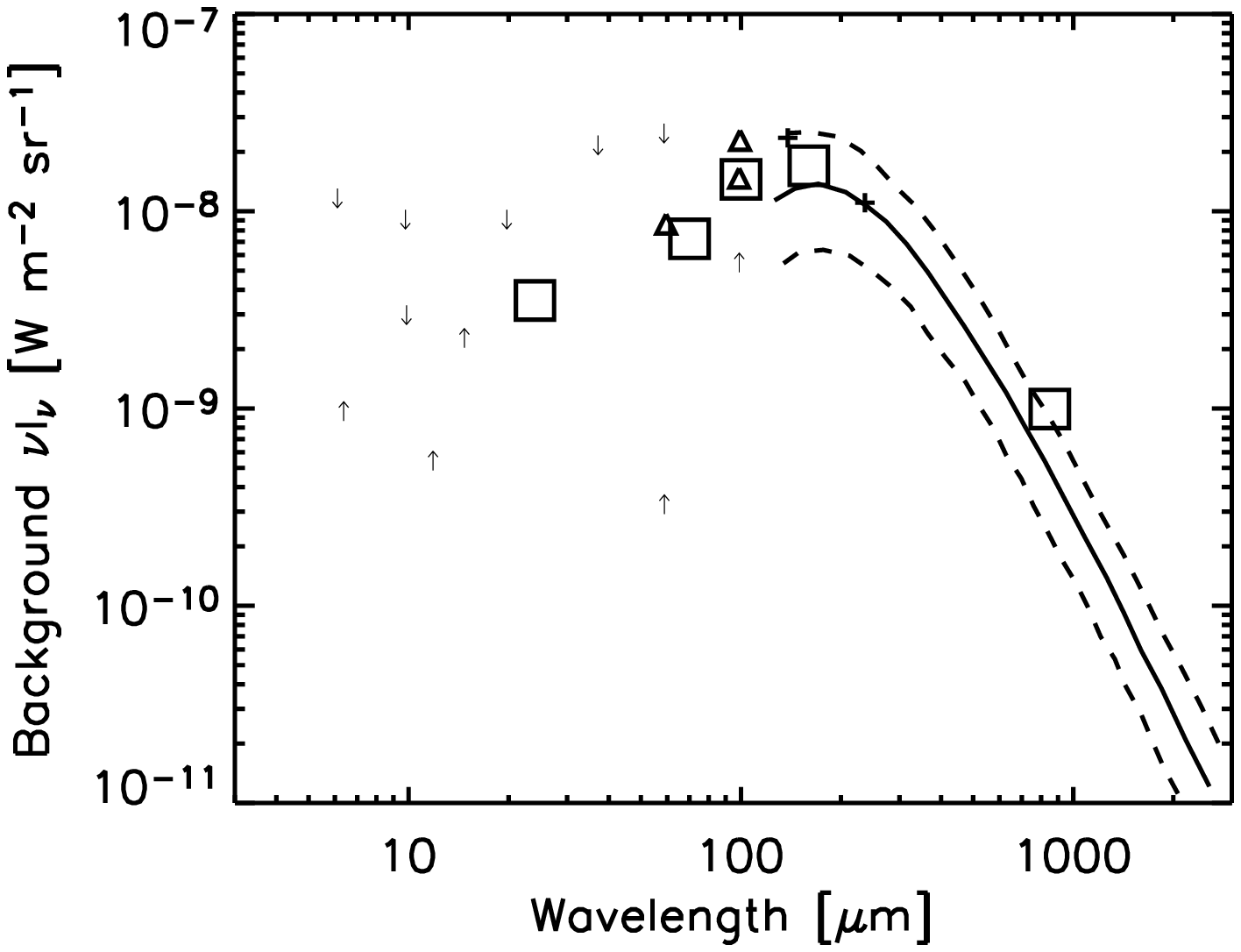}
\caption{Cosmic background from mid-IR to millimeter wavelengths for model M5. The CIB derived from the models at 24, 70, 100, 160 and $850\,\mu{\rm m}$ is shown by big open squares. The analytic form of the CIB at the FIRAS wavelengths is from \citet{fixsen98} ({\itshape solid and dashed lines}). Measurements are at DIRBE wavebands (140 and $240\,\mu\rm{m}$, {\itshape plus signs}, \citealt{lagache00}), at $100\,\mu\rm{m}$ ({\itshape triangles}, \citealt{lagache00,renault01}) and at $60\,\mu\rm{m}$ ({\itshape triangles}, \citealt{mivilledeschenes02}). See \citet{lagache03} for references about the limits at shorter wavelengths. 
}
\label{cobeplot}
\end{figure}

\begin{figure}
\centering

\includegraphics[width=13.cm,keepaspectratio,angle=90]{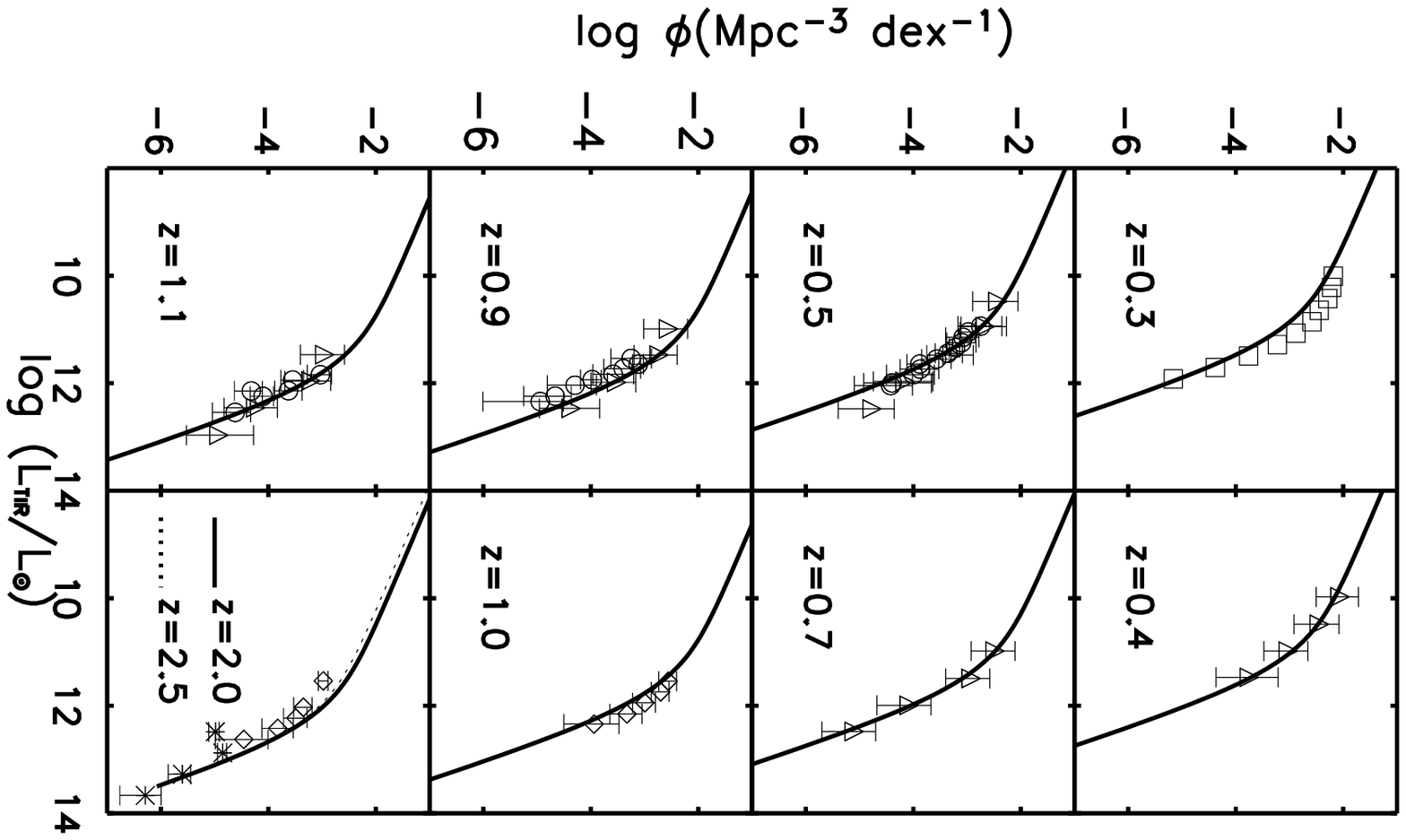}
\caption{Luminosity function of model M5 calculated at different redshifts, compared with available measurements. Data are from \citet{huang07} ({\itshape squares}), \citet{lefloch05} ({\itshape triangles}), \citet{magnelli09} ({\itshape circles}), \citet{caputi07} ({\itshape diamonds}), \citet{chapman05} ({\itshape stars}).
}
\label{luminosity_func}
\end{figure}

A further constraint on our evolutionary model comes from the comparison with the current CIB measurements. Our model predictions for the CIB are derived by summing up the flux densities of all sources for the bands in question. Fig~\ref{cobeplot} shows the predicted CIB intensity at specific wavelengths together with the comparison with present observations for the model M5: it agrees with all the available measurements and limits. We also compare the IR luminosity function calculated at different redshifts with measurements. Fig.~\ref{luminosity_func} shows the luminosity function of our best model M5 fitting all data from low redshifts $z=0.3$ \citep{huang07} to $z=0.4,0.5,0.7,0.9,1.1$ measured by \citet{lefloch05} and \citet{magnelli09}, $z=1,2$ by \citet{caputi07} and SMGs at $z\sim 2.5$ \citep{chapman05}. 

\section{Star formation history}



There has been much interest in the star formation rate and metal production history of the Universe since \citet{madau96} showed that they could be derived from deep UV/optical cosmological surveys. Since then the so-called {\itshape Madau diagram} has been revised many times through various improvements, including (1) the corrections for the effect of dust extinction on UV/optical luminosities (e.g.~\citealt{steidel99}), (2) results from less extinction-sensitive Balmer line surveys (e.g.~\citealt{yan99}), and (3) results from submillimeter SCUBA surveys (e.g.~\citealt{chapman05}).

\begin{figure}
\centering

\includegraphics[width=13.cm,height=9.3cm]{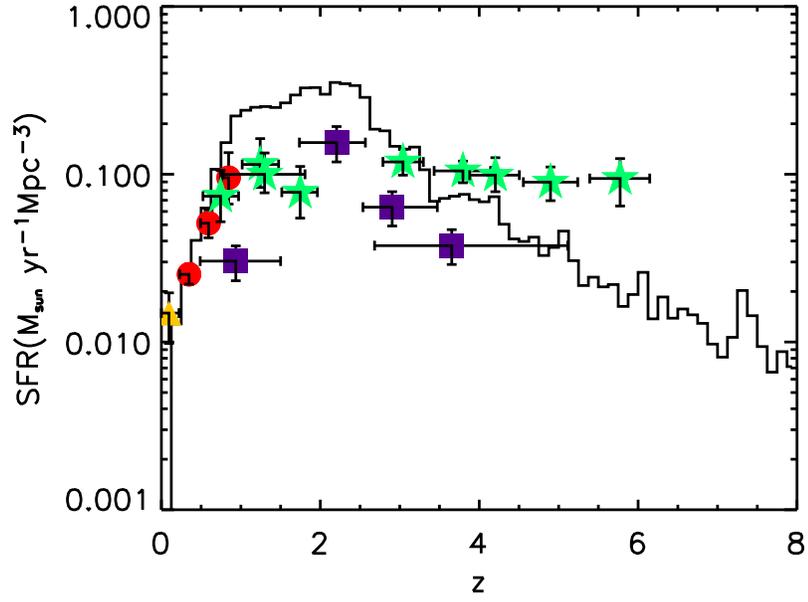}
\caption{Histogram showing the SFR evolution of the model M5 compared with the observational results shown by \citet{chapman05} (see their Fig.~12). {\itshape Yellow triangle}: radio surveys; {\itshape red circles}: mid-IR surveys; {\itshape green stars} UV and optical surveys, corrected for dust extinction; {\itshape violet squares}: SMGs corrected for completeness. See \citet{chapman05} for a full list of references.
}
\label{SFR}
\end{figure}

Adopting the conversion factor of \citet{kennicutt98}, 
\begin{equation}
\rm{SFR}\,[M_{\odot}\,\rm{yr}^{-1}]=4.5\cdot 10^{-44}\,\times\,L_{\rm TIR}\,[\rm{erg}\;\rm{s}^{-1}],
\end{equation} 
we can convert the cosmic luminosity density evolution of our model to a SFR curve. In the calculation, we take into account only the fraction on infrared luminosity due to star formation ($L_{\rm SB}$). In Fig.~\ref{SFR} the results of the simulation using model M5 are compared with the survey data already shown by \citet{chapman05} (see their Fig.~12). The model is in very good agreement with the observations for the full redshift range. In particular, for $z\gtrsim 4$, the model results are similar to the results from SCUBA surveys, while it is slightly lower than the measurements obtained from the UV/optical surveys. This does not surprising, since the model itself is constrained by the infrared/submillimeter data.

\section{AGN contribution to the infrared emission}\label{AGNcontrib}

As discussed earlier, LIRGs and ULIRGs, with their huge infrared luminosities, correspond to an extremely active phase of dust enshrouded star formation and/or AGN activity. They become an increasingly significant population at high redshift, representing an important phase in the buildup of massive galaxy bulges and in the growth of their central supermassive black holes. To understand these processes, it is essential to separate the relative contributions of AGN and starburst activity to the infrared luminosity of LIRGs and ULIRGs.

Recent work has shown how the IRAC color-color diagram and the MIPS 24 to $8\,\mu{\rm m}$ color can be used to identify AGN-dominated sources \citep{lacy04,sajina05,stern05,yan04,brand06}. \citet{brand06}, in particular, have demonstrated how the 24 to $8\,\mu{\rm m}$ flux ratio ($\zeta \equiv \log[\nu f_{nu}(24\,\mu{\rm m})/\nu f_{\nu}(8\,\mu{\rm m})]$) can be used to disentangle the contribution of AGN and starbursts to the total reprocessed mid-IR ($\sim 5-25\,\mu{\rm m}$) emission as a function of the $24\,\mu{\rm m}$ flux. We use this study as a first example of how to compare our model with observations that constrain the AGN content of high$-z$ sources.

Nevertheless, a photometric study of this type is subject to several caveats. Many broad emission- and absorption-line features are known to be present in the mid-IR spectrum of ULIRGs (e.g.~\citealt{houck05,yan05}), and this may affect $\zeta$ as function of redshift. For example, at $1.1\lesssim z\lesssim 1.7$, the $24\,\mu{\rm m}$ observed emission can be strongly
attenuated by the silicate absorption feature at $9.7\,\mu{\rm m}$. It is also possible that an AGN could be heavily embedded in large amounts of cooler dust and could remain undetected in the $8\,\mu{\rm m}$ band even though it dominates the bolometric infrared emission. In practice: on one hand, the galaxies, which in \citet{brand06} are classified as ``AGN-dominated'', do host an AGN, but it is not necessarily the dominant source powering the infrared emission; on the other hand, objects that are known to be dominated by AGN in the mid-IR, like NGC1068, do not present the color ($\zeta \sim 0$) that is considered ``typical'' for AGN-dominated sources in the \citet{brand06} study.

\begin{figure}
\centering
{\includegraphics[width=12.cm,height=8.6cm]{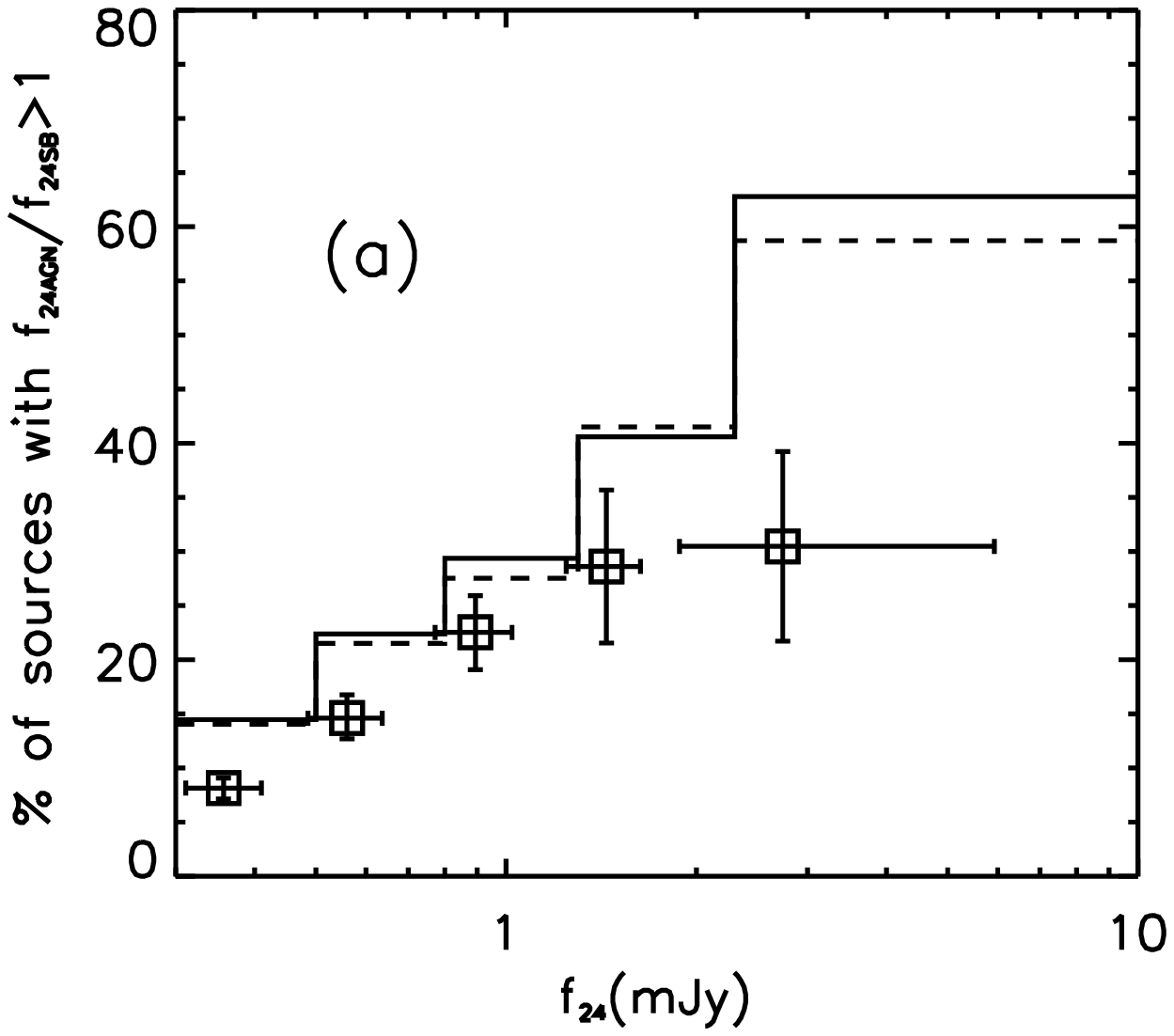}}
{\includegraphics[width=12.cm,height=8.6cm]{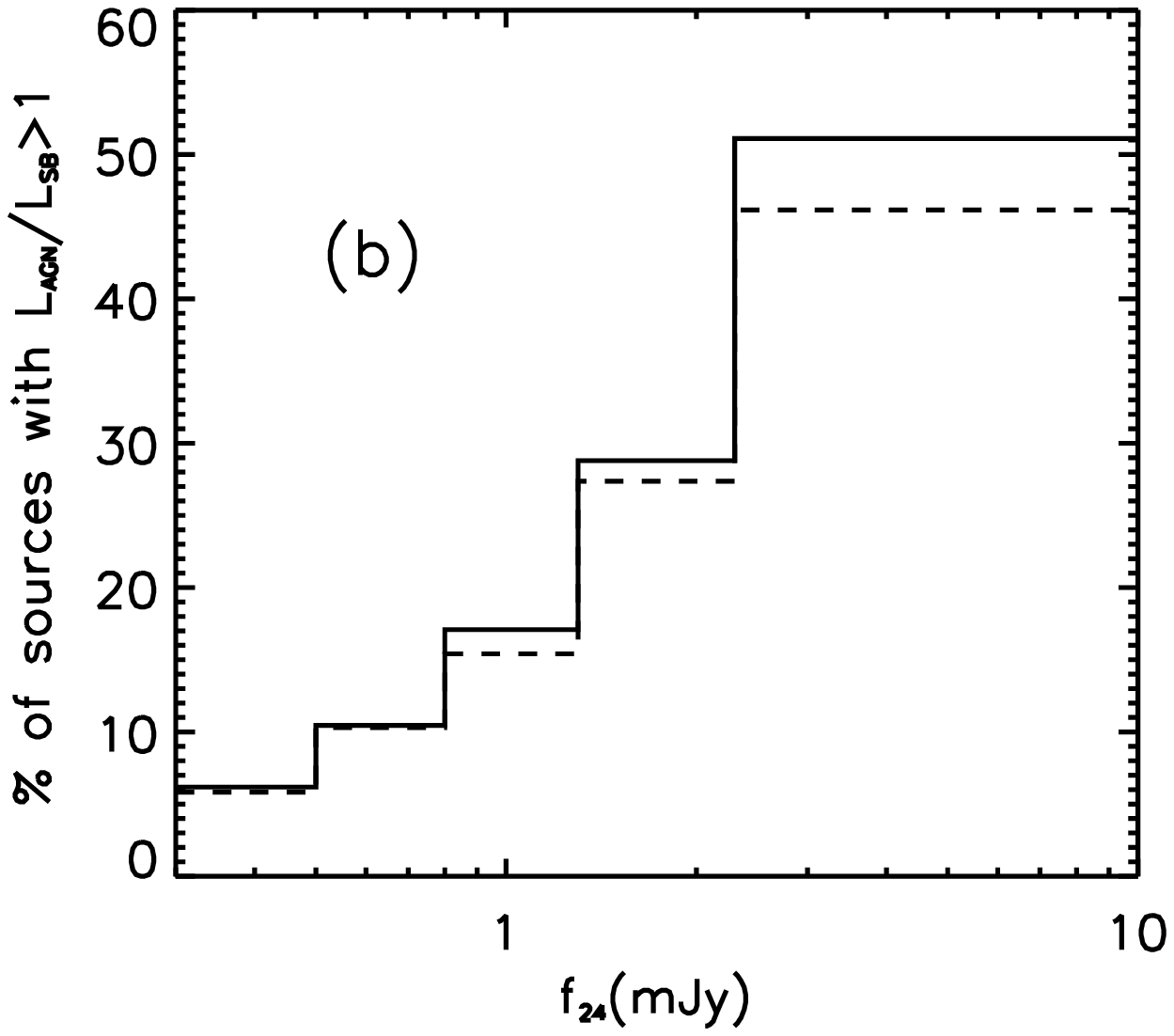}}
\caption{Fraction of all sources whose mid-IR emission ($a$) and total infrared emission ($b$) are dominated by AGN as a function of $f_{24}$ for model M5 ({\itshape solid histogram}) and M6 ({\itshape dashed histogram}). Redshift range of the sources is $0\leq z\leq8$. {\itshape Open squares} represent the results from \citet{brand06}.
}
\label{AGNfraction}
\end{figure}

Obtaining mid-IR spectroscopy of complete unbiased $24\,\mu{\rm m}$ selected samples of infrared sources will be important in putting more constraints on the AGN contribution to the total infrared emission and in characterization of the $24\,\mu{\rm m}$ selected population. Large {\itshape Spitzer}-IRS observing programs covering sources with $24\,\mu\rm{m}$ fluxes around and below $1\,\rm{mJy}$ are under analysis and will be well suited for this task. In combination with IRS surveys at brighter fluxes, they will trace the evolution of the AGN/SB ratio, strength of PAH emission and mid-IR opacities as a function of $L_{\rm TIR}$ and $z$. In the meantime, results of photometric studies like the one of \citet{brand06} can be compared to our models (see Fig.~\ref{AGNfraction}$a$).

The models developed here allow us to predict the fraction of AGN-dominated sources as a function of the mid-IR flux. In particular, it is possible to distinguish between sources dominated by AGN at $24\,\mu{\rm m}$ and sources AGN-dominated with respect to the total infrared luminosity. This distinction is important when comparing the results of the simulation with the observations: the information about $f_{24}$ will be useful in comparison with the upcoming results from the IRS observations cited above, while the predictions related to $L_{\rm TIR}$ will be confronted with {\itshape Herschel} results.

At the moment, as already explained, the shape of the AGN distribution is not well constrained at the highest ratios of AGN to total infrared luminosity. As an experiment, in order to reduce the number of luminous AGN, we change the assumed gaussian distribution introducing a cut for large values of $\nu L_{6\,\mu{\rm m}}/L_{{\rm TIR}}$: we reject all values of $\nu L_{6\,\mu{\rm m}}/L_{{\rm TIR}}$ more than $3\,\sigma$ away from its expected value. This model, again using the same parameter set ($n_1=3.4$, $n_2=-1$, $z_1=2.3$, $m_1=1$, $m_2=-1.5$, $z_2=1$, model M6), does not significantly change the number counts and redshift distributions with respect to the best model M5, so we do not show them here.

Differences between the two models are evident when doing different kinds of analysis. Fig.~\ref{AGNfraction}$a$ shows the fraction of all sources whose mid-IR emission is dominated by AGN ($f_{24\rm {AGN}}/f_{24\rm{SB}}>1$) as a function of the $24\,\mu{\rm m}$ flux density for model M5 ({\itshape solid histogram}) and M6 ({\itshape dashed histogram}). The trend and the values obtained are similar to those measured by \citet{brand06} ({\itshape open squares}), even if the measurements have to be compared carefully for the reasons explained above. Our predictions will be easily compared instead with the upcoming IRS results, that will help us to put more constraints on the modeled AGN contribution. In fact, the AGN fraction of model M6, where the AGN contribution has been slightly modified, is a bit lower compared to model M5: only more constraints coming from mid-IR spectroscopic observations can discriminate between the two models M5 and M6. Fig.~\ref{AGNfraction}$b$ shows the fraction of all sources whose total infrared emission is dominated by AGN ($L_{\rm AGN}/L_{\rm SB}>1$) as a function of the $24\,\mu{\rm m}$ flux density for the two models M5 and M6. As expected, the values of the fraction are lower, because the AGN starts to dominate the mid-IR part of the SED before being dominant in the entire infrared range (see e.g.~Fig.~\ref{sedsAGN}). 

\section{Open issues}\label{open}

Our best-fit model M5 is able to reproduce most of the measurements available. Nevertheless, there are still some issues that need a deeper analysis and discussion and will be dealt with in future studies.

The first point concerns the so called ``local cold population''. Even if there are several clues about the existence of a population of colder IR galaxies not detected by IRAS, only forthcoming surveys at longer wavelengths carried out with {\itshape Herschel} will be able to reveal and characterize this kind of objects. 

The second point regards the distribution of the AGN contribution and the fraction of AGN-dominated sources. We already pointed out that the models M5 and M6 predict similar $24\mu{\rm m}$ number counts, but different amount of AGN-dominated sources. Galaxies at $z\sim 2$ presenting a ``mid-IR excess'' have redder $K-5.8\,\mu{\rm m}$ colours than normal galaxies: this is evidence for an AGN contribution to the mid-IR continuum due to warm dust. The presence of Compton-thick AGN is confirmed by the stacked {\itshape Chandra} X-ray data \citep{daddi07}. The sky density and the volume density of this population of obscured AGN agree reasonably well with those predicted by the background synthesis models of \citet{gilli07}. In order to discriminate between our best models, M5 and M6, we need to compare the AGN-dominated sources predicted with upcoming results with IRS. This will introduce a further constraint, in addition with the bounds already used to the total number counts, and it will enable the final model to reproduce the co-evolution of AGN and starbursts in more detail.

The third point regards the evolutions of the AGN$-$ and Color$-L_{\rm TIR}$ relations. It is clear that some evolution is needed in order to reproduce existing observations, nevertheless the evolution implemented in this model is quite simple. More information regarding the AGN content of high$-z$ far-IR selected sources and the shape of the SED of SMGs will enable a better characterization of these details. Understanding evolution is intimately related to the black hole growth and to the star-formation rate history of massive galaxies.

\section{Predictions for multiband {\itshape Herschel} surveys}\label{predictions}


In this section, we show predictions using the best-fit model M5. We concentrate on future surveys with {\itshape Herschel}, which will be launched in early 2009, in particular on the PEP survey\footnote{http://www.mpe.mpg.de/ir/Research/PEP/}. All three PACS bands (70, 100, $160\,\mu{\rm m}$) are considered. Predicted number counts at 70 and $160\,\mu{\rm m}$ are already shown in Fig.~\ref{M4} and compared with the currently available data in those wavebands.

\begin{figure}
\centering
{\includegraphics[width=7.cm,height=5.cm]{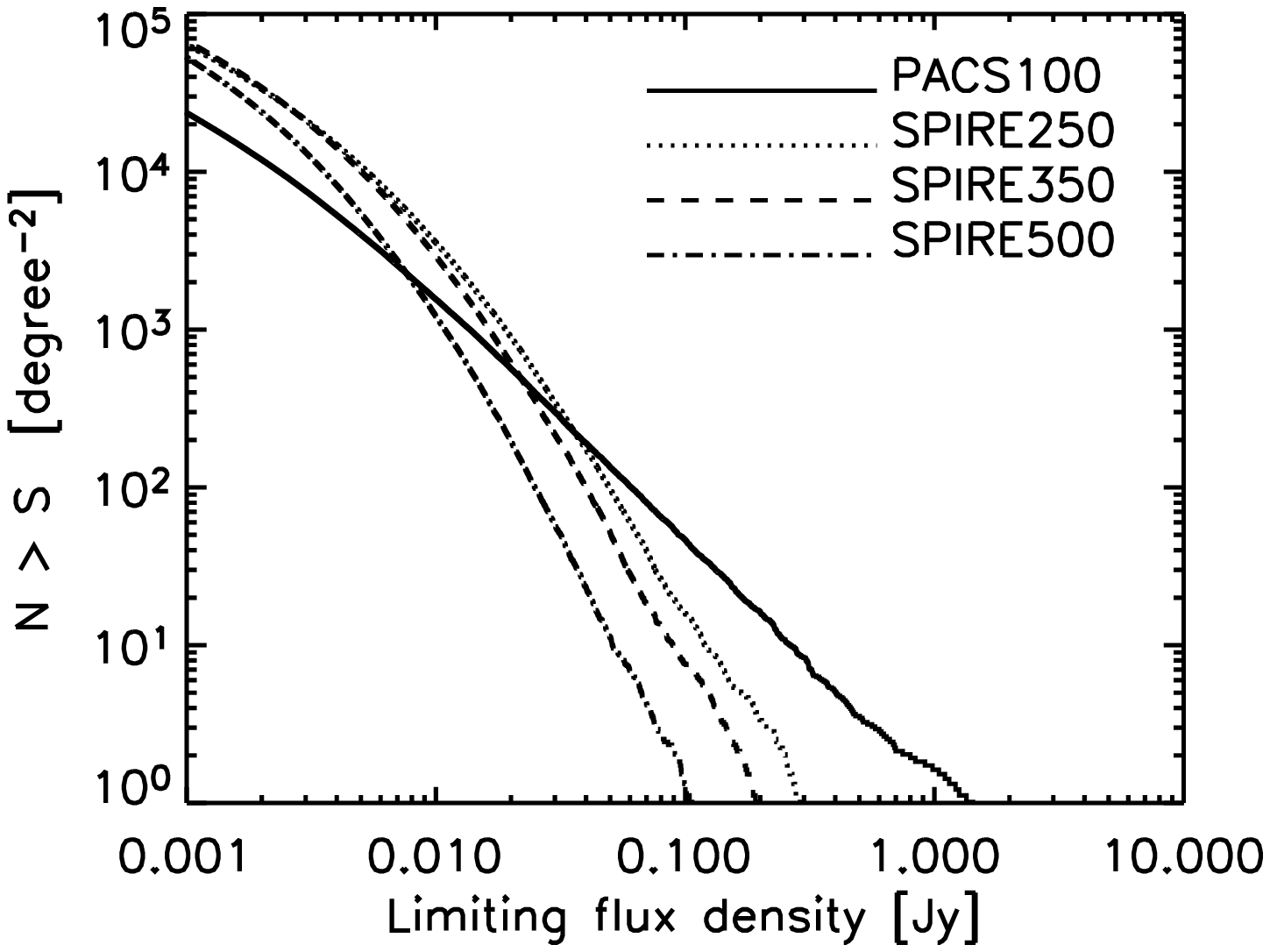}}\\
{\includegraphics[width=7.cm,height=5.cm]{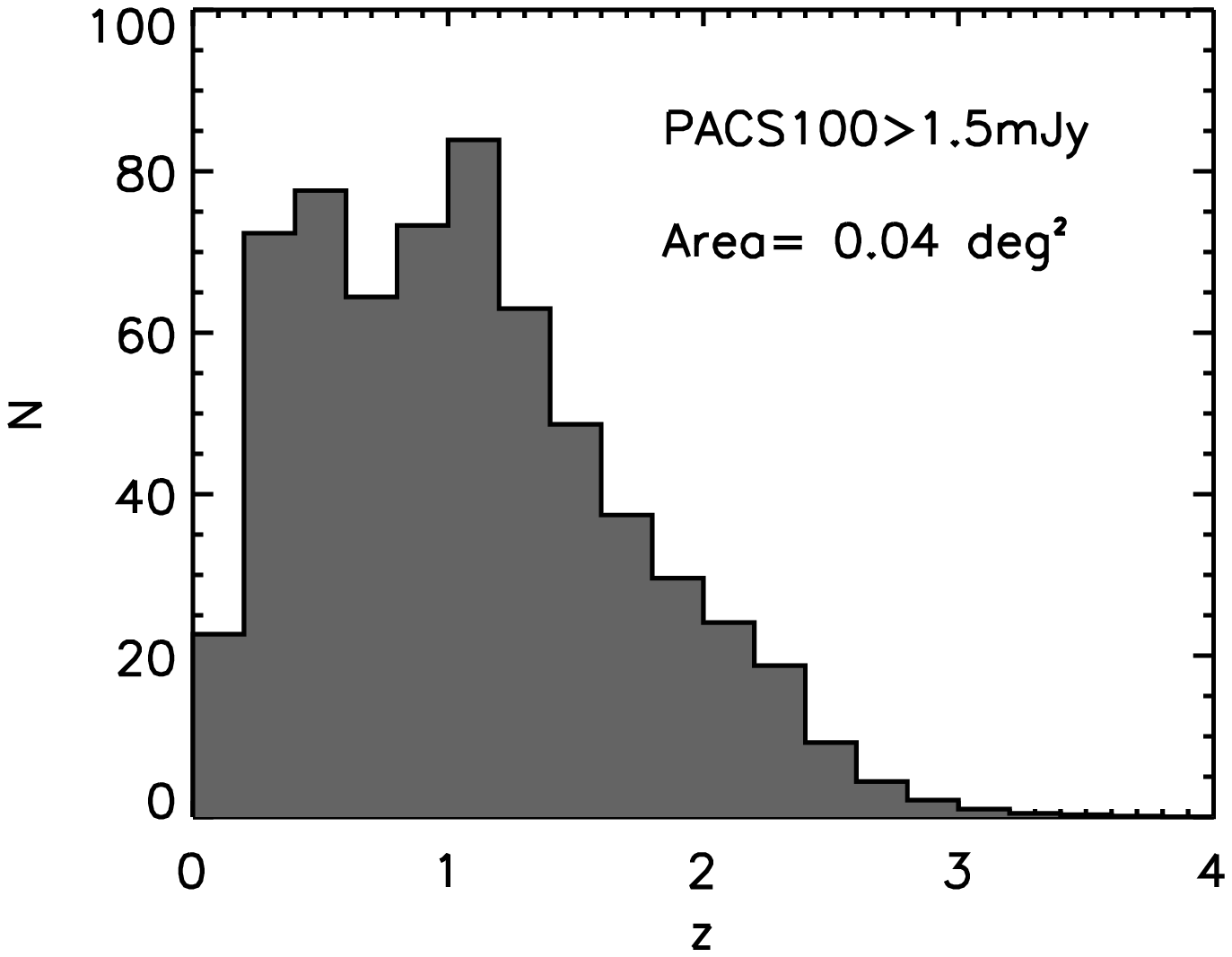}}
{\includegraphics[width=7.cm,height=5.cm]{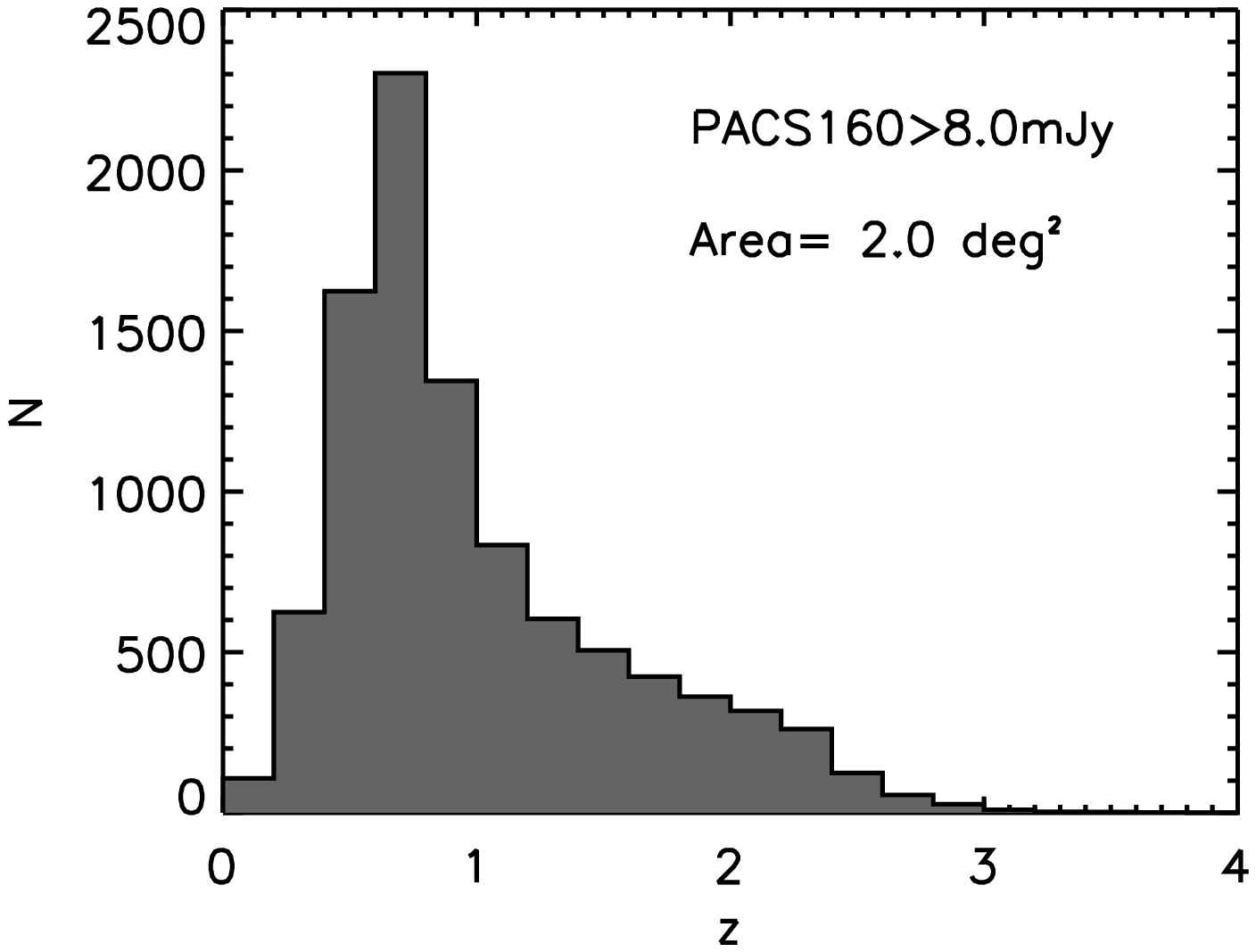}}\\
{\includegraphics[width=7.cm,height=5.cm]{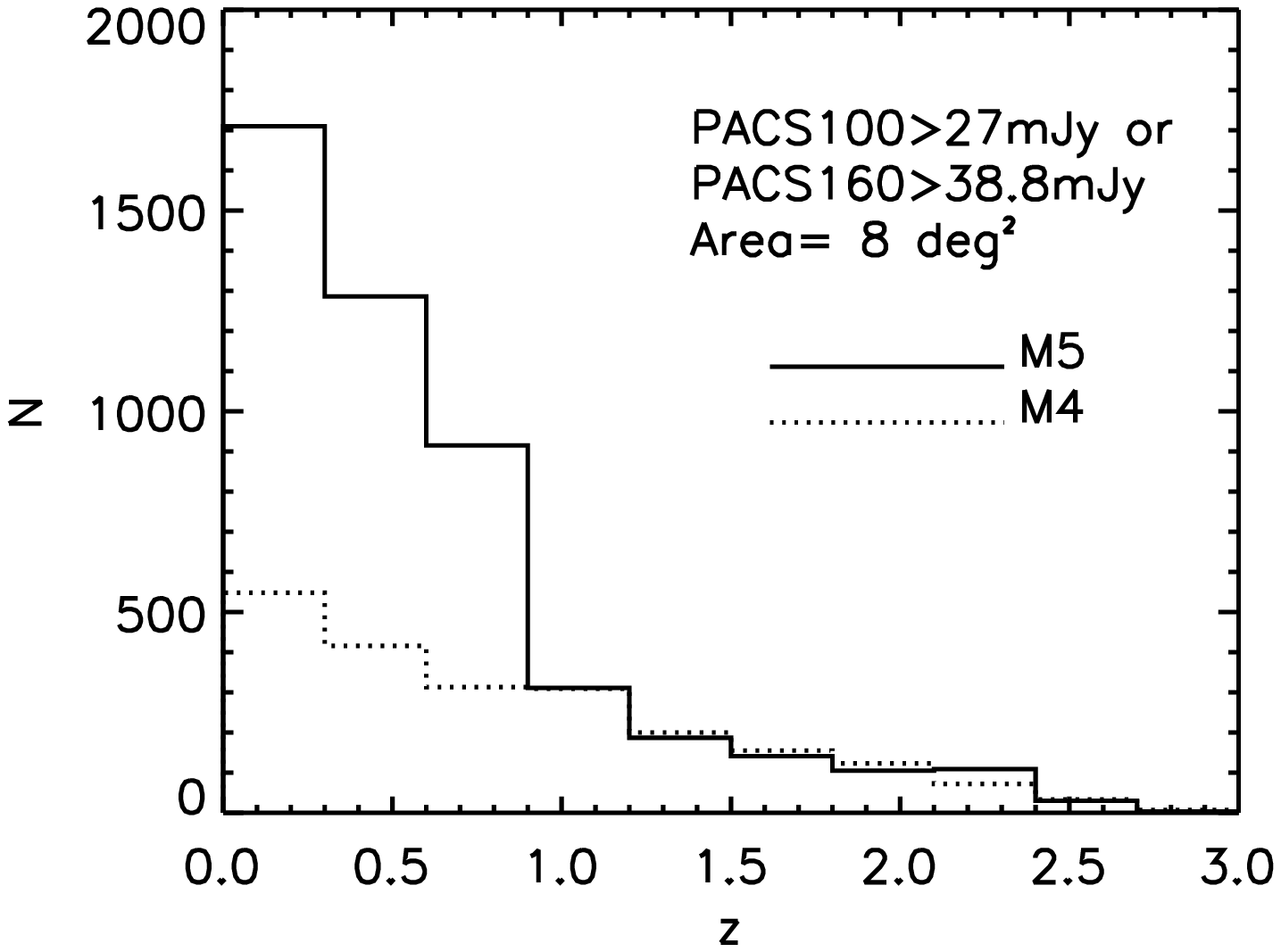}}
{\includegraphics[width=7.cm,height=5.cm]{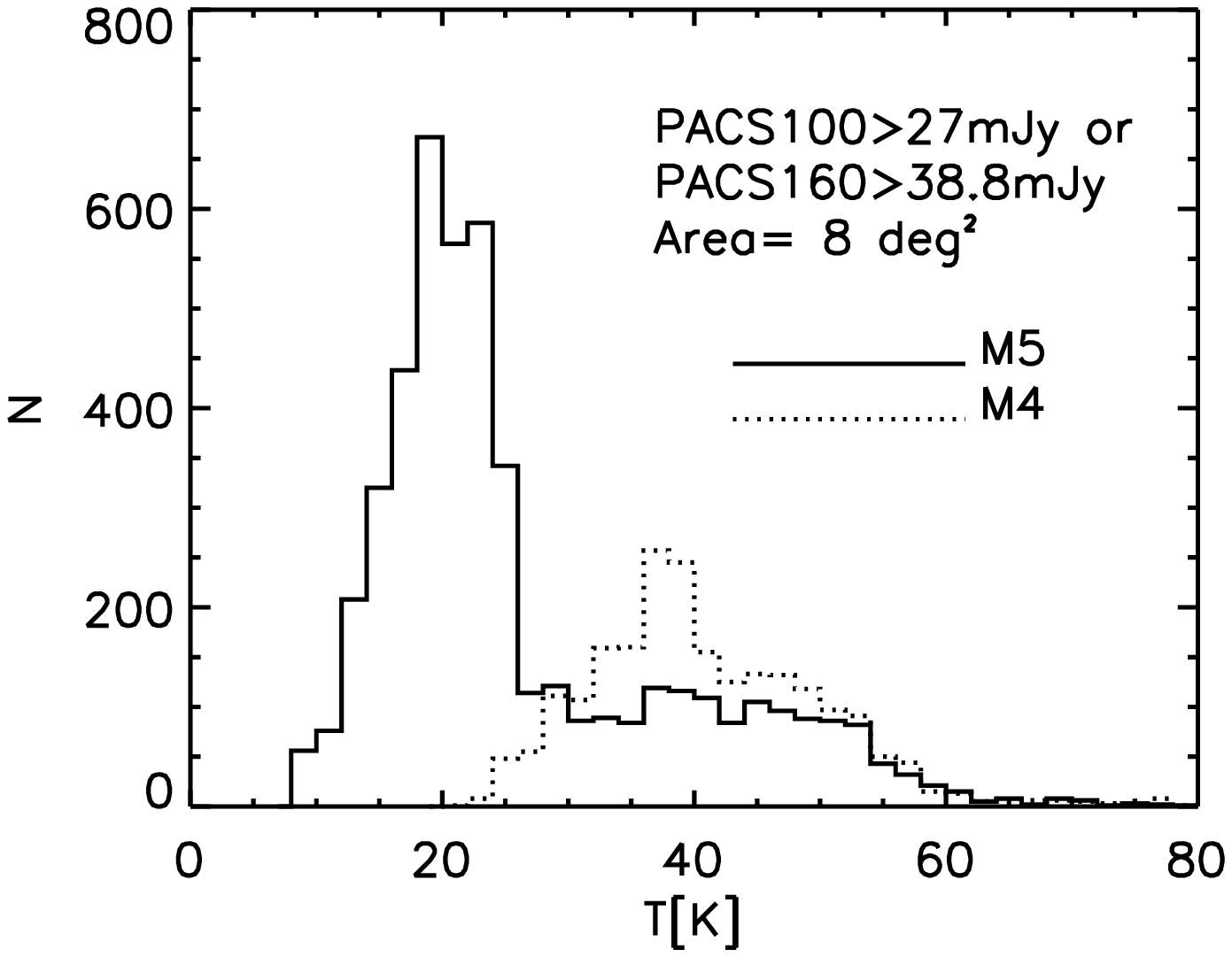}}
\caption{Predictions using the best-fit model M5 for the upcoming {\itshape Herschel} surveys. {\itshape Top}: integral number counts at $100\,\mu{\rm m}$ ({\itshape solid line}), $250\,\mu{\rm m}$ ({\itshape dotted line}), $350\,\mu{\rm m}$ ({\itshape dashed line}), $500\,\mu{\rm m}$ ({\itshape dot-dashed line}). {\itshape Middle}: redshift distributions of PEP surveys in the deep ($10^{\prime}\times 15^{\prime}$, $f_{100}>1.5\,{\rm mJy}$, {\itshape left}) and large ($85^{\prime}\times 85^{\prime}$, $f_{160}>8\,{\rm mJy}$, {\itshape right}) fields. {\itshape Bottom}: expected distributions for the survey in the CDFS field for model M4 ({\itshape dotted line}) and M5 ({\itshape solid line}) of redshifts ({\itshape left}) and temperatures ({\itshape right}).
}
\label{PEPpredict}
\end{figure}

In Fig.~\ref{PEPpredict}{\itshape top} the predicted integral number counts at 100, 250, 350, $500\,\mu{\rm m}$ are plotted. While at $70\,\mu{\rm m}$ the AGN-dominated sources can give a non-negligible contribution to the total number counts, in particular for $f_{70}\gtrsim 50\,{\rm mJy}$, in the other {\itshape Herschel} bands the number counts are exclusively dominated by the starburst component.

Fig.~\ref{PEPpredict}{\itshape middle} shows predicted redshift distributions for the deep ($f_{100}>1.5\,{\rm mJy}$) and large ($f_{160}>8\,{\rm mJy}$) PEP fields. The deep ($10^{\prime}\times 15^{\prime}$) and the large ($85^{\prime}\times 85^{\prime}$) fields are centred respectively on the GOODS-S and on the COSMOS fields, in order to have the maximum availability of multi-wavelength data and follow up opportunities. The chosen fields are in fact fully covered in several of the following: deep X-ray surveys, UV/optical/near-IR/IRAC imaging, HST imaging, {\itshape Spitzer} mid-IR surveys, submillimeter surveys and radio mapping. The predicted distributions show a prominent peak at $z\sim 1$ and a shoulder reaching $z\sim 2$ and beyond. Both the surveys will observe sources up to $z\sim 3$, allowing a consistent step forward in the study of the cosmic evolution of dusty star formation and of the infrared luminosity function, as well as in the determination of the overall SEDs of active galaxies.

In Fig.\ref{PEPpredict}{\itshape bottom} the predicted redshift and dust temperature distributions of galaxies selected in the PACS bands (100 and $160\,\mu{\rm m}$) for models M4 (no local cold population) and M5 (with local cold population) are plotted. The selection is assumed to be the same as expected in the CDFS field ($\sim 8\,\rm{deg}^2$, $f_{100}>27\,{\rm mJy}$, $f_{160}>38.8\,{\rm mJy}$). Redshift distributions are similar in shape, but M5 presents an increase in the number of detected objects at $z\lesssim 1$ of a factor $\sim 3$. Dust temperature distributions are completely different: M4 predicts a mean temperature of $T_{\rm d} \sim 38\,\rm{K}$, while M5 presents a skewed distribution with a prominent peak at $T_{\rm d} \sim 23\,\rm{K}$ and a tail up to $T_{\rm d} \sim 60\,\rm{K}$. Multiband {\itshape Herschel} surveys will measure these distributions.

\section{Conclusions}

The aim of this work has been to find a ``backward evolution'' model that can viably fit the galaxy source counts and redshift distributions from the mid-infrared to submillimeter wavelengths whilst not violating the constraints set by the Cosmic Infrared Background measurements. Increasingly detailed observations are characterizing high redshift galaxies simultaneously at many infrared wavelengths, and are starting to unravel the role of AGN within them, suggesting a refinement of previous models. We adopted a Monte-Carlo-based approach that considers the evolution of a coherent population of infrared galaxies with distribution functions describing the contribution of AGN and the spread around the luminosity-temperature relation for the far-infrared emission.

As a necessary ingredient of this model, we have characterized the local distribution of AGN contributions as a function of total infrared luminosity. We have applied spectral decomposition to a large sample of {\itshape Spitzer}-IRS spectra of ULIRGs and LIRGs with $L_{\rm TIR} > 10^{11}\,L_{\odot}$ to isolate the AGN $6\,\mu\rm{m}$ continua. The distribution of $L_{\rm 6AGN}/L_{\rm TIR}$ changes with $L_{\rm TIR}$. We compare the AGN detections and limits to simple Monte-Carlo simulations, making the assumption of an intrinsically gaussian distribution, to quantify this increase with luminosity of the AGN contribution to the infrared luminosity. The best fit, $\nu L_{6\,\mu{\rm m}}/L_{{\rm TIR}} \propto L_{{\rm TIR}}^{\alpha}$, gives $\alpha=1.4 \pm 0.6$. The relation does not hold any more at high redshifts. The most luminous high redshift infrared galaxies, $L_{\rm TIR}\sim10^{13}L_{\odot}$, show a small contribution from AGN, being mainly starburst powered.

The local start of our backward evolution models combines this result with the luminosity function and temperature-luminosity relation of \citet{chapin09}. We have explored several models considering evolutionary variation of increasing subsets of the parameters within this framework. We find that we can satisfactorily reproduce number counts and redshift constraints from the literature with such a population of infrared galaxies evolving in luminosity and density. The contribution of starbursts and AGN varies with redshift and luminosity, as does the far-infrared dust temperature. The luminosity evolution is mainly constrained by number counts at 24 and $850\,\mu\rm{m}$, while the density evolution is strongly influenced by the $24\,\mu\rm{m}$ redshift distribution. While this evolution in luminosity, density, AGN contribution and dust temperature is clearly suggested by this sequence of models, the detailed functional forms and parameters are necessarily uncertain with current data.

As in previous work (e.g.~\citealt{lagache03}), we have realized a need to invoke a population of local cold galaxies in order to reproduce the number counts at $160\,\mu\rm{m}$. Even with such clues about the existence of this population, its observational characterization remains incomplete.

The coherent consideration of both AGN and starburst in our model allows for direct comparison to observations constraining the role of AGN in high redshift infrared galaxies. Trends with modelled mid-infrared flux from our best fit model are consistent with photometric estimates \citep{brand06}. Forthcoming IRS observations of large flux-limited samples will both reduce the observational uncertainties in characterizing the role of AGN and allow to better constrain the role of the most luminous AGN in our models.

Characterization of the SEDs of $z\sim 0.5-3$ galaxies near their rest frame far-infrared peak will make a big step forward with the Herschel mission, and advance the understanding of the AGN-galaxy co-evolution. The model type presented here is prepared for the substantial tests and refinements of assumptions on the evolution of role of AGN and luminosity-temperature relation that will be possible. As a first step, we have presented predictions from our current best-fit model for redshift distributions and the role of AGN in deep surveys with {\itshape Herschel}-PACS.

\clearpage

\begin{deluxetable*}{l c c r c}
\label{AGNdata}
\tablecolumns{5}
\tablewidth{0pt}
\tablenum{1}
\tabletypesize{\footnotesize}
\tablecaption{Full sample of infrared galaxies: results of the fit and corrected $f_{6{\rm AGN}}$ fluxes}
\tablehead{
\colhead{Name} & \colhead{$z$} & \colhead{$\log(L_{\rm TIR})$} & \colhead{$\tilde{f}_{6{\rm AGN}}$} & \colhead{$f_{6{\rm AGN}}$}  \\
\colhead{} &\colhead{} &\colhead{$L_{\bigodot}$} &\colhead{mJy} &\colhead{mJy}
}
\startdata
\multicolumn{5}{c}{\small Sources from RBGS catalog \citep{sanders03}} \\
NGC0023 & 0.015 & 11.05 & 13.9$\pm$1.0 & $<$23.8\\
NGC0034 & 0.020 & 11.44 & 48.8$\pm$3.1 & $<$62.9\\
MCG-02-01-051/2 & 0.027 & 11.41 & 6.8$\pm$0.8 & $<$19.9\\
ESO350-IG038 & 0.021 & 11.22 & 57.8$\pm$0.5 & 53.4$\pm$4.0\\
NGC0232 & 0.020 & 11.30 & 20.5$\pm$1.5 & $<$31.4\\
MCG+12-02-001 & 0.016 & 11.44 & 20.5$\pm$1.7 & $<$48.4\\
NGC0317B & 0.018 & 11.11 & 14.6$\pm$0.7 & $<$23.6\\
IC1623A/B & 0.020 & 11.65 & 196.4$\pm$2.5 & 162.0$\pm$30.5\\
MCG-03-04-014 & 0.035 & 11.63 & 8.3$\pm$0.7 & $<$24.1\\
ESO244-G012 & 0.023 & 11.39 & 12.6$\pm$1.3 & $<$34.3\\
CGCG436-030 & 0.031 & 11.63 & 35.8$\pm$0.6 & 26.5$\pm$8.2\\
ESO353-G020 & 0.016 & 11.00 & 20.2$\pm$1.1 & $<$30.1\\
ESO297-G011/012 & 0.017 & 11.09 & 16.0$\pm$1.1 & $<$31.1\\
IRASF01364-1042 & 0.048 & 11.76 & 4.0$\pm$0.2 & $<$5.2\\
IIIZw035 & 0.028 & 11.56 & 2.1$\pm$0.1 & $<$2.7\\
NGC0695 & 0.033 & 11.63 & 5.5$\pm$0.5 & $<$13.5\\
UGC01385 & 0.019 & 10.99 & 7.1$\pm$0.7 & $<$16.5\\
NGC0828 & 0.018 & 11.31 & 17.3$\pm$1.1 & $<$29.2\\
NGC0838 & 0.013 & 11.00 & 5.6$\pm$1.5 & $<$36.0\\
IC0214 & 0.030 & 11.37 & 3.9$\pm$0.5 & $<$11.6\\
NGC0877 & 0.013 & 11.04 & 3.7$\pm$0.1 & 3.3$\pm$0.4\\
UGC01845 & 0.016 & 11.07 & 25.2$\pm$2.0 & $<$42.5\\
NGC0958 & 0.020 & 11.17 & 5.0$\pm$0.1 & 4.4$\pm$0.6\\
NGC0992 & 0.014 & 11.02 & 5.0$\pm$1.0 & $<$27.6\\
NGC1068 & 0.003 & 11.27 & 10493.6$\pm$30.6 & 10482.0$\pm$32.3\\
UGC02238 & 0.021 & 11.26 & 16.4$\pm$1.4 & $<$35.9\\
IRASF02437+2122 & 0.023 & 11.11 & 18.3$\pm$0.4 & 15.8$\pm$2.3\\
UGC02369 & 0.031 & 11.60 & 9.3$\pm$0.6 & $<$15.3\\
UGC02608 & 0.023 & 11.35 & 67.0$\pm$1.1 & 60.0$\pm$6.3\\
NGC1275 & 0.018 & 11.20 & 120.2$\pm$0.8 & 119.7$\pm$0.9\\
IRASF03217+4022 & 0.023 & 11.28 & 12.9$\pm$0.9 & $<$18.9\\
NGC1365 & 0.005 & 11.00 & 227.8$\pm$1.7 & 213.9$\pm$12.4\\
IRASF03359+1523 & 0.035 & 11.47 & 0.4$\pm$0.1 & $<$0.7\\
CGCG465-012 & 0.022 & 11.15 & 6.3$\pm$0.5 & $<$12.2\\
IRAS03582+6012 & 0.030 & 11.37 & 189.7$\pm$1.5 & 188.7$\pm$1.7\\
UGC02982 & 0.017 & 11.13 & 6.0$\pm$0.8 & $<$19.4\\
ESO420-G013 & 0.012 & 11.02 & 86.6$\pm$1.1 & 69.5$\pm$15.2\\
NGC1572 & 0.020 & 11.24 & 18.6$\pm$0.6 & $<$24.9\\
IRAS04271+3849 & 0.019 & 11.06 & 9.3$\pm$0.7 & $<$21.4\\
NGC1614 & 0.016 & 11.60 & 24.8$\pm$2.9 & $<$56.7\\
UGC03094 & 0.025 & 11.35 & 21.2$\pm$0.5 & $<$27.2\\
ESO203-IG001 & 0.053 & 11.79 & 12.9$\pm$0.4 & 12.6$\pm$0.5\\
MCG-05-12-006 & 0.019 & 11.12 & 13.9$\pm$0.5 & $<$20.3\\
NGC1797 & 0.015 & 11.00 & 12.8$\pm$0.9 & $<$24.8\\
CGCG468-002 & 0.017 & 11.10 & 41.1$\pm$0.6 & 33.8$\pm$6.5\\
VIIZw031 & 0.054 & 11.94 & 8.5$\pm$0.7 & $<$21.0\\
IRAS05083+2441 & 0.023 & 11.21 & 5.1$\pm$0.9 & $<$28.2\\
IRAS05129+5128 & 0.027 & 11.36 & 11.4$\pm$0.4 & $<$17.5\\
IRASF05187-1017 & 0.028 & 11.23 & 6.2$\pm$0.3 & $<$8.5\\
IRASF05189-2524 & 0.043 & 12.11 & 204.8$\pm$0.9 & 201.3$\pm$3.2\\
IRAS05223+1908 & 0.030 & 11.59 & 367.8$\pm$1.1 & 366.7$\pm$1.5\\
NGC1961 & 0.013 & 11.02 & 8.6$\pm$0.3 & 7.8$\pm$0.8\\
MCG+08-11-002 & 0.019 & 11.41 & 26.3$\pm$1.1 & $<$38.6\\
UGC03351 & 0.015 & 11.22 & 22.3$\pm$0.9 & $<$31.8\\
IRAS05442+1732 & 0.019 & 11.25 & 13.0$\pm$1.5 & $<$33.2\\
UGC03410 & 0.013 & 11.04 & 5.2$\pm$0.5 & $<$11.3\\
IRASF06076-2139 & 0.037 & 11.59 & 9.7$\pm$0.3 & 7.1$\pm$2.3\\
NGC2146 & 0.003 & 11.07 & 45.6$\pm$6.2 & $<$150.3\\
ESO255-IG007 & 0.039 & 11.84 & 11.7$\pm$1.3 & $<$36.6\\
ESO557-G002 & 0.021 & 11.19 & 5.3$\pm$0.3 & $<$9.0\\
UGC3608 & 0.022 & 11.30 & 8.0$\pm$0.4 & $<$12.4\\
IRASF06592-6313 & 0.023 & 11.17 & 11.7$\pm$0.5 & $<$16.5\\
AM0702-601 & 0.031 & 11.58 & 100.2$\pm$0.4 & 98.0$\pm$2.0\\
NGC2342 & 0.018 & 11.25 & 5.6$\pm$0.4 & $<$11.5\\
NGC2369 & 0.011 & 11.10 & 30.0$\pm$1.4 & $<$42.5\\
IRAS07251-0248 & 0.088 & 12.32 & 16.4$\pm$1.1 & 15.9$\pm$1.2\\
NGC2388 & 0.014 & 11.23 & 22.0$\pm$1.1 & $<$34.5\\
MCG+02-20-003 & 0.016 & 11.08 & 53.4$\pm$1.7 & 46.2$\pm$6.6\\
IRASF08339+6517 & 0.019 & 11.05 & 2.9$\pm$0.8 & $<$15.7\\
NGC2623 & 0.018 & 11.54 & 22.3$\pm$0.8 & $<$28.8\\
IRAS08355-4944 & 0.026 & 11.56 & 78.7$\pm$0.6 & 69.1$\pm$8.5\\

\enddata
\end{deluxetable*}

\begin{deluxetable*}{l c c r c}
\tablecolumns{5}
\tablewidth{0pt}
\tablenum{1}
\tabletypesize{\footnotesize}
\tablecaption{{\em continued} -- Full sample of infrared galaxies: results of the fit and corrected $f_{6{\rm AGN}}$ fluxes}
\tablehead{
\colhead{Name} & \colhead{$z$} & \colhead{$\log(L_{\rm TIR})$} & \colhead{$\tilde{f}_{6{\rm AGN}}$} & \colhead{$f_{6{\rm AGN}}$}  \\
\colhead{} &\colhead{} &\colhead{$L_{\bigodot}$} &\colhead{mJy} &\colhead{mJy}
}
\startdata
ESO432-IG006 & 0.016 & 11.02 & 18.7$\pm$1.0 & $<$28.8\\
ESO60-IG016 & 0.046 & 11.76 & 56.4$\pm$0.6 & 53.0$\pm$3.1\\
IRASF08572+3915 & 0.058 & 12.10 & 291.5$\pm$2.9 & 290.5$\pm$3.0\\
IRAS09022-3615 & 0.060 & 12.26 & 83.2$\pm$0.5 & 75.9$\pm$6.5\\
IRASF09111-1007 & 0.054 & 12.00 & 5.8$\pm$0.3 & $<$8.4\\
UGC04881 & 0.040 & 11.69 & 9.0$\pm$0.3 & $<$13.6\\
UGC05101 & 0.039 & 11.95 & 46.7$\pm$1.0 & 40.3$\pm$5.8\\
MCG+08-18-013 & 0.026 & 11.28 & 0.3$\pm$0.1 & $<$0.3\\
IC0563/4 & 0.020 & 11.19 & 6.0$\pm$0.4 & $<$10.0\\
NGC3110 & 0.017 & 11.31 & 9.1$\pm$0.7 & $<$21.0\\
IC2545 & 0.034 & 11.73 & 311.7$\pm$6.2 & 308.8$\pm$6.7\\
IRASF10173+0828 & 0.049 & 11.80 & 2.1$\pm$0.1 & 1.5$\pm$0.5\\
NGC3221 & 0.013 & 11.00 & 0.5$\pm$0.2 & $<$5.9\\
NGC3256 & 0.009 & 11.56 & 66.1$\pm$3.1 & $<$109.3\\
ESO264-G036 & 0.023 & 11.35 & 10.6$\pm$1.0 & 6.9$\pm$3.4\\
IRASF10565+2448 & 0.043 & 12.02 & 16.6$\pm$0.6 & $<$24.7\\
ESO264-G057 & 0.017 & 11.08 & 7.3$\pm$0.5 & $<$13.0\\
MCG+07-23-019 & 0.035 & 11.61 & 0.7$\pm$0.1 & $<$1.5\\
CGCG011-076 & 0.025 & 11.37 & 29.9$\pm$0.8 & 20.0$\pm$8.8\\
IC2810 & 0.034 & 11.59 & 5.9$\pm$0.3 & $<$9.1\\
ESO319-G022 & 0.016 & 11.04 & 9.4$\pm$0.3 & 6.4$\pm$2.7\\
NGC3690/IC694 & 0.011 & 11.88 & 409.0$\pm$5.6 & 359.1$\pm$44.5\\
ESO320-G030 & 0.011 & 11.10 & 24.1$\pm$1.2 & $<$35.3\\
ESO440-IG058 & 0.023 & 11.36 & 10.3$\pm$0.8 & $<$21.1\\
IRASF12112+0305 & 0.073 & 12.28 & 4.6$\pm$0.4 & $<$6.9\\
ESO267-G030 & 0.018 & 11.19 & 14.1$\pm$0.7 & $<$20.9\\
NGC4194 & 0.009 & 11.06 & 50.2$\pm$3.1 & $<$77.6\\
IRAS12116-5615 & 0.027 & 11.59 & 42.3$\pm$0.8 & 29.3$\pm$11.6\\
IRASF12224-0624 & 0.026 & 11.27 & 6.2$\pm$0.3 & 6.1$\pm$0.3\\
NGC4418 & 0.007 & 11.08 & 202.3$\pm$12.1 & 198.1$\pm$12.7\\
UGC08058 & 0.042 & 12.51 & 670.0$\pm$2.8 & 665.9$\pm$4.6\\
NGC4922 & 0.024 & 11.32 & 43.6$\pm$0.5 & 39.6$\pm$3.6\\
CGCG043-099 & 0.037 & 11.62 & 9.3$\pm$0.4 & $<$15.1\\
MCG-02-33-098/9 & 0.016 & 11.11 & 12.0$\pm$0.9 & $<$23.1\\
ESO507-G070 & 0.022 & 11.49 & 13.6$\pm$0.7 & $<$20.3\\
IRAS13052-5711 & 0.021 & 11.34 & 7.1$\pm$0.4 & $<$12.5\\
NGC5010 & 0.021 & 11.50 & 29.2$\pm$4.6 & 22.9$\pm$7.2\\
IRAS13120-5453 & 0.031 & 12.26 & 45.3$\pm$1.4 & $<$60.2\\
IC0860 & 0.013 & 11.17 & 6.4$\pm$0.2 & 4.4$\pm$1.8\\
VV250a & 0.031 & 11.74 & 14.1$\pm$1.1 & $<$24.9\\
UGC08387 & 0.023 & 11.67 & 19.5$\pm$0.9 & $<$32.7\\
NGC5104 & 0.019 & 11.20 & 16.1$\pm$0.8 & $<$23.4\\
MCG-03-34-064 & 0.017 & 11.24 & 211.0$\pm$1.3 & 210.3$\pm$1.4\\
NGC5135 & 0.014 & 11.17 & 33.8$\pm$1.8 & $<$45.1\\
ESO173-G015 & 0.010 & 11.34 & 62.4$\pm$2.6 & $<$85.0\\
IC4280 & 0.016 & 11.08 & 6.5$\pm$0.5 & $<$10.8\\
NGC5256 & 0.028 & 11.49 & 14.0$\pm$0.5 & $<$22.1\\
NGC5257/8 & 0.023 & 11.55 & 5.3$\pm$0.3 & $<$8.2\\
UGC08696 & 0.038 & 12.14 & 52.1$\pm$1.8 & 46.6$\pm$5.2\\
UGC08739 & 0.017 & 11.08 & 9.8$\pm$0.5 & $<$13.5\\
ESO221-IG010 & 0.010 & 11.17 & 7.7$\pm$0.9 & $<$19.5\\
NGC5331 & 0.033 & 11.59 & 6.4$\pm$0.4 & $<$11.7\\
NGC5394/5 & 0.012 & 11.00 & 14.3$\pm$1.2 & $<$23.6\\
CGCG247-020 & 0.026 & 11.32 & 10.3$\pm$0.5 & $<$16.4\\
NGC5653 & 0.012 & 11.06 & 7.4$\pm$0.5 & $<$12.2\\
IRASF14348-1447 & 0.082 & 12.30 & 7.4$\pm$0.4 & 5.2$\pm$2.0\\
IRASF14378-3651 & 0.068 & 12.15 & 6.1$\pm$0.4 & $<$8.0\\
NGC5734 & 0.014 & 11.06 & 12.8$\pm$0.6 & $<$17.2\\
VV340a & 0.033 & 11.67 & 13.4$\pm$0.6 & $<$20.4\\
VV705 & 0.041 & 11.89 & 1.9$\pm$0.3 & $<$3.0\\
ESO099-G004 & 0.029 & 11.67 & 15.3$\pm$0.5 & $<$22.9\\
IRASF15250+3608 & 0.055 & 12.02 & 35.6$\pm$2.5 & 34.4$\pm$2.7\\
NGC5936 & 0.013 & 11.07 & 11.8$\pm$0.9 & $<$21.6\\
UGC09913 & 0.018 & 12.21 & 44.7$\pm$1.9 & 34.9$\pm$8.9\\
NGC5990 & 0.013 & 11.06 & 97.2$\pm$0.5 & 89.9$\pm$6.5\\
NGC6052 & 0.016 & 11.02 & 3.6$\pm$0.9 & $<$10.5\\
NGC6090 & 0.030 & 11.51 & 1.5$\pm$0.4 & $<$8.1\\
IRASF16164-0746 & 0.027 & 11.55 & 17.3$\pm$2.2 & $<$24.9\\
CGCG052-037 & 0.024 & 11.38 & 10.5$\pm$0.7 & $<$22.8\\
NGC6156 & 0.011 & 11.07 & 16.0$\pm$0.3 & 12.4$\pm$3.2\\
ESO069-IG006 & 0.046 & 11.92 & 12.4$\pm$1.4 & $<$28.8\\

\enddata
\end{deluxetable*}

\begin{deluxetable*}{l c c r c}
\tablecolumns{5}
\tablewidth{0pt}
\tablenum{1}
\tabletypesize{\footnotesize}
\tablecaption{{\em continued} -- Full sample of infrared galaxies: results of the fit and corrected $f_{6{\rm AGN}}$ fluxes}
\tablehead{
\colhead{Name} & \colhead{$z$} & \colhead{$\log(L_{\rm TIR})$} & \colhead{$\tilde{f}_{6{\rm AGN}}$} & \colhead{$f_{6{\rm AGN}}$}  \\
\colhead{} &\colhead{} &\colhead{$L_{\bigodot}$} &\colhead{mJy} &\colhead{mJy}
}
\startdata
IRASF16399-0937 & 0.027 & 11.56 & 13.5$\pm$0.4 & 9.7$\pm$3.4\\
ESO453-G005 & 0.021 & 11.29 & 3.6$\pm$0.2 & $<$5.1\\
NGC6240 & 0.024 & 11.85 & 73.5$\pm$3.5 & 57.4$\pm$14.7\\
IRASF16516-0948 & 0.023 & 11.24 & 1.7$\pm$0.3 & $<$8.5\\
NGC6286 & 0.019 & 11.32 & 14.3$\pm$0.6 & $<$22.1\\
IRASF17132+5313 & 0.051 & 11.89 & 3.1$\pm$0.2 & $<$4.8\\
IRASF17138-1017 & 0.017 & 11.42 & 5.9$\pm$0.9 & $<$10.5\\
IRASF17207-0014 & 0.043 & 12.39 & 19.6$\pm$1.1 & $<$27.4\\
ESO138-G027 & 0.021 & 11.34 & 12.2$\pm$0.4 & $<$17.4\\
UGC11041 & 0.016 & 11.04 & 8.2$\pm$0.6 & $<$14.3\\
CGCG141-034 & 0.020 & 11.13 & 12.5$\pm$0.5 & $<$17.9\\
IRAS17578-0400 & 0.013 & 11.35 & 8.9$\pm$0.7 & $<$19.1\\
IRAS18090+0130 & 0.029 & 11.58 & 8.5$\pm$0.6 & $<$17.6\\
NGC6621 & 0.021 & 11.23 & 8.9$\pm$0.6 & $<$14.6\\
CGCG142-034 & 0.019 & 11.11 & 9.6$\pm$0.4 & $<$13.4\\
IRASF18293-3413 & 0.018 & 11.81 & 41.1$\pm$2.8 & $<$86.5\\
NGC6670A/B & 0.029 & 11.60 & 9.6$\pm$0.6 & $<$21.3\\
IC4734 & 0.016 & 11.30 & 20.2$\pm$0.8 & $<$28.5\\
NGC6701 & 0.013 & 11.05 & 16.9$\pm$1.0 & $<$26.0\\
ESO593-IG008 & 0.049 & 11.87 & 7.1$\pm$0.5 & $<$12.3\\
NGC6786/UGC11415 & 0.025 & 11.43 & 22.1$\pm$0.7 & $<$29.1\\
IRASF19297-0406 & 0.086 & 12.37 & 7.7$\pm$0.4 & $<$9.9\\
IRAS19542+1110 & 0.065 & 12.04 & 11.1$\pm$0.4 & 9.3$\pm$1.6\\
ESO339-G011 & 0.019 & 11.12 & 28.5$\pm$0.6 & 23.5$\pm$4.5\\
NGC6907 & 0.011 & 11.03 & 7.5$\pm$0.6 & $<$13.0\\
MCG+04-48-002 & 0.014 & 11.06 & 22.6$\pm$1.0 & $<$34.4\\
NGC6926 & 0.020 & 11.26 & 5.3$\pm$0.1 & 4.9$\pm$0.4\\
IRAS20351+2521 & 0.034 & 11.54 & 5.0$\pm$0.5 & $<$10.7\\
CGCG448-020 & 0.036 & 11.87 & 20.2$\pm$0.5 & 14.3$\pm$5.3\\
ESO286-IG019 & 0.043 & 12.00 & 39.5$\pm$1.0 & 37.3$\pm$2.2\\
ESO286-G035 & 0.017 & 11.13 & 8.1$\pm$0.8 & $<$23.6\\
IRAS21101+5810 & 0.039 & 11.75 & 5.6$\pm$0.3 & $<$9.2\\
ESO343-IG013 & 0.019 & 11.07 & 12.9$\pm$0.5 & $<$19.7\\
NGC7130 & 0.016 & 11.35 & 33.4$\pm$0.5 & 27.0$\pm$5.7\\
ESO467-G027 & 0.018 & 11.02 & 3.3$\pm$0.3 & $<$7.0\\
ESO602-G025 & 0.025 & 11.27 & 29.9$\pm$0.7 & $<$40.3\\
UGC12150 & 0.021 & 11.29 & 17.5$\pm$0.8 & $<$25.7\\
ESO239-IG002 & 0.043 & 11.78 & 10.2$\pm$0.3 & $<$13.2\\
IRASF22491-1808 & 0.077 & 12.11 & 3.3$\pm$0.2 & $<$4.5\\
NGC7469 & 0.016 & 11.59 & 171.8$\pm$1.9 & 146.7$\pm$22.3\\
CGCG453-062 & 0.025 & 11.31 & 6.0$\pm$0.3 & $<$9.6\\
ESO148-IG002 & 0.045 & 12.00 & 24.7$\pm$0.4 & 19.1$\pm$4.9\\
NGC7552 & 0.005 & 11.03 & 114.9$\pm$3.5 & $<$156.9\\
IC5298 & 0.027 & 11.54 & 44.1$\pm$1.7 & 40.2$\pm$3.9\\
NGC7591 & 0.017 & 11.05 & 12.2$\pm$0.5 & $<$16.5\\
NGC7592 & 0.024 & 11.33 & 28.4$\pm$0.6 & $<$37.8\\
ESO077-IG014 & 0.042 & 11.70 & 11.2$\pm$0.6 & $<$16.5\\
NGC7674 & 0.029 & 11.50 & 138.5$\pm$1.2 & 137.0$\pm$1.8\\
NGC7679 & 0.017 & 11.05 & 13.6$\pm$1.2 & $<$24.8\\
IRASF23365+3604 & 0.064 & 12.13 & 5.1$\pm$0.3 & 3.5$\pm$1.5\\
MCG-01-60-022 & 0.023 & 11.21 & 5.3$\pm$0.5 & $<$12.8\\
IRAS23436+5257 & 0.034 & 11.51 & 11.4$\pm$0.3 & 8.0$\pm$3.0\\
NGC7752/3 & 0.017 & 11.01 & 5.3$\pm$0.4 & $<$9.5\\
NGC7771 & 0.014 & 11.34 & 17.8$\pm$1.0 & $<$26.8\\
MRK0331 & 0.018 & 11.41 & 22.3$\pm$1.5 & $<$46.9\\
\hline
\vspace{0.2cm}\\
\multicolumn{5}{c}{\small Sources from \citet{tran01}}\\
IRASF00183-7111 & 0.327 & 12.77 & 45.1$\pm$0.6 & 44.9$\pm$0.6\\
IRAS00188-0856 & 0.129 & 12.31 & 11.8$\pm$0.3 & 11.1$\pm$0.7\\
IRAS00275-2859 & 0.279 & 12.46 & 35.8$\pm$0.1 & 35.4$\pm$0.3\\
IRAS00406-3127 & 0.342 & 12.64 & 11.6$\pm$0.3 & 11.5$\pm$0.3\\
IRAS02113-2937 & 0.194 & 12.29 & 1.7$\pm$0.1 & $<$2.3\\
IRASF02115+0226 & 0.400 & 12.48 & 1.2$\pm$0.1 & 1.0$\pm$0.2\\
IRASF02455-2220 & 0.296 & 12.57 & 1.8$\pm$0.1 & 1.6$\pm$0.2\\
IRAS03000-2719 & 0.221 & 12.41 & 2.7$\pm$0.1 & 2.2$\pm$0.5\\
IRAS03538-6432 & 0.310 & 12.65 & 8.9$\pm$0.2 & 8.4$\pm$0.5\\
IRAS03521+0028 & 0.152 & 12.46 & 2.1$\pm$0.1 & 1.5$\pm$0.6\\
IRAS04384-4848 & 0.213 & 12.32 & 3.8$\pm$0.3 & 3.3$\pm$0.5\\
IRAS17463+5806 & 0.309 & 12.48 & 2.2$\pm$0.1 & 1.9$\pm$0.3\\
IRAS18030+0705 & 0.146 & 12.18 & 0.5$\pm$0.1 & $<$2.7\\
IRASF23529-2119 & 0.430 & 12.55 & 8.9$\pm$0.1 & 8.8$\pm$0.1\\

\enddata
\end{deluxetable*}

\clearpage

\begin{deluxetable}{c c c c c c c c c c c c c c}
\tablecolumns{14}
\tablewidth{0pt}
\tablenum{2}
\tabletypesize{\footnotesize}
\tablecaption{Distributions parameters of measurements and simulations}
\tablehead{
\colhead{$\log(L_{\rm TIR})$} & \colhead{} &
\multicolumn{3}{c}{Measurements} & \colhead{} & \colhead{Detect.limit} & \colhead{} &
\multicolumn{3}{c}{Simulations} & \colhead{} & \colhead{} & \colhead{}\\
\cline{3-5}
\cline{9-11}
\colhead{$L_{\odot}$} & \colhead{} & 
\colhead{$m_{\rm det}$} &\colhead{$\sigma_{\rm det}$} & \colhead{d.r.$[\%]$} & \colhead{} & \colhead{mJy} & \colhead{} &
\colhead{$m_{\rm det}$} &\colhead{$\sigma_{\rm det}$} & \colhead{d.r.$[\%]$} & \colhead{} & \colhead{{\itshape{\bfseries m}}} & \colhead{$\boldsymbol{\sigma}$}
}
\startdata
10.9$\div$11.3   & &  -1.9  &  0.6  &  23.3  & & 17.9 & & -2.0  &  0.6  &  23.4  & &  -3.1  &  1.0  \\
11.3$\div$11.7   & &  -1.6  &  0.5  &  26.8  & & 17.5 & & -1.9  &  0.6  &  26.9  & &  -3.0  &  0.9  \\
11.7$\div$12.1   & &  -1.8  &  0.5  &  48.0  & & 10.5 & & -2.0  &  0.6  &  48.2  & &  -2.6  &  0.8  \\
12.1$\div$12.5   & &  -1.7  &  0.6  &  65.2  & & 8.3  & & -1.9  &  0.7  &  65.2  & &  -2.3  &  0.9  \\
12.5$\div$12.9   & &  -1.0  &  0.5  & 100.0  & & 0.9  & & -1.0  &  0.5  & 100.0  & &  -1.0  &  0.5  \\
\enddata
\label{AGNresults}
\end{deluxetable}

\begin{deluxetable}{l c c r r c}
\tablecolumns{6}
\tablewidth{0pt}
\tablenum{3}
\tabletypesize{\footnotesize}
\tablecaption{High redshift sample of infrared galaxies: results of the fit and corrected $f_{6{\rm AGN}}$ fluxes}
\tablehead{
\colhead{Name} & \colhead{$z$} & \colhead{$\log(L_{\rm TIR})$} & \colhead{$\tilde{f}_{6{\rm AGN}}$} &\colhead{$\tilde{f}_{6.2{\rm peakSB}}$} & \colhead{$f_{6{\rm AGN}}$}  \\
\colhead{} &\colhead{} &\colhead{$L_{\bigodot}$} &\colhead{mJy} &\colhead{mJy} &\colhead{mJy}
}
\startdata
\multicolumn{6}{c}{\small Sources from \citet{brand08}} \\
70Bootes1 & 0.50 & 12.15 & 0.21$\pm$0.04 & 0.30$\pm$0.08 & 0.18$\pm$0.05\\
70Bootes2 & 0.37 & 11.79 & 0.12$\pm$0.04 & 0.50$\pm$0.07 & $<$0.12\\
70Bootes3 & 0.99 & 13.05 & 0.15$\pm$0.04 & 1.10$\pm$0.08 & $<$0.25\\
70Bootes4 & 0.98 & 12.95 & 0.14$\pm$0.04 & 1.15$\pm$0.07 & $<$0.14\\
70Bootes5 & 1.21 & 13.24 & 1.98$\pm$0.08 & 0.28$\pm$0.14 & 1.96$\pm$0.08\\
70Bootes6 & 0.94 & 13.01 & 1.81$\pm$0.07 & 0.25$\pm$0.12 & 1.79$\pm$0.07\\
70Bootes7 & 0.66 & 12.61 & 0.32$\pm$0.04 & 1.30$\pm$0.07 & $<$0.43\\
70Bootes8 & 0.81 & 12.88 & 0.74$\pm$0.06 & 0.30$\pm$0.10 & 0.71$\pm$0.06\\
70Bootes9 & 0.67 & 12.81 & 0.23$\pm$0.05 & 2.38$\pm$0.09 & $<$0.42\\
70Bootes10 & 0.51 & 12.41 & 0.72$\pm$0.06 & 0.29$\pm$0.10 & 0.69$\pm$0.07\\
70Bootes11 & 0.48 & 12.49 & 2.01$\pm$0.07 & 0.31$\pm$0.13 & 1.98$\pm$0.08\\
\hline
\vspace{0.2cm}\\
\multicolumn{6}{c}{\small Sources from \citet{pope08}}\\
C3 & 1.88 & 12.78 & 0.15$\pm$0.03 & 0.10$\pm$0.05 & 0.14$\pm$0.03\\
GN39 & 1.98 & 12.70 & 0.08$\pm$0.04 & 0.54$\pm$0.07 & $<$0.08\\
GN07 & 1.99 & 12.84 & 0.09$\pm$0.03 & 0.30$\pm$0.06 & $<$0.09\\
GN06 & 2.00 & 12.81 & 0.06$\pm$0.03 & 0.51$\pm$0.05 & $<$0.06\\
C1 & 2.01 & 12.98 & 0.48$\pm$0.07 & 0.30$\pm$0.12 & 0.45$\pm$0.07\\
GN26 & 1.23 & 12.56 & 0.05$\pm$0.02 & 0.81$\pm$0.03 & $<$0.14\\
GN17 & 1.73 & 12.30 & 0.17$\pm$0.03 & 0.39$\pm$0.06 & 0.13$\pm$0.05\\
GN19 & 2.48 & 13.08 & 0.13$\pm$0.04 & 0.26$\pm$0.06 & $<$0.13\\
GN04 & 2.55 & 12.82 & 0.24$\pm$0.04 & 0.15$\pm$0.08 & 0.23$\pm$0.05\\
\hline
\vspace{0.2cm}\\
\multicolumn{6}{c}{\small Sources from \citet{menendez07}}\\
SMMJ030228+000654 & 1.41 & 13.44 & 0.07$\pm$0.03 & 0.27$\pm$0.05 & $<$0.12\\
SMMJ163639+405636 & 1.50 & 12.81 & 0.09$\pm$0.02 & 0.17$\pm$0.04 & 0.07$\pm$0.03\\
SMMJ163650+405735 & 2.38 & 13.52 & 0.41$\pm$0.05 & 0.33$\pm$0.07 & 0.38$\pm$0.06\\
\hline
\vspace{0.2cm}\\
\multicolumn{6}{c}{\small Sources from \citet{valiante07}}\\
SMMJ00266+1708 & 2.73 & 12.70 & 0.15$\pm$0.05 & 0.30$\pm$0.09 & $<$0.15\\
SMMJ02399-0136 & 2.81 & 13.40 & 0.42$\pm$0.04 & 0.21$\pm$0.07 & 0.40$\pm$0.04\\
SMMJ09429+4659 & 2.38 & 13.80 & 0.08$\pm$0.04 & 0.43$\pm$0.07 & $<$0.08\\
SMMJ09431+4700 & 3.36 & 12.95 & 0.58$\pm$0.04 & 0.29$\pm$0.07 & 0.55$\pm$0.04\\
SMMJ10519+5723 & 2.67 & 13.07 & 0.11$\pm$0.03 & 0.14$\pm$0.05 & 0.10$\pm$0.03\\
SMMJ10521+5719 & 2.69 & 13.25 & 0.13$\pm$0.03 & 0.13$\pm$0.05 & 0.11$\pm$0.03\\
MMJ154127+6616 & 2.79 & 12.95 & 0.14$\pm$0.01 & 0.19$\pm$0.02 & 0.12$\pm$0.02\\
SMMJ16369+4057 & 1.21 & 12.23 & 0.15$\pm$0.02 & 0.23$\pm$0.03 & 0.12$\pm$0.03\\
SMMJ16371+4053 & 2.38 & 13.04 & 0.14$\pm$0.03 & 0.18$\pm$0.05 & 0.12$\pm$0.03\\
\enddata
\label{highz_AGNdata}
\end{deluxetable}

\begin{deluxetable}{c c c c c c c c}
\tablecolumns{8}
\tablewidth{0pt}
\tablenum{4}
\tabletypesize{\normalsize}
\tablecaption{Evolution models}
\tablehead{
\colhead{Model} & \colhead{Description} & 
\colhead{$n_1$} &\colhead{$n_2$} & \colhead{$z_1$} & \colhead{$m_1$} & \colhead{$m_2$} &\colhead{$z_2$} 
}
\startdata

M1   & SB                  &   3.6  &  0.0  &  2.0  &  1.0  &   0.0  &  2.0   \\
M2   & SB$+$AGN            &   3.6  &  0.0  &  2.0  &  1.0  &   0.0  &  2.0   \\
M3   & SB$+$AGN            &   3.4  & -1.0  &  2.3  &  2.0  &  -1.5  &  1.0   \\
M4   & SB$+$AGN$+\,L-T$ evolution &   3.4  & -1.0  &  2.3  &  2.0  &  -1.5  &  1.0   \\
M5   & SB$+$AGN$+\,L-T$ evolution$+$``cold spread''&   3.4 &  -1.0 &  2.3  &  2.0  &  -1.5  &  1.0   \\
M6   & SB$+$``cut'' AGN$+\,L-T$ evolution$+$``cold spread''&   3.4  & -1.0  &  2.3  & 2.0 &  -1.5  &  1.0   \\

\enddata
\label{tab:models}
\end{deluxetable}

\clearpage

\clearpage

\end{document}